\documentclass[aps, pra, a4paper, amssymb, reprint, showpacs,superscriptaddress,footinbib]{revtex4-1}

\usepackage{amssymb}
\usepackage{graphicx}
\usepackage{amsmath}
\usepackage[colorlinks=true,linkcolor=blue,citecolor=blue]{hyperref}
\usepackage{bbold}
\usepackage{xcolor}

\newcommand{\sabs}[1]{|#1|}
\newcommand{\ket}[1]{| #1 \rangle}
\newcommand{\bra}[1]{\langle #1|}
\newcommand{\eff}{\textrm{eff}}

\newcommand{\hc}{\textrm{h.c.}}
\newcommand{\rmL}{\textrm{L}}
\newcommand{\f}{\textrm{f}}
\newcommand{\fib}{\textrm{fib}}
\newcommand{\dB}{\textrm{dB/km}}
\newcommand{\cav}{\textrm{cav}}
\newcommand{\om}{\omega}
\newcommand{\FSR}{\textnormal{FSR}}
\newcommand{\figref}[1]{Fig.\,\ref{#1}}
\newcommand{\tabref}[1]{Table \,\ref{#1}}
\newcommand{\eeqref}[1]{Eq.\,\eqref{#1}}
\newcommand{\eqsref}[1]{Eqs.\,\eqref{#1}}
\newcommand{\secref}[1]{Sec.\,\ref{#1}}
\newcommand{\appref}[1]{Appendix \ref{#1}}

\newenvironment{psmallmatrix}{\left(\begin{smallmatrix}}{\end{smallmatrix}\right)}
  
\begin{document}
\date{\today}

\title{Deterministic quantum state transfer between remote qubits in cavities}

\author{B. Vogell}
\affiliation{Institute for Theoretical Physics, University of Innsbruck, A-6020 Innsbruck, Austria}
\affiliation{Institute for Quantum Optics and Quantum Information of the Austrian Academy of Sciences, A-6020 Innsbruck, Austria}
\author{B. Vermersch}
\affiliation{Institute for Theoretical Physics, University of Innsbruck, A-6020 Innsbruck, Austria}
\affiliation{Institute for Quantum Optics and Quantum Information of the Austrian Academy of Sciences, A-6020 Innsbruck, Austria}
\author{T. E. Northup}
\affiliation{Institute for Experimental Physics, University of Innsbruck, A-6020 Innsbruck, Austria}
\author{B. P. Lanyon}
\affiliation{Institute for Quantum Optics and Quantum Information of the Austrian Academy of Sciences, A-6020 Innsbruck, Austria}
\author{C. A. Muschik}
\affiliation{Institute for Theoretical Physics, University of Innsbruck, A-6020 Innsbruck, Austria}
\affiliation{Institute for Quantum Optics and Quantum Information of the Austrian Academy of Sciences, A-6020 Innsbruck, Austria}

\begin{abstract}
Performing a faithful transfer of an unknown quantum state is a key challenge for enabling quantum networks. The realization of networks with a small number of quantum links is now actively pursued, which calls for an assessment of different state transfer methods to guide future design decisions. Here, we theoretically investigate quantum state transfer between two distant qubits, each in a  cavity, connected by a waveguide, e.g., an optical fiber. We evaluate the achievable success probabilities of state transfer for two different protocols: standard wave packet shaping and adiabatic passage. The main loss sources are transmission losses in the waveguide and absorption losses in the cavities. 
While special cases studied in the literature indicate that adiabatic passages may be beneficial in this context, it remained an open question under which conditions this is the case and whether their use will be advantageous in practice. We answer these questions by providing a full analysis, showing that state transfer by adiabatic passage -- in contrast to wave packet shaping --  can mitigate the effects of undesired cavity losses, far beyond the regime of coupling to a single waveguide mode and the regime of lossless waveguides, as was proposed so far. 
Furthermore, we show that the photon arrival probability is in fact bounded in a trade-off between losses due to non-adiabaticity and due to coupling to off-resonant waveguide modes. 
We clarify that neither protocol can avoid transmission losses and discuss how the cavity parameters should be chosen to achieve an optimal state transfer.
\end{abstract}

\maketitle

%%%%%%%%%%%%%%%%%%%%%%%%%%%%%%%%%%%%
%% Introduction %%
%%%%%%%%%%%%%%%%%%%%%%%%%%%%%%%%%%%%

\section{Introduction}
\label{sec:introduction}

%% Quantum Networks %%%%
The ability to faithfully transmit an unknown quantum state between remote locations is a key primitive for the development of various quantum technologies. The quest to create {\it long-distance links} that can connect multiple nodes into a quantum internet~\cite{KimbleReview,TracyReview,Manifesto} is motivated by applications such as unconditionally secure communication~\cite{Bruss2000,Lo2014}, distributed quantum computing~\cite{Beals2013}, quantum fingerprinting~\cite{QuFingerprinting1,QuFingerprinting2}, quantum credit cards~\cite{QuCreditCards}, quantum secret voting~\cite{QuSecretVoting}, quantum secret sharing~\cite{QSecretSharing}, secure quantum cloud computing~\cite{QuCloudComputing1,QuCloudComputing2}, quantum time and frequency metrology~\cite{Komar14}, and tests of the foundations of quantum physics~\cite{BellTest1,BellTest2}. 
{\it Short-distance links} acting as `quantum USB cables', on the other hand, allow the connection of different types of quantum hardware and are a promising approach to scalable quantum computing architectures~\cite{ScalableArchitecture1}.

\begin{figure}[h!]
\centering
\includegraphics[width=0.5\textwidth]{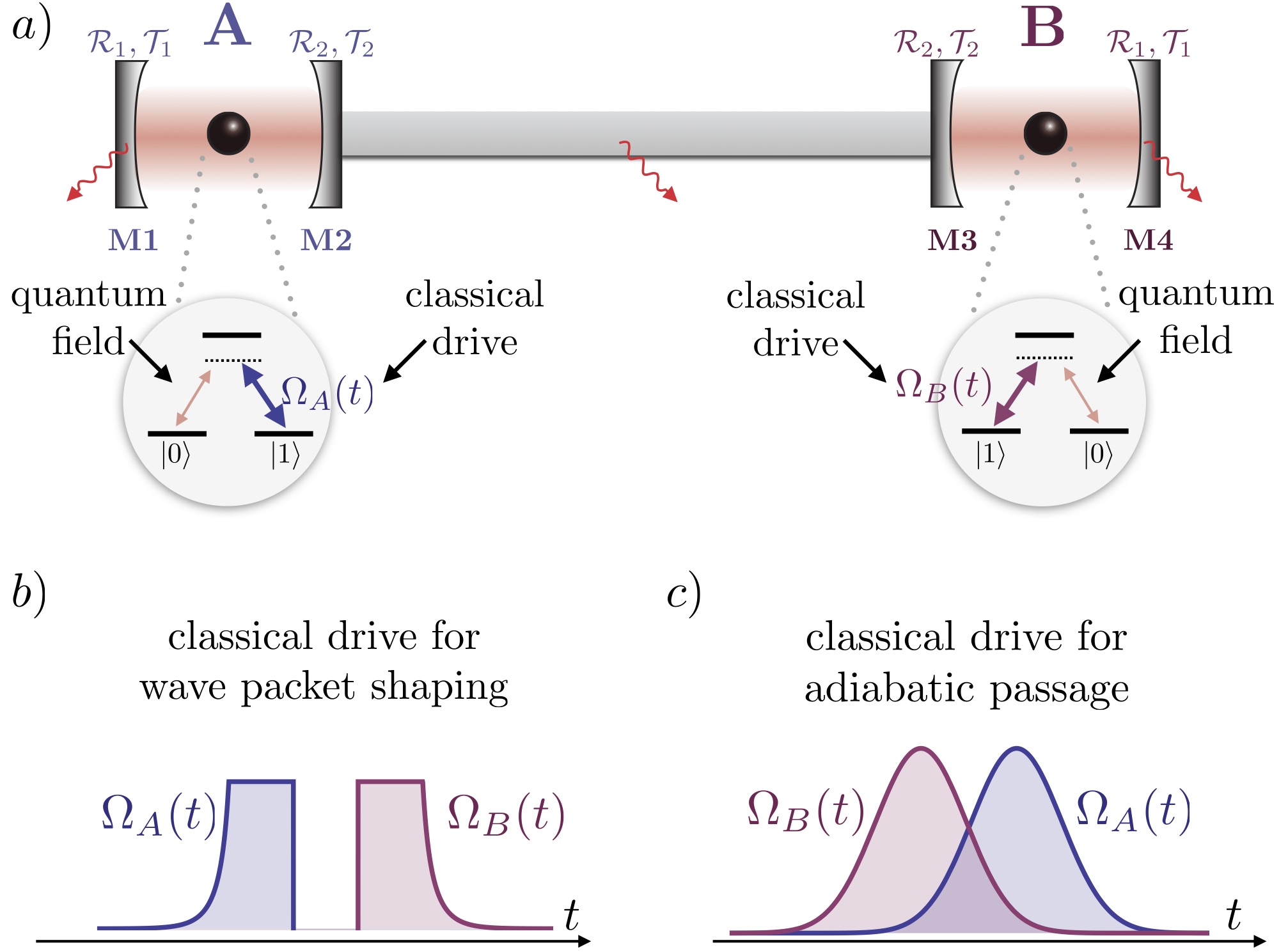}
\caption{Basic quantum network for a deterministic state transfer. (a) Matter qubits ($|0\rangle$, $|1\rangle$) in cavities $A$ and $B$ are coupled via a waveguide. Both cavities are asymmetric with reflectivities $\mathcal{R}_1\gg \mathcal{R}_2$. The qubits consist of off-resonantly driven three level systems, where one transition is coupled to the cavity field, while the other is driven externally by a time-dependent classical drive with Rabi frequency $\Omega_{A,B}(t)$. The classical drive leads to emission of a photon from the qubit into the surrounding cavity.
(b) The temporal shape of classical drive for wave packet shaping is designed such that in the absence of losses, the wave packet created in the waveguide by the first qubit is perfectly absorbed by the second qubit.
(c) The temporal shape of the applied classical drive for adiabatic passage transfer is given by Gaussian pulses with $\Omega_B(t)$ in cavity $B$ being turned on before $\Omega_A(t)$ in cavity $A$.}
\label{fig:mainresults}
\end{figure}

%% Deterministic vs Probabilistic %%%
A quantum state transfer can be accomplished probabilistically or deterministically. Probabilistic state transfer protocols use a two-step procedure: First, an entangled state between two network nodes is generated in a probabilistic and heralded fashion~\cite{Cabrillo99,Chou2005,Moehring07,Hofmann2012,Bernien2013}. Second, this entangled link is used for transferring a quantum state by teleportation~\cite{Olmschenk2009,Bao2012,Krauter2013,Noelleke2013,Pfaff2014}. 
For deterministic state transfer~\cite{detST}, a qubit state at one node is mapped onto a photon wave packet, which propagates across the desired distance and is then absorbed by the qubit in the receiving cavity~\cite{Mabuchi1997,ExpDetTransfer}.  
Deterministic approaches, as discussed here, do not rely on the availability of entangled resource states. Hence, such protocols are particularly well suited for the implementation of time-continuous schemes for quantum information processing~\cite{ContTeleportation,Hofer2013,Vollbrecht2011} and also dispense with the photon counters typically required for probabilistic approaches, e.g.,~\cite{Duan2010}.

%% Elementary Link %%%
We are interested in the simple scenario depicted in \figref{fig:mainresults}a, in which two nodes are connected by a waveguide.  Each node consists of a qubit placed in a cavity.  
In this context, relevant to, e.g., atoms, ions, and superconducting circuits coupled by waveguides, we study the task of transmitting quantum information \textit{deterministically} between the two nodes. 
We focus on evaluating the performance of two deterministic protocols. First, we consider the standard approach based on wave packet shaping~\cite{Mabuchi1997}, in which a classical qubit drive, as depicted in Fig.\ref{fig:mainresults}b, is designed such that the photon emitted by the first qubit is entirely captured by the second qubit, without reflections.
The second approach uses the techniques of adiabatic passage to perform a quantum state transfer~\cite{Pellizzari1997} in the same setup but with classical driving fields 
in a counterintuitive order, in which the receiving drive is turned on before the emitting drive, as shown in \figref{fig:mainresults}c. 
While experiments performing state transfer by wave packet shaping have already been carried out~\cite{ExpDetTransfer}, state transfer between remote nodes by adiabatic passage has yet to be realized experimentally. 

%% Losses: fiber losses & cavity losses are main problems %%%
The central problem for deterministic state transfer protocols is photon loss. Photon losses mainly occur either during transmission in the waveguide or locally in the cavities. Note that even for links spanning hundreds of meters, state-of-the-art cavity setups with mirror absorption losses of only a few parts per million per round-trip nevertheless operate in a `cavity loss' dominated regime (see \tabref{tab:exp}).

%% main results of the paper %%%
%% Long Photon Limit %%%
In this article, we analyze the limitations and prospects for transferring quantum information in the presence of the aforementioned photon losses, leading to two main results: 
First, we show that neither wave packet shaping nor adiabatic passage can mitigate waveguide transmission losses. It has been stated in the literature (e.g., in Refs.~\cite{Serafini2006, Yin2007,Chen2007,Ye2008,Lu2008,Zhou2009,Clader2014,Chen2015,Hua2015,Huang2016}) that waveguide losses can be avoided in the single mode or short-fiber limit~\cite{Pellizzari1997}, in which the cavities couple effectively only to a single mode of the waveguide. 
We show that this is incorrect since the photon arrival probability is bounded by a trade-off between losses due to non-adiabaticity and losses due to coupling to off-resonant waveguide modes. Taking this trade-off into account will be important for optimizing the experimental design parameters of future quantum networks.

Second, we derive an analytical solution of the achievable state transfer success probability for adiabatic passages and provide a full numerical analysis. With this analysis, we show that, in contrast to wave packet shaping, quantum state transfer by adiabatic passage can mitigate losses due to absorption in the cavities far beyond the regime of lossless waveguides introduced in the original proposal~\cite{Pellizzari1997} and the single mode limit introduced in Ref.~\cite{vanEnk1999}.
We show, however, that the single mode limit imposes far stronger constraints on the parameters of the system than is necessary: in order to mitigate cavity losses it is sufficient to be in the `long photon limit', in which the effective length of the photon is longer than the distance between the two nodes of the setup.
%% Concrete Context %%%
The long photon regime is naturally reached for short transmission links and can be realized for distances up to thousands of kilometers, using slowly varying classical driving fields, by state-of-the-art experiments. 

%% Structure of the paper %%%
The paper is organized as follows: First, we provide a brief overview of the setup and the main results in \secref{sec:mainresults}. In \secref{sec:setupprotocol} we describe the setup and the two quantum state transfer protocols under consideration in detail. In \secref{sec:fiberlosses} we treat the influence of waveguide losses; \secref{sec:cavitylosses} then also includes the influence of cavity losses. Finally, we discuss further experimental imperfections in \secref{sec:robustness} and give our conclusion and outlook in \secref{sec:outlook}.

%%%%%%%%%%%%%%%%%%%%%%%%%%%%%%%%%%%%
\section{Overview and main results}
\label{sec:mainresults}
%%%%%%%%%%%%%%%%%%%%%%%%%%%%%%%%%%%%

In the following we provide a brief overview of the setup, the two quantum state transfer protocols considered and the main results. The rest of the paper provides the detailed explanation and derivations of these main results.\\
\\ \textbf{Setup:} 
We consider two emitters (matter qubits) placed in distant cavities $A$ and $B$ that are connected by a waveguide (for example, an optical fiber) of length $L$,  as displayed in \figref{fig:mainresults}a and detailed in \secref{sec:basicmodel}. 
We consider the transfer of a quantum state encoded in the ground state levels of the emitters from cavity $A$ to cavity $B$, 
\begin{align}
\ket{\Psi^{\text{in}}}_{A}\ket{\Psi^{\text{in}}}_{B}&=\left(a_0 |0\rangle_{A}+a_1|1\rangle_{A}\right)\ket{0}_B, \label{eq:statetransfer}\\
\ket{\Psi^{\text{out}}}_{A}\ket{\Psi^{\text{out}}}_{B}&=\ket{0}_A\left(a_0 |0\rangle_{B}+a_1|1\rangle_{B}\right),\notag
\end{align}
where $a_0$ and $a_1$ are the normalized amplitudes and the indices $A$ and $B$ refer to the qubits in cavities $A$ and $B$, respectively.
The state of the emitter qubit is mapped to flying photonic Fock states such that $|0\rangle_A\rightarrow|0\rangle_P$, $|1\rangle_A\rightarrow|1\rangle_P$ (and conversely  $|0\rangle_P\rightarrow|0\rangle_B$, $|1\rangle_P\rightarrow|1\rangle_B$)~\cite{Boozer2007}, where the quantum state with index $P$ refers to the photonic state in both cavity and waveguide.
Note that this setting is not limited to the use of photonic Fock states $|0\rangle_P$, $|1\rangle_P$. Quantum information can also be encoded in the light field using polarization qubits (our results apply to specific types of polarization encoded state transfer protocols, as explained in~\cite{polEncoding}).

We assume here two identical cavities $A$ and $B$ of length $l$, with $l\ll L$. The outer mirrors (M1 and M4 in \figref{fig:mainresults}a), not coupled to the waveguide, have a reflectivity $\mathcal{R}_1$. The two inner mirrors (M2 and M3 in \figref{fig:mainresults}a), adjoined to the waveguide, have a reflectivity $\mathcal{R}_2$. The cavities are asymmetric with reflectivities  $\mathcal{R}_1 \gg \mathcal{R}_2$, such that photons leave predominantly through the inner mirrors.
The rate $\kappa_\cav$ of this desired photon coupling between waveguide and either cavity is proportional to the transmission $\mathcal{T}_2$ of the interfacing mirrors (M2 and M3).
 
We consider waveguide losses parameterized by a loss rate $\gamma_\fib$, and cavity losses at a rate $\gamma_\cav$. The latter rate refers to photons leaving through the outer mirrors (M1 and M4) with transmission $\mathcal{T}_1$ (with $\mathcal{T}_2 \gg \mathcal{T}_1$), and absorption and scattering losses in the mirrors at rate $\mathcal{L}$. The effect of other experimental imperfections such as spontaneous decay of the emitters, timing errors of the classical drive and in- and outcoupling losses due to imperfect coupling of cavity and waveguide will be discussed in \secref{sec:robustness}.\\
\\ \textbf{Classification of the relevant regimes:}
We distinguish between two different regimes, the single mode limit and the long photon regime.
To this end, we introduce two length scales: $L_{\text{eff}}$ refers to the {\it natural spatial length of a photon} that got emitted by a cavity (in the absence of an atom). 
$L_\textrm{ph}$ refers to the {\it spatial length of a photon} that got emitted by a qubit driven by a classical field mediated by a cavity (see inset in \figref{fig:mainresults}a).
The single mode limit refers to the parameter regime in which the cavity linewidth $\kappa$ is much smaller than the free spectral range of the waveguide $\FSR_\fib=\pi c_\f/L$ (with $c_\f$ the speed of light in the waveguide)~ \cite{Pellizzari1997}. This regime is characterized by the single mode parameter
\begin{align}
\mathfrak{n}\equiv\frac{2\kappa}{\FSR_\fib} \ll 1,
\label{eq:SMLp}
\end{align}
see \figref{fig:spectrum}a-b.
Note that this condition is equivalent to $L\ll L_{\text{eff}}$ (short-fiber limit as in Ref.~\cite{vanEnk1999}), where the  natural spatial photon length is defined by $L_{\text{eff}}=c_\f/\kappa$.

We define the long photon limit through two main conditions. First, the desired coupling rate of the cavity to the waveguide $\kappa_\cav$ is assumed to be much larger than the effective coupling of the qubit to the cavity $G_{A/B}$ (defined in \eeqref{eq:ioncavcoupl}) such that the cavities' photon population is always much less than one.
Under this assumption, the cavity can be eliminated, leading to an effective qubit-waveguide coupling rate $\gamma_{A/B}=\kappa_\cav (G_{A/B}/\kappa)^2$ (\secref{sec:WPS}). In analogy to the natural spatial length of the photon $L_{\text{eff}}$ as defined above, the length of the photon $L_{\text{ph}}$ is defined by $L_\textrm{ph}=c_\f/\gamma_{A/B}$.
Second, the length of the photon is assumed to be larger than the link, such that $L_\textrm{ph}\gg L$.

While $L_\eff$ is a fixed quantity for a given setup (see \tabref{tab:exp} for typical values), $L_\textrm{ph}$ can be varied via the effective coupling $G_{A/B}$ between qubit and cavity. 
Current experiments can access the long photon limit by choosing a small amplitude of the effective coupling $G_{A/B}$ and applying the classical driving field for a long exposure time.
In particular, they can reach the regime $L_\textrm{ph}\gg L_\eff$, 
in which they can operate in the long photon limit but not in the single mode limit for a given fiber length $L$. \\ 
\\ \textbf{Quantum state transfer by wave packet shaping:}
The standard protocol for transferring a quantum state deterministically between two cavities is based on wave packet shaping~\cite{Mabuchi1997,ExpDetTransfer, Stannigel2011}.
The main idea behind wave packet shaping is to choose a temporal variation of the classical driving field applied to the atoms in cavities $A$ and $B$ such that in the absence of losses, the photon emitted by the first cavity is perfectly absorbed by the second cavity. This approach avoids the reflection of the photon by the highly reflective mirror M3 of the second cavity due to a quantum interference effect, as studied in~\cite{Gorshkov2007,Fleischhauer2000,Dilley2012}.
For simplicity and concreteness, we discuss a time-symmetric wave packet emitted by the first qubit~\cite{Mabuchi1997} due to a classical coupling $\Omega_A(t)$, which can be reabsorbed by the second qubit under a time-reversed coupling $\Omega_B(t)=\Omega_A(\tau-t)$. Here $\tau=L/c_\f$ is the time delay between the first and the second coupling; see \figref{fig:mainresults}b and \secref{sec:WPS}. Note that the wave packet is not required to be symmetric: any choice of shaping pulses that avoids the reflection of the wave packet from cavity $B$ yields the limitations discussed below.\\
\\ \textbf{Quantum state transfer by adiabatic passage:}
Adiabatic passage as a protocol to transfer a quantum state between two remote qubits in cavities~\cite{Pellizzari1997} uses the methods known from STIRAP in atoms~\cite{RMPAP2016} within the setup shown in \figref{fig:mainresults}a. The principal idea is to perform a coherent transfer through a dark state with respect to the photon fields. This transfer is accomplished by temporally shaping the intensity of the classical driving fields of both atoms with a Gaussian shape in a counterintuitive order; see \figref{fig:mainresults}c and \secref{sec:AP}. 
Importantly, adiabatic passage state transfer has to be performed in the long photon limit.\\
\\ \textbf{Limitations of wave packet shaping:} 
We find that, by using the method of wave packet shaping, the maximal success probability $F$ (formally defined in \secref{sec:modelEOM}) of quantum state transfer is strictly limited by $P_1$ (below), i.e.,  $F\leq P_1$; see \secref{sec:fiberlosses} and \secref{sec:cavitylosses}. Here, $P_1$ is given by
\begin{align}
P_{1}=P_\textrm{out} \,P_{\fib} \, P_\textrm{in},
\label{eq:Ptot}
\end{align}
and denotes the probability of a photon to propagate through a waveguide of length $L$ 
\begin{align}
P_{\fib}=e^{-\gamma_\fib L/c_\f},
\label{eq:Pfib}
\end{align}
multiplied by the probability of a photon being emitted from the cavity into the desired output mode 
\begin{align}
P_\textrm{out}=\frac{\mathcal{T}_2}{\mathcal{T}_1+\mathcal{T}_2+\mathcal{L}},
\label{eq:Pout0}
\end{align}
and being absorbed by the second cavity $P_{\text{in}}$. Due to symmetry reasons, the probability for a photon to enter the second cavity equals the emitting probability, i.e., $P_{\text{out}}=P_{\text{in}}$.\\
\\ \textbf{Waveguide losses:}
It has been stated~\cite{Serafini2006, Yin2007,Chen2007,Ye2008,Lu2008,Zhou2009,Clader2014,Chen2015,Hua2015,Huang2016} that limitations due to waveguide losses can be overcome in the single mode limit. These results are based on a description that takes only a single waveguide mode into account and in which, in analogy to stimulated Raman adiabatic passages (STIRAP), the corresponding success probability of state transfer is given by
\begin{align}
F_{\text{STIRAP}}=\exp\left(-\frac{ \gamma_\fib}{{g_{0}^2 T}} \frac{\pi}{2}\right),
\label{eq:fidelitystirapINTRO}
\end{align}
as detailed in~\appref{app:atomiclimit}. The effective atom-waveguide
coupling is denoted by $g_0$ (see \secref{sec:fiberanalytics}) and the pulse width of the driving laser by $T$ (see \secref{sec:AP}).
In the adiabatic limit $g_{0}^2 T/\gamma_\fib \rightarrow \infty$ the success probability $F_{\text{STIRAP}}$ reaches unity, corresponding to a perfect state transfer.
We provide an analytical example that demonstrates why the coupling to far-detuned waveguide modes can in fact not be neglected. As explained in \secref{sec:fiberanalytics}, including three waveguide modes already leads to non-negligible effects, even deep in the single mode limit. The corresponding amended success probability of state transfer is given by
\begin{align}
F_\textnormal{fib} &=\exp\left(- \frac{\gamma_\fib \pi}{2}\left[\frac{ 1}{{g_0^2 T}}+\frac{{g_0^2 T}}{ \FSR_\fib^2} \right] \right),
\label{eq:fidelity3MLINTRO}
\end{align}
revealing a clear trade-off (see \secref{sec:fiberanalytics} for details). While the first summand in \eeqref{eq:fidelity3MLINTRO} recovers the dependency seen in previous work~\cite{Serafini2006, Yin2007,Chen2007,Ye2008,Lu2008,Zhou2009,Clader2014,Chen2015,Hua2015,Huang2016}, the second summand in \eeqref{eq:fidelity3MLINTRO} arises due to the coupling to detuned waveguide modes. As a result, choosing the adiabatic limit as done in previous work is in fact incompatible with obtaining a high success probability of state transfer. Optimizing $F_\textnormal{fib}$ with respect to $g_0^2 T$  leads to 
\begin{align}
F_\textnormal{fib}^\textnormal{opt}&=\exp\left(- \gamma_\fib \pi/\FSR_\fib \right)=\exp\left(- \gamma_\fib L/c_\f \right)= P_\fib.
\label{eq:fidelityMAXsecII}
\end{align}
These results are also shown numerically for an even larger parameter space, taking a large number of waveguide modes into account (see \secref{sec:fibernumerics}). We show that, in the absence of cavity losses ($P_\textnormal{out}=1$), the success probability of state transfer is strictly limited by $F \leq P_\fib$.\\
\\\textbf{Cavity losses:}
In contrast to wave packet shaping, quantum state transfer by adiabatic passage allows one to outperform limitations due to cavity losses imposed by $P_\textrm{out}$. 
Note that for high-finesse cavities, cavity losses can play a significant role due to the high number of round-trips of the photon. Experimental values of $P_\textnormal{out}$ for current optical setups are given in \tabref{tab:exp}.

It has already been shown that limitations due to $P_{\text{out}}$ can be mitigated for perfect waveguides ($\gamma_\fib=0$)~\cite{Pellizzari1997} and in the single mode limit for imperfect waveguides~\cite{vanEnk1999}. 
In this article, we show that quantum state transfer by adiabatic passage can in fact mitigate cavity losses 
for both $\gamma_\fib>0$ and well beyond the single mode limit; see \secref{sec:cavitylosses} and \figref{fig:cavitylosses}. We find that the parameter regime over which cavity losses can be mitigated is determined by the long photon limit.
The figure of merit determining the maximal success probability of state transfer for a given waveguide with length $L$ and loss rate $\gamma_\fib$ is the probability of the photon to leave the cavity $P_\textrm{out}$, as demonstrated in \secref{sec:cavitylosses}. 
Extending the analytics for waveguide losses only, we show that the success probability of state transfer in the presence of both cavity and waveguide losses is given by
\begin{align}
F_\textnormal{1}&=\exp\left(- \frac{ \pi}{2}\left[\frac{\gamma_\fib}{{g_0^2 T}}+\frac{{\gamma_\fib g_0^2 T }}{ \FSR_\fib^2} +\frac{\tilde{\gamma}_\cav T}{4} \right] \right),\label{eq:PAPunoptsecII}
\end{align}
with effective cavity decay rate $\tilde{\gamma}_\cav $ (see \appref{app:beyondSML}). 
We show (\secref{sec:cavitylossesanalytics}) that the achievable success probability of state transfer by adiabatic passage can be optimized to
\begin{align}
F_\textnormal{AP}\equiv F_1^\textnormal{opt}=\exp\left(- P^{\text{fib}}_{\text{loss}} \sqrt{1+\frac{\pi^2}{2 P^{\text{fib}}_{\text{loss}}} \frac{1-P_\text{out}}{P_\text{out}}}\ \right),
\label{eq:FAPsecII}
\end{align}
where $P^{\text{fib}}_{\text{loss}}=\gamma_\fib L/c_\f$ (see \appref{app:beyondSML}).
As a result, we find (\secref{sec:cavitylossesnumerics} and \figref{fig:cavitylosses}) that the success probability of state transfer by adiabatic passage exceeds $P_1$, i.e., $F_\textnormal{AP}\geq P_1$, and thus adiabatic passage allows for better state transfer performance than wave packet shaping, which is limited by $P_1$, cf. last two columns of \tabref{tab:exp}.

For adiabatic passages, the same state transfer success probability can therefore be obtained using a cm-long slowly emitting cavity with a linewidth of tens of kHz or a $\mu$m-short fast emitting cavity with a linewidth of tens of MHz, as long as their probabilities $P_\textrm{out}$ of emitting the photon into the desired output mode are equal. 
Experimental values for the state transfer probability $F_\textnormal{AP}^{\textnormal{exp}}$ that can be reached by adiabatic passages are given in \tabref{tab:exp}.

\begin{table}[h]
\caption{Overview of a selection of current experimental realizations of ions/atoms coupled to cavities. 
The cavity emission rate into the desired output mode $\kappa_\cav$ is given by  $\kappa_\cav=2\kappa-\gamma_\cav$, where $\kappa$ is the total linewidth of the cavity and $\gamma_\cav$ is the undesirable cavity loss rate. The latter includes mirror absorption losses $\mathcal{L}$ as well as the undesired transmission $\mathcal{T}_1$ of photons through the outer mirrors (M1 and M4 in \figref{fig:mainresults}a). 
$L_\eff=c_\f/\kappa$ is the natural spatial length of a photon leaking out of a cavity, 
$P_\textnormal{out}$ is the probability that a photon in the cavity is emitted into the desired output mode (into the fiber). 
For comparison: the transmission probability of a photon through a telecom fiber with absorption of $0.2$\,dB/km and length of $500$\,m is $P_\fib^{0.2\dB,500\textnormal{m}}=97.7\%$ such that the
total transmission probability of the photon is $P_1^{\textnormal{exp}}=P_\textnormal{out} P_\fib^{0.2\dB,500\textnormal{m}} P_\textnormal{in}$ with $P_\textnormal{in}=P_\textnormal{out}$. The success probability of state transfer by adiabatic passage $F_\textnormal{AP}^{\textnormal{exp}}$ is defined in \eeqref{eq:FAPsecII} and evaluated for the same fiber parameters as $P_1^{\textnormal{exp}}$.
Typical fiber loss rates are $\gamma_\fib^{0.2 \dB}/2 \pi =1.5$\,kHz (telecom wavelengths) and $\gamma_\fib^{3 \dB}/2 \pi= 22$\,kHz (optical wavelengths). }
\begin{center}
    \begin{tabular}{l |c |c | c | c | c|c}
   % \hline
    \bf{Experiment} & $\kappa_\cav/2\pi$& $\gamma_\cav/2\pi$& $L_\eff$& $P_\textnormal{out}$ & $P_1^{\textnormal{exp}}$ & $F_\textnormal{AP}^{\textnormal{exp}}$  \\ 
    & [MHz] &  [MHz]  & [m]  & [\%] & [\%] & [\%]\\ \hline
    Mainz~\cite{Pfister2016} & $4.77$  &  $31.5$ & $1.74$ & $13.2 $   & $1.7$ & $42.1$ \\ \hline
   Innsbruck~\cite{Stute2012} & $ 0.02$  &  $0.08$& $ 636$ & $15.8 $  &$2.4$ & $45.9$ \\ \hline
Paris~\cite{Hunger2010} & $ 19.2$ &  $88.4$& $0.6$  & $17.8 $ &$3.1$  & $48.5$   \\ \hline
 Bonn K~\cite{Steiner2014} & $14.0$  &  $29.5$  & $1.27$ & $32.3 $& $10.2$  & $61.3$  \\ \hline
 Caltech~\cite{Hood1998,vanEnk1999} & $38.2$ &  $43.0$ & $0.8$   & $47.1 $ &$21.6$ & $69.9$   \\ \hline
$\textnormal{MPQ}_1$~\cite{Hamsen2016,Chibani2016} & $2.12$  &  $1.67$ & $15.9$   & $56.0$ & $30.6$ & $74.1$\\ \hline
Bonn M~\cite{Gallego2016,WAlt} & $ 32.2$  &  $16.9$ & $1.3$  & $65.6 $ & $41.1$  & $78.3$  \\ \hline
Aarhus~\cite{Herskind2008} & $ 3.03$  &  $1.22$  & $15.2$ & $71.3 $ & $49.6$  & $80.6$ \\ \hline
Sussex~\cite{Begley2016,MKeller}& $0.45$ &  $0.07$  & $135$   & $87.0 $ & $73.9$ & $87.6$ \\ \hline
$\textnormal{MPQ}_2$~\cite{Reiserer2014} & $4.52$  &  $0.5$ & $ 12.7$  & $90.2$ & $79.5$ & $89.3$ \\ \hline
     \end{tabular}
\end{center}
 \label{tab:exp}
\end{table}

%%%%%%%%%%%%%%%%%%%%%%%%%%%%%%%%%%%%
\section{Setup and Transfer Protocols}
\label{sec:setupprotocol}
%%%%%%%%%%%%%%%%%%%%%%%%%%%%%%%%%%%%

In this section, we provide a detailed description of the setup (\secref{sec:basicmodel}) and the state transfer protocols (\secref{sec:protocols}) under consideration.
In the following, we use the language of optical platforms, considering atoms as matter qubits and an optical fiber as a waveguide. 
Note that our derivations also apply to other platforms such as, e.g., superconducting qubits.

%%%%%%%%%%%%%%%%%%%%%%%%%%%%%%%%%%%%
\subsection{Basic Model}
\label{sec:basicmodel}
%%%%%%%%%%%%%%%%%%%%%%%%%%%%%%%%%%%%
The Hamiltonian of our system consists of two parts: $H_\textnormal{sys}$ describing the coherent interactions  (\secref{sec:modelhamiltonian}) and $H_\textnormal{loss}$ describing the couplings to undesired dissipative channels (\secref{sec:modeldissipation}).

%%%%%%%%%%%%%%%%%%%%%%%%%%%%%%%%%%%%
\subsubsection{Hamiltonian}
\label{sec:modelhamiltonian}
%%%%%%%%%%%%%%%%%%%%%%%%%%%%%%%%%%%%
We model the cavity-fiber-cavity system as three linearly coupled cavities with the field modes being represented by independent annihilation and creation operators of the corresponding cavity or fiber mode. As explained in \appref{app:methodCFC}, we also employed an alternative description for our numerical simulations in which the system is described by the eigenmodes of the cavity-fiber-cavity system. Both descriptions yield the same results in the regime of high finesse cavities and for time scales that are long compared to the round-trip time $2\tau$ of a photon. Throughout the main text, we will use the former choice of basis states.

\begin{figure}[h]
\centering
\includegraphics[width=01.0\columnwidth, angle=0]{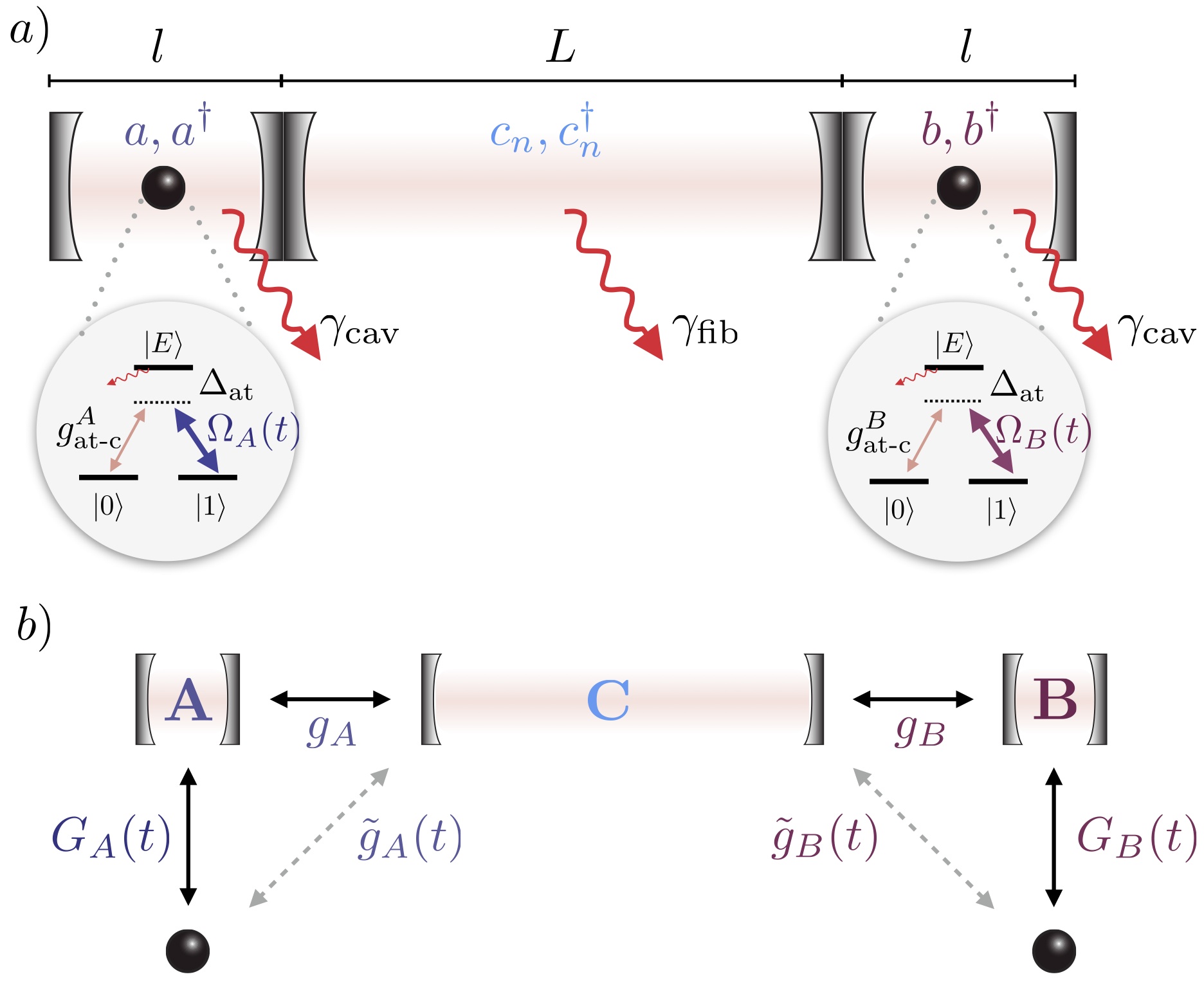}
\caption{Quantum network setup for a deterministic state transfer. (a) 
The cavities $A$ and $B$ (length $l$) are coupled by a fiber $C$ (length $L$). Each cavity contains an atom that is modeled as a three-level system with ground states $\ket{0}$ and $\ket{1}$ and excited state $\ket{E}$. The transition between $\ket{0}$ and $\ket{E}$ is coupled to the cavity field with coupling strength $g_\textnormal{at-c}^{A/B}$, while the transition from $\ket{1}$ to $\ket{E}$ is driven by an external classical time-dependent field with Rabi frequency $\Omega_{A/B}(t)$. Both transitions are coupled off-resonantly with a detuning $\Delta_\textnormal{at}$. 
The rate of photon losses in the fiber and in the cavities is given by $\gamma_{\fib}$ and $\gamma_{\cav}$ respectively.
(b) Coupling scheme: the qubit in cavity $A$ couples with a time-dependent coupling $G_{A}(t)=g_\textnormal{at-c}^{A} \Omega_{A}(t) / \Delta_\textnormal{at}$ to the cavity field, which is coupled to the fiber with coupling strength $g_{A}$. The fiber couples with strength $g_{B}$ to the cavity field in $B$, 
which couples to the second atom with a time-dependent coupling strength $G_{B}(t)=g_\textnormal{at-c}^{B} \Omega_{B}(t) / \Delta_\textnormal{at}$. In the regime in which cavities $A$ and $B$ can be eliminated, the effective coupling strength between atom and fiber is given by $\tilde{g}_{A/B}(t)$ (dotted arrows); see \eeqref{eq:TMLEOM}.}
\label{fig:APM} 
\end{figure}

The full Hamiltonian for the setup under consideration in \figref{fig:APM} is given by
\begin{align}
H=H_\cav+H_\fib+H_\textnormal{cav-fib}+H_\textnormal{at}+H_\textnormal{at-cav}.
\label{eq:Htot}
\end{align}
The Hamiltonian $H_\cav$ describes the bare evolution of both cavities $A$ and $B$ and is given by
\begin{align}
H_\cav=\hbar \omega_0 \left( a^{\dag}a +b^{\dag}b\right),
\label{eq:Hcav}
\end{align}
where the annihilation operator $a$ ($b$) refers to the cavity mode of cavity $A$ ($B$). 
In \eeqref{eq:Hcav} we consider only a single cavity resonance for each cavity, with frequency $\omega_0$ for both cavities. 
The restriction to a single cavity mode is well justified in the limit in which the cavity length is much smaller than the fiber length, $l \ll L$.

The fiber modes are described by the Hamiltonian
\begin{align}
H_\fib=\hbar \sum_{n=-\infty}^{\infty}  \, \omega_n \, c_n^{\dag} c_n,
\label{eq:Hfib}
\end{align}
where the annihilation operator $c_n$ denotes the $n$th fiber mode with frequency $\omega_n=\omega_0+n\cdot \FSR_\fib$. We assume the fiber mode $c_0$ with frequency $\omega_0$ to be resonant with the cavity modes $a$ and $b$, which translates into the condition $L=m\cdot l$ with integer $m$. 
Note that the fiber can alternatively be modeled by using spatially localized modes, allowing for a more intuitive representation of a travelling photon~\cite{Ramos2016}.

\begin{figure}[h]
\centering
\includegraphics[width=0.5\textwidth]{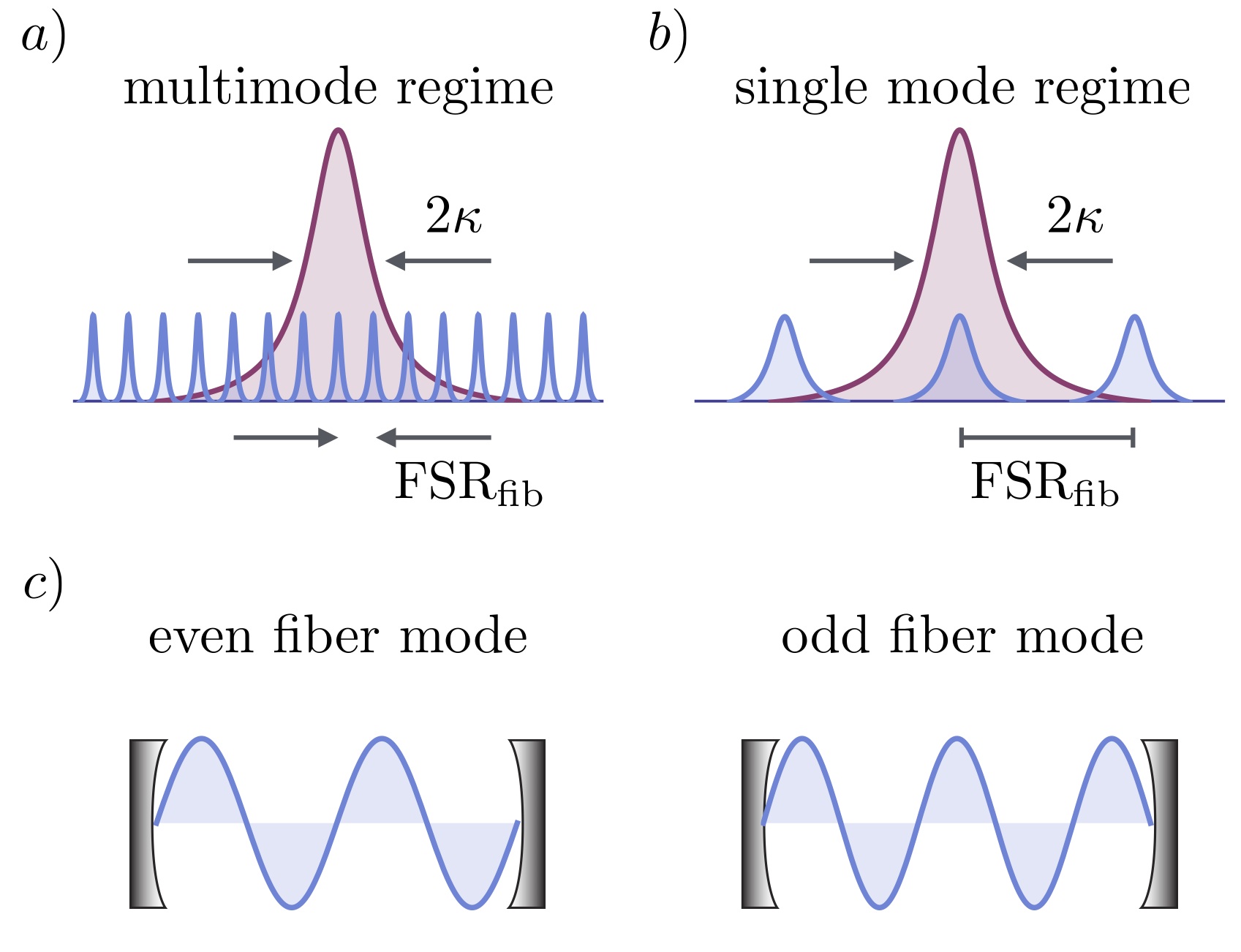}
\caption{Cavity and fiber properties. (a) Spectra of cavity $B$ (magenta) and the fiber (blue) in the multimode limit. Here, the cavity linewidth $2\kappa$ is much larger than the free spectral range of the fiber $\FSR_\fib$. (b) Spectra for cavity and fiber in the single mode limit. The cavity linewidth $2\kappa$ is smaller than the free spectral range of the fiber $\FSR_\fib$ and therefore the cavity effectively couples to only a single fiber mode. (c) Even and odd fiber modes correspond to modes with an even or odd number of maxima in the intensity profile.}
\label{fig:spectrum}
\end{figure}

The interaction Hamiltonian $H_\textnormal{cav-fib}$ is given by the coupling between cavity and fiber modes~\cite{Pellizzari1997}
\begin{align}
H_\textnormal{cav-fib}&=\hbar \sum_{n}\Big[g_{A} \,a^{\dag}+(-1)^n g_{B} \,b^{\dag}\Big]c_n+\hc \ ,
\label{eq:Hcavfib}
\end{align}
 where $\hc$ is the Hermitian conjugate. The coupling strengths $g_{A}$ and $g_{B}$ of the cavity modes $a$ and $b$ to the fiber modes $c_n$ are related to the effective decay rates of the cavities $A$ or $B$ coupled to the fiber given by $\kappa_\cav=2 \pi g_{A/B}^2/\FSR_\fib$ \cite{Vermersch2016}.
For optical implementations, the cavity emission rate $\kappa_\cav$ into the desired output mode is given by $\kappa_\cav \equiv c |\mathfrak{t}|^2/(2 l)$~\cite{vanEnk1999}. Therefore, the coupling strengths between cavity and fiber modes are given by
\begin{align}
g_{A/B}=\sqrt{\frac{\kappa_\cav \FSR_\fib}{2 \pi}}=\sqrt{\frac{c \, c_\f \, |\mathfrak{t}|^2}{4\, L\, l}},
\label{eq:gAB}
\end{align}
where $\sabs{\mathfrak{t}}^2=\mathcal{T}_2$ is the transmission coefficient of the identical inner mirrors (M2 and M3 in \figref{fig:mainresults}) and $c_\f=2 c/3$ is the speed of light in the fiber. 
Note that the coupling $g_{A/B}$ is equally strong for all fiber modes. 
The phase factor $(-1)^n$ in \eeqref{eq:Hcavfib} introduces alternating signs for the coupling to even or odd modes in the fiber. As illustrated in \figref{fig:spectrum}c, even and odd fiber modes correspond to wave functions with an even or odd number of nodes in the intensity profile. 

Each atom is modeled as a three-level system with degenerate ground states $|0\rangle$ and $|1\rangle$ of equal energy and an excited state $\ket{E}$ with energy $\hbar \omega_E$ with respect to the ground states (see \figref{fig:APM}). The bare atomic Hamiltonian is hence given by 
\begin{align}
H_\textnormal{at}=\hbar \omega_E (\ket{E}\bra{E}_A+\ket{E}\bra{E}_B).
\label{eq:Hat}
\end{align}
The transition between the atomic ground state $|0\rangle$ and the excited state $|E\rangle$ is coupled to the cavity field $a$ ($b$) in cavity $A$ ($B$) with coupling strength $g_\textnormal{at-c}^{A}$ ($g_\textnormal{at-c}^{B}$). The transition between the atomic ground state $|1\rangle$ and the excited state $|E\rangle$ is driven by a time-dependent classical field with Rabi frequency $\Omega_{A/B}(t)$ and frequency $\om_\rmL$. Note that the atom-cavity coupling  $g_\textnormal{at-c}^{A/B}$ depends on the mode volume of the cavity and on the position and dipole moment of the atom~\cite{AtomPhotonInt}. 
In order to avoid populating the excited state $|E\rangle$, which suffers from spontaneous emission at rate $\Gamma$, the classical drive and cavity are strongly detuned from the atomic transition, i.e., $\sabs{\Delta_\textnormal{at}}=\omega_\textnormal{L} - \omega_{E} \gg \Gamma$ such that photon loss due to atomic decay is strongly suppressed. The effect of spontaneous emission is discussed in  \secref{sec:robustness}.
Due to the strongly detuned laser drive, the excited state $|E\rangle$ can be eliminated such that the effective atom-cavity interaction Hamiltonian~\cite{Mabuchi1997} is given by 
\begin{align}
H_\textnormal{at-cav}=&\hbar G_A \left( \sigma_A^+ a+ a^\dag \sigma_A^- \right) + \hbar G_B \left( \sigma_B^+ b+ b^\dag \sigma_B^- \right),
\end{align}
where $\sigma^+_{A/B}=|1\rangle\langle0|_{A/B}$ ($\sigma^-_{A/B}=|0\rangle\langle1|_{A/B}$) is the raising (lowering) operator for the qubit in cavity $A/B$.
The effective atom-cavity coupling is given by
\begin{align}
 G_{A/B}(t)= \frac{g_\textnormal{at-c}^{A/B} \Omega_{A/B}(t)}{ \Delta_\textnormal{at}}.
 \label{eq:ioncavcoupl}
\end{align}
After the excited state $\ket{E}$ is eliminated, the bare atomic Hamiltonian $H_\textnormal{at}$ in \eeqref{eq:Hat} vanishes.
Note that eliminating the excited state $\ket{E}$ also results in effective Stark shifts for both ground states $\ket{0}$ and $\ket{1}$, which however can be compensated; see Ref.~\cite{Pellizzari1997}.

The full Hamiltonian given in \eeqref{eq:Htot} can be expressed in an interaction picture with respect to $H_0=\hbar \omega_0( a^{\dag}a + b^{\dag}b +c_0^{\dag}c_0)$ such that
\begin{align} 
H_\textnormal{sys}= \,&\hbar\, G_A(t) \left( \sigma_A^+ a+ a^\dag \sigma_A^- \right) \label{eq:HAPM}  \\
&+\hbar \, \sum_n \,n \, \FSR_\fib \ c_n^{\dag}c_n \notag \\
&+\hbar \, \sum_{n}\Big[g_{A}\, a^{\dag}+(-1)^n g_{B} \, b^{\dag}\Big]c_n+\hc  \notag  \\
&+ \hbar \, G_B(t) \left( \sigma_B^+ b+ b^\dag \sigma_B^- \right). \notag
\end{align}

%%%%%%%%%%%%%%%%%%%%%%%%%%%%%%%%%%%%
\subsubsection{Dissipation}
\label{sec:modeldissipation}
%%%%%%%%%%%%%%%%%%%%%%%%%%%%%%%%%%%%

Here, we discuss the two main sources of imperfection in deterministic state transfer: fiber and cavity losses. The influence of other imperfections will be discussed in \secref{sec:robustness}.

The loss Hamiltonian is given by $H_\textnormal{loss}=V_{\cav,a}+V_{\cav,b}+\sum_n V_{\fib,n}$. 
To model losses in the fiber, we consider each fiber mode to couple in a Markovian way to a bath of bosonic modes with annihilation (creation) operators $\tilde{c}_{\omega,n}$ ($\tilde{c}_{\omega,n}^\dag$) described by the Hamiltonian
\begin{align}
V_{\fib,n}=\sqrt{\frac{\gamma_\fib}{2\pi}} \int \textnormal{d}\omega  \left(c_n^\dag \tilde{c}_{\omega,n}+ \tilde{c}_{\omega,n}^\dag c_{n} \right).
 \label{eq:Vfib}
\end{align}
The fiber loss rate $\gamma_\fib=\alpha c_\f$, where $\alpha$ is the absorption coefficient in the fiber \cite{vanEnk1999}. 
The absorption coefficient $\alpha$ is defined by the fraction absorbed inside a fiber of length $L$
\begin{align}
&\exp(-\alpha L)= 10^{- \frac{X}{10} \cdot \frac{L}{1000}}, \label{eq:fiberloss}\\
&\, \Rightarrow \alpha = X \cdot \ln(10)10^{-4}, \notag
\end{align}
where $X$ is the attenuation coefficient of the fiber in decibels per kilometer. For telecom wavelength fibers, a typical attenuation is $0.2$\,dB/km, yielding a fiber loss rate of $\gamma_\fib^{0.2 \dB}/2\pi=1.5$\,kHz, and for optical wavelengths, a typical attenuation is $3$\,dB/km, with rate $\gamma_\fib^{3 \dB}/2\pi=22$\,kHz.
Note that frequency conversion from optical to telecom wavelengths has been achieved with efficiencies up to $80\%$, e.g., \cite{Pelc2011}.
The probability of a photon to propagate through a fiber of length $L$ is given by $P_\fib$, as defined in \eeqref{eq:Pfib}.

Equivalently, we model cavity losses by considering each cavity $A$ and $B$ to decay to free space, with the interaction given by the coupling of cavity modes $a$  and $b$ to a frequency bath with annihilation (creation) operator $\tilde{a}_\om$ ($\tilde{a}_\om^\dag$) and $\tilde{b}_\omega$ ($\tilde{b}_\omega^\dag$):
\begin{align}
V_{\cav,a}&= \sqrt{\frac{\gamma_\cav}{2\pi}}\int \textnormal{d}\omega \left(a^\dag \tilde{a}_{\omega}+ \tilde{a}_{\omega}^\dag a \right), \label{eq:Vcava}\\
V_{\cav,b}&=\sqrt{\frac{\gamma_\cav}{2\pi}} \int \textnormal{d}\omega  \left(b^\dag \tilde{b}_{\omega}+ \tilde{b}_{\omega}^\dag b \right).\label{eq:Vcavb}
\end{align}
Here, cavity losses at rate $\gamma_\cav$ include the losses through the outer mirrors (M1 and M4 in \figref{fig:mainresults}a) with transmission $\mathcal{T}_1$ as well as absorption losses $\mathcal{L}$ in the cavities. The total linewidth of the cavity $2\kappa$ consists of the rate of coupling into the fiber $ \kappa_\cav$ as well as the total loss rate $\gamma_\cav$ such that
\begin{align}
2\kappa&= \kappa_\cav + \gamma_\cav, \label{eq:kappagoodandbad}\\
&\equiv \frac{c |\mathfrak{t}|^2}{2 l} +  \frac{c |{\ell}|^2}{2 l}, \notag 
\end{align}
where $\gamma_\cav$ contains both transmission and absorption losses: $\sabs{\ell}^2=\mathcal{T}_1+\mathcal{L}$. In \tabref{tab:exp} we summarize the cavity losses of a selection of experiments.
The probability of a photon to be emitted into the desired output mode $P_\textnormal{out}$ as defined in \eeqref{eq:Pout0} can be rephrased as
\begin{align}
P_\textnormal{out}=\frac{\kappa_\cav}{\kappa_\cav+\gamma_\cav}=\frac{\sabs{\mathfrak{t}}^2}{\sabs{\mathfrak{t}}^2+\sabs{\ell}^2},
\label{eq:Pout}
\end{align}
which is equivalent to the probability $P_\textnormal{in}$ of the photon being absorbed by the (second) cavity.\\

Accordingly, we expect the total success probability of a photon transfer between two cavities through a fiber to be limited by 
\begin{align}
P_1&=P_\textnormal{out} P_\fib P_\textnormal{in} =\left(\frac{\sabs{\mathfrak{t}}^2}{\sabs{\mathfrak{t}}^2+\sabs{\ell}^2}\right)^2  \exp\left(-\gamma_\fib \frac{L}{c_\f}\right),
\label{eq:Ploss}
\end{align}
cf. Ref.~\cite{vanEnk1999}.
We use this limit $P_1$ later in \secref{sec:fiberlosses} and \secref{sec:cavitylosses} as a benchmark for the  success probability of state transfer.

%%%%%%%%%%%%%%%%%%%%%%%%%%%%%%%%%%%%
\subsubsection{Equations of Motion}
\label{sec:modelEOM}
%%%%%%%%%%%%%%%%%%%%%%%%%%%%%%%%%%%%

As we are interested in performing a quantum state transfer, we solve the dynamics of the full system according to the Hamiltonian in \eeqref{eq:HAPM}, taking into account the loss mechanisms described  in \eqsref{eq:Vfib}, \eqref{eq:Vcava} and \eqref{eq:Vcavb} using a single-excitation Wigner Weisskopf ansatz. 
The wave function of the full model in this single excitation ansatz is given by
\begin{align}
|\Psi\rangle=&c_{A} |1\rangle_A+ c_{B} |1\rangle_B +c_a |a\rangle+c_b |b\rangle\label{eq:wavefunction}\\
&+\sum_n c_{c_n} |c_n\rangle  + \int \! \textnormal{d} \om \, \left( c_{\tilde{a}_\om} \tilde{a}_\om^\dag +c_{\tilde{b}_\om} \tilde{b}_\om^\dag\right) \ket{\textnormal{vac}} \notag \\
&+ \sum_n \int \!\textnormal{d} \om \, c_{\tilde{c}_{\om,n}} \tilde{c}_{\om,n}^\dag \ket{\textnormal{vac}}+c_\textnormal{vac} \ket{\textnormal{vac}},\notag
\end{align}
where $|1\rangle_{A/B}$ denotes the state of system with the excitation in the atom in cavity $A/B$ with amplitude $c_{A/B}$, $\ket{a/b}$ the state with the excitation in the cavity $A/B$ with amplitude $c_{a/b}$ and $\ket{c_n}$ the state with the excitation in the $n$th fiber mode with amplitude $c_{c_n}$. The sixth and seventh term in \eeqref{eq:wavefunction} describe the baths associated with the cavity and fiber losses as modeled in the previous section, where $\ket{\textnormal{vac}}$ is the vacuum state of light field and $c_{\tilde{x}_\om}$ are the amplitudes of the baths $x\in (a,b,c_n)$. Lastly, the amplitude $c_\textnormal{vac}$ denotes the state of the system without an excitation, i.e., $\ket{\textnormal{vac}}$ corresponds to the state in which both atoms are in the ground state $|0\rangle_{A/B}$, while the cavities and the fiber are empty.

Starting from the Schr\"odinger equation $i\hbar\dot{\ket{\Psi}}=(H_\textnormal{sys}+H_\textnormal{loss} )\ket{\Psi}$, we obtain the time evolution of the amplitudes of the system
\begin{align}
i \dot{c}_{A}&= G_A(t)  c_{a},\label{eq:EOM} \\
i\dot{c}_{a}&=-i \frac{\gamma_\cav}{2}c_{a}+G_A(t)  c_{A}+g_A \sum_n  c_{c_n}, \notag \\
i\dot{c}_{c_{n}}&=-\left(i \frac{\gamma_\fib}{2} -\, n \, \FSR_\fib \right) c_{c_{n}}+ g_A c_{a} +  g_B (-1)^n c_{b}, \notag \\
i\dot{c}_{b}&=-i \frac{\gamma_\cav}{2}c_{b}+G_B(t)  c_{B}+g_B \sum_n (-1)^n c_{c_n}, \notag  \\
i\dot{c}_{B}&= G_B(t) c_{b}, \notag
\end{align}
where the amplitudes of the lossy channels have been intregrated out~\cite{Dorner2002}.
Finally, we solve \eeqref{eq:EOM} for the initial state $\ket{\Psi(t=0)}=\ket{1_A}$ to obtain the success probability of the state transfer, which we define as the probability
\begin{align}
F=\sabs{c_B(t\rightarrow \infty)}^2.
\label{eq:Fidelity}
\end{align}
This probability provides a measure for a successful transmission of the photonic excitation through the setup.
The error $1-F$ denotes the probability to emit a photon into an undesired channel $\tilde{a}_\omega$, $\tilde{b}_\omega$ or $\tilde{c}_{\omega,n}$.
\figref{fig:APM}b illustrates the coupling scheme corresponding to \eeqref{eq:EOM}.  

%%%%%%%%%%%%%%%%%%%%%%%%%%%%%%%%%%%%
\subsection{Quantum State Transfer Protocols}
\label{sec:protocols}
%%%%%%%%%%%%%%%%%%%%%%%%%%%%%%%%%%%%

%%%%%%%%%%%%%%%%%%%%%%%%%%%%%%%%%%%%
\subsubsection{Wave Packet Shaping}
\label{sec:WPS}
%%%%%%%%%%%%%%%%%%%%%%%%%%%%%%%%%%%%

As explained in \secref{sec:mainresults}, one possibility to realize quantum state transfer by wave packet shaping is to produce a time-symmetric photon wave packet inside the fiber by the first combined atom-cavity system such that the back reflection of the wave packet at the inner mirror of the second cavity (M3 in \figref{fig:APM}) is prevented \cite{Mabuchi1997}. 

For the time-symmetric wave packet shaping we consider here, it is essential to use a temporal profile for the classical drive of the first atom that produces a time-symmetric wave packet in the fiber. The classical drive for the second atom is then given by the time-reversed temporal profile with time delay $\tau=L/c_\f$.
We consider the regime in which the maximal coupling between atom and cavity $G_\textnormal{max}\equiv g_\textnormal{at-c}^{A/B} \Omega_{A/B}^\textnormal{max}/\Delta_\textnormal{at}$ is much smaller than the cavity decay rate $\kappa$ (and equal for both cavities). 
In this regime, we can effectively eliminate the cavity, 
which results in an effective coupling rate between the atoms and the fiber modes given by $\gamma_{A/B}(t)=\kappa_\cav (G_{A/B}(t)/\kappa)^2$~\cite{Habraken2012}. 

In this case, a possible classical drive sequence for the atom-fiber coupling rate $\gamma_{A/B}$ is given by Ref.~\cite{Stannigel2011}
\begin{align}
\gamma_A(t)&=
   \begin{cases}
 \gamma_\textnormal{max} \frac{\exp(\gamma_\textnormal{max} t)}{2-\exp(\gamma_\textnormal{max} t)}   & \text{if } t \textless 0 \\
    \gamma_\textnormal{max}  & \text{if } t \ge 0 ,
   \end{cases}
   \label{eq:WPSpulsesequence} \\
\gamma_B(t)&= \gamma_A\left(\tau-t\right),  \notag
\end{align}
where the maximal atom-fiber coupling rate is given by $\gamma_\textnormal{max}=\kappa_\cav (G_\textnormal{max}/\kappa)^2$. This drive sequence generates a time-symmetric wave packet with exponential shape and of length $L_\textnormal{ph}$. 
 Experimentally, the relevant parameter to vary the coupling rate is given by the classical laser drive $\Omega_{A/B}(t)$, as shown in \figref{fig:mainresults}b. This classical drive relates to the effective atom-fiber drive sequence $\gamma_{A/B}(t)$ in \eeqref{eq:WPSpulsesequence} through the expression
 \begin{align}
 \Omega_{A/B}(t)=\frac{\Delta_\textnormal{at} \kappa}{g_\textnormal{at-c}^{A/B}  \sqrt{\kappa_\cav}}  \,\sqrt{ \gamma_{A/B}(t)}.
 \end{align}
Note that the wave packet shaping approach works for both limits, the long photon limit ($L_\textnormal{ph}>L$) and also the limit of short wave packets with $L_\textnormal{ph}<L$.

%%%%%%%%%%%%%%%%%%%%%%%%%%%%%%%%%%%%
\subsubsection{Adiabatic Passage}
\label{sec:AP}
%%%%%%%%%%%%%%%%%%%%%%%%%%%%%%%%%%%%

The general idea of performing quantum state transfer by adiabatic passage is to use the methods known from STIRAP \cite{RMPAP2016} for atoms to perform a coherent transfer by using a dark state with respect to the photon fields~\cite{Pellizzari1997,vanEnk1999}. 

The time-dependent coupling of the atoms to the cavity modes $G_{A/B}(t)$ in \eeqref{eq:HAPM} is varied via the classical laser drive $\Omega_{A/B}(t)$ of the atoms. The classical laser drive of both atoms realizes a counterintuitive pulse sequence~\cite{Vitanov1997b}, in which the classical field in the receiving cavity $B$ is switched on before the driving field of the sending cavity $A$:
\begin{align}
\lim_{t\rightarrow - \infty} \frac{\Omega_A(t)}{\Omega_B(t)}=0, \qquad \lim_{t\rightarrow  \infty} \frac{\Omega_B(t)}{\Omega_A(t)}=0.
\label{eq:adiabaticcondition}
\end{align}
We choose the temporal profiles of both pulses to be Gaussian functions of equal maximal strength $\Omega_\textnormal{max}=\Omega_{A}^\textnormal{max}=\Omega_{B}^\textnormal{max}$ with a retardation $\tau_\textnormal{spl}$ between them
\begin{align}
\Omega_A(t)&= \Omega_{A}^\textnormal{max} \exp\left(-[t-\tau_\textnormal{spl}]^2/T^2\right),   \label{eq:APpulsesequence} \\
\Omega_B(t)&=\Omega_{B}^\textnormal{max} \exp\left(-t^2/T^2\right), \notag
\end{align}
where $T$ is the pulse width, cf. \figref{fig:mainresults}c. In general, the temporal separation of the pulses $\tau_\textnormal{spl}$ is on the order of the pulse width \cite{Vitanov1997b}, such that we introduce the relative temporal separation $x_\textnormal{spl}=\tau_\textnormal{spl}/T$.

Performing a quantum state transfer by adiabatic passage requires the optimization of three parameters: the coupling strength $G^\textnormal{max}_{A/B}$, the temporal width of the classical drive $T$ and the relative temporal separation of the two pulses $x_\textnormal{spl}$. The maximal coupling strength $G^\textnormal{max}_{A/B}= g_\textnormal{at-c}^{A/B} \Omega^\textnormal{max}_{A/B}/ \Delta_\textnormal{at}$ is optimized for fixed atom-cavity coupling $g_\textnormal{at-c}^{A/B}$.

In contrast to wave packet shaping, quantum state transfer by adiabatic passage only works in the long photon limit. The reason lies in the mechanism of adiabatic passage, for which a standing wave of the photon field is required to perform the transfer.

%%%%%%%%%%%%%%%%%%%%%%%%%%%%%%%%%%%%
\section{Fiber Losses}
\label{sec:fiberlosses}
%%%%%%%%%%%%%%%%%%%%%%%%%%%%%%%%%%%%

In this section, we evaluate and discuss the influence of fiber losses on the achievable quantum state transfer success probability by means of a numerical analysis (\secref{sec:fibernumerics}) and an analytical example (\secref{sec:fiberanalytics}). The effect of cavity losses will be addressed later in \secref{sec:cavitylosses}. 

%%%%%%%%%%%%%%%%%%%%%%%%%%%%%%%%%%%%
\subsection{Numerical Treatment}
\label{sec:fibernumerics}
%%%%%%%%%%%%%%%%%%%%%%%%%%%%%%%%%%%%

We numerically study a wide parameter range, including the concrete regime (single mode limit) that has been identified in the literature~\cite{Serafini2006, Yin2007,Chen2007,Ye2008,Lu2008,Zhou2009,Clader2014,Chen2015,Hua2015,Huang2016} as the regime in which limitations $F<P_1$ can be overcome (see \secref{sec:fiberanalytics} and \appref{app:atomiclimit}).
As a result, we find that even deep in the single mode limit, the success probability of the state transfer is always limited by $P_1$ (which is equal to $P_\fib$ for $\gamma_\cav=0$). 
This result holds for both state transfer methods. 

\begin{figure}[h]
\centering
\includegraphics[width=1.0\columnwidth, angle=0]{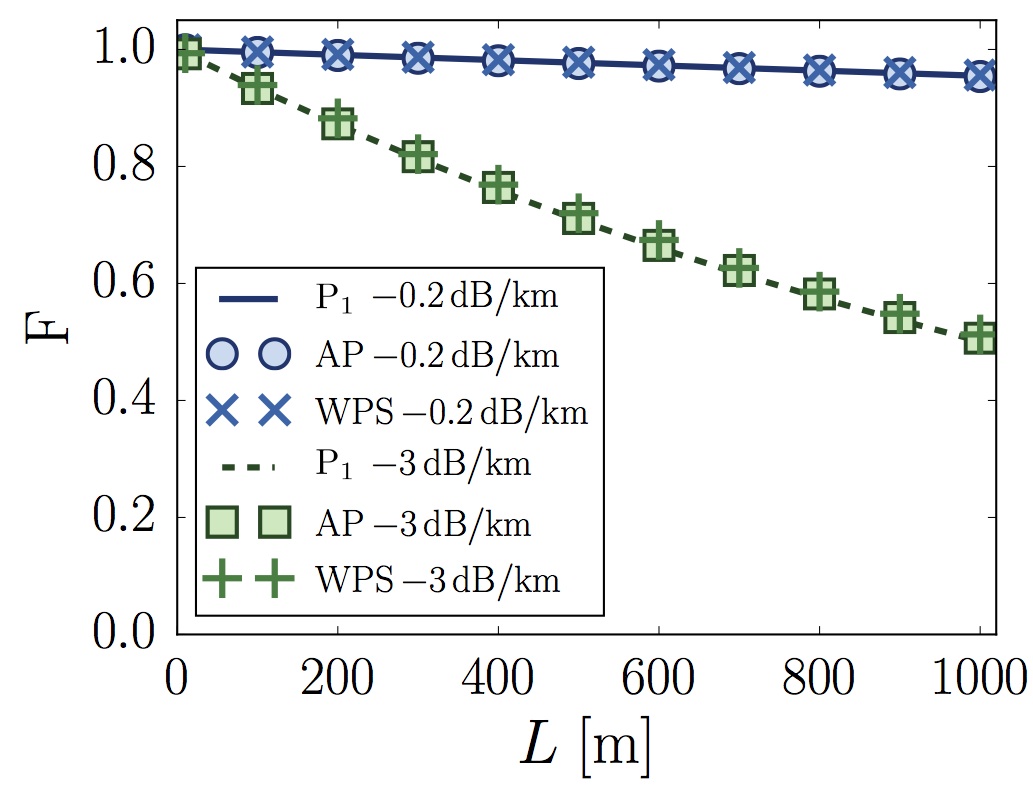}
\caption{State transfer success probability $F$ for two fiber absorption coefficients $0.2$\,dB/km (blue) and $3$\,dB/km (green) with $\gamma_\cav=0$ as a function of the fiber length $L$: the numerical results for an adiabatic state transfer (AP, filled circles and squares) coincide with those for a state transfer via wave packet shaping (SH, tilted crosses and crosses). 
The numerical results agree with the state transfer limit  $P_1$ given by \eeqref{eq:Ploss} in the main text (solid and dashed line). The considered cavity has a length of $l=0.02$\,m and a transmissivity of $\mathcal{T}_2=|\mathfrak{t}|^2=13$\,ppm.} 
\label{fig:fiberlosses} 
\end{figure}

\figref{fig:fiberlosses} provides an example of the state transfer success probability $F$, defined in \eeqref{eq:Fidelity}, as a function of the fiber length $L$ based on the experimental parameters in Ref.~\cite{Stute2012} for $\gamma_\cav=0$ (the effect of cavity losses will be included in \secref{sec:cavitylosses} below). In this example, the identical cavities $A$ and $B$ have a length of $l=0.02$\,m and an inner mirror (M2 \& M3 in \figref{fig:mainresults}) with transmissivity of $|\mathfrak{t}|^2=13$\,ppm. 

We consider two fibers with different loss rates $\gamma_\fib$: first, absorption losses of $0.2$\,dB/km corresponding to fibers at telecom wavelengths and second, absorption losses of $3$\,dB/km corresponding to optical wavelengths. 

In addition, in \figref{fig:fiberlosses} we compare both state transfer methods. 
For every simulation, we ensure that we consider sufficiently many fiber modes in \eeqref{eq:HAPM} that our results converge with respect to the number of fiber modes included.

For wave packet shaping, the state transfer success probability is optimized in the regime $\kappa_\cav \gg G_\textnormal{max}$, in which the cavity can be eliminated~\cite{WPS}. 
In the case of adiabatic passage, every plot point in \figref{fig:fiberlosses} is optimized with respect to the pulse length $T$, the relative temporal separation of the pulses $x_\textnormal{spl}$ and the pulse area $\Omega_\textnormal{max} T$. In particular, the optimized values for adiabatic passage lie in the same regime as those for wave packet shaping, i.e., $\kappa_\cav \gg G_\textnormal{max}$, in which the cavity is barely populated. In order to reach high success probabilities, the pulse area must be large ($\Omega^2_\textnormal{max} T\gg1$), and due to the necessity of a weak coupling $G_\textnormal{max}$, this is achieved for long pulse durations $T$. 
The pulse length is varied in the range $T=(10 \dots 1000) / \kappa_\cav$, which translates for the chosen parameters into  pulse lengths of $T\propto10^{-3}-10^{-5}$\,s.  
The relative separation of the pulses and the coupling strength ratio are optimized in the ranges $x_\textnormal{spl}=(0.8 \dots 2.1)$ and $\Omega_\textnormal{max}/\Delta_\textnormal{at}=(0.0001 \dots 0.1)$.
As a benchmark for the state transfer we plot the expected survival probability $P_1$ of a photon through a lossy fiber as given in \eeqref{eq:Ploss}.

We find that the numerically optimized success probability for both state transfer methods and both fiber absorption losses is in excellent agreement with the limit given by $P_1$ (see \figref{fig:fiberlosses}). For wave packet shaping, this agreement seems natural because, in the optimal case, the photon only passes through the fiber once.
In the case of adiabatic passages, the agreement implies that the state transfer success probability is limited by $P_1$ and therefore fiber losses cannot be overcome. 
Note that the numerical results shown in \figref{fig:fiberlosses} are obtained for a parameter set deep in the single mode limit (see \figref{fig:spectrum}), for which the single mode parameter $\mathfrak{n}$ (defined in \eeqref{eq:SMLp}) for  $L=1000$\,m is $\mathfrak{n}=0.31$. 
Numerical results were however also derived for different parameter sets ($|\mathfrak{t}|^2,|\ell|^2,  l$) beyond the single mode limit. 
We find that for all regimes the quantum state transfer is strictly limited by $P_1$.

%%%%%%%%%%%%%%%%%%%%%%%%%%%%%%%%%%%%
\subsection{Analytical Example}
\label{sec:fiberanalytics}
%%%%%%%%%%%%%%%%%%%%%%%%%%%%%%%%%%%%

It has been stated in the literature~\cite{Serafini2006, Yin2007,Chen2007,Ye2008,Lu2008,Zhou2009,Clader2014,Chen2015,Hua2015,Huang2016} that the Hamiltonian in \eeqref{eq:HAPM} can be simplified in the single mode limit by neglecting all fiber modes but the resonant mode $c_0$. 
This argument is based on the weak coupling assumption $g_{A/B}\ll \FSR_\fib$,
under which the fiber modes $n_{n\neq 0}$ are far detuned from the cavities.
However, as we show below, these off-resonant contributions integrated over long times lead to non-negligible effects.

As detailed in \appref{app:atomiclimit}, the simplified Hamiltonian that includes only a single fiber mode as used in Refs.~\cite{Serafini2006, Yin2007,Chen2007,Ye2008,Lu2008,Zhou2009,Clader2014,Chen2015,Hua2015,Huang2016} leads, in complete analogy to STIRAP in a three-level atomic system, to a success probability of state transfer by adiabatic passage that reaches unity in the adiabatic limit $g_{0}^2 T/\gamma_\fib \rightarrow \infty$,
\begin{align}
F_{\text{STIRAP}}=\exp\left(-\frac{ \gamma_\fib}{{g_{0}^2 T}} \frac{\pi}{2}\right),
\label{eq:fidelitystirap}
\end{align}
where $g_0$ denotes the maximal coupling of atom and fiber.

In the following, we give an illustrative analytical example that points out a crucial difference between adiabatic passages in atoms compared to adiabatic passage state transfer in a fiber. \\ 

\begin{figure}[h]
\centering
\includegraphics[width=01.0\columnwidth, angle=0]{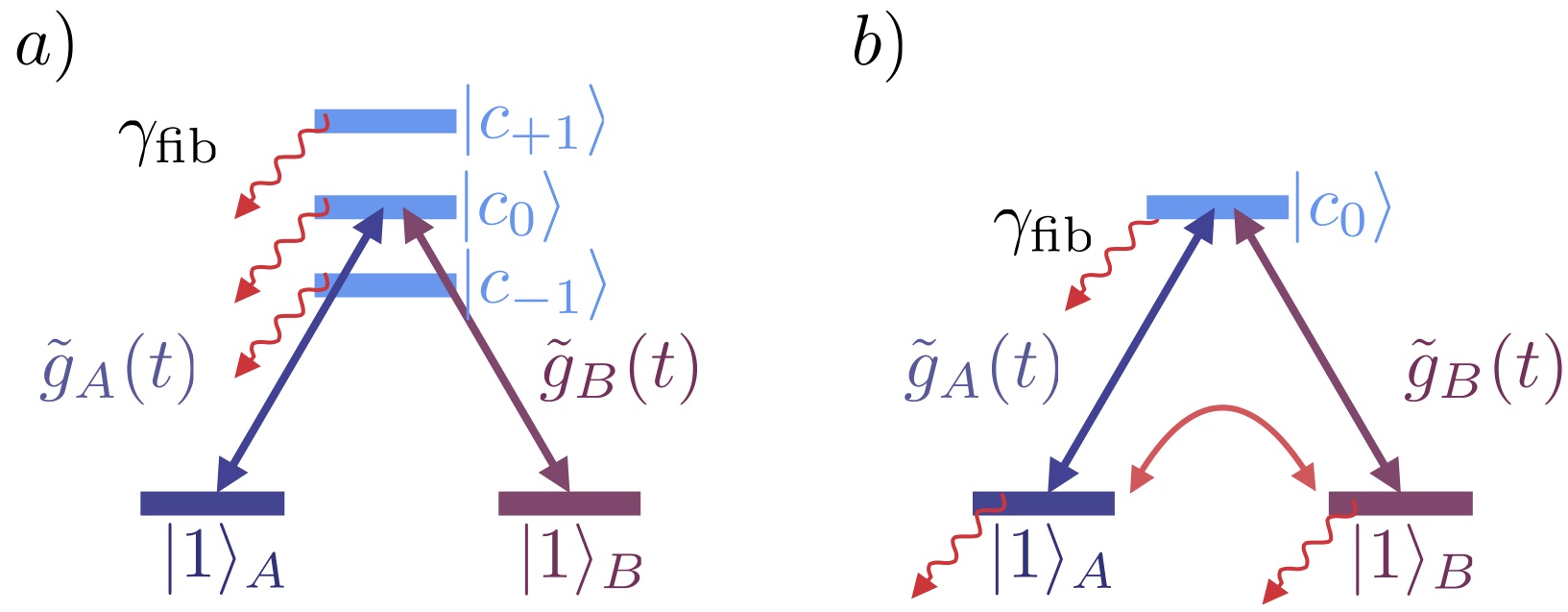}
\caption{Level scheme analogue for adiabatic passage state transfer. 
(a) Effectively coupled atom-fiber-atom description in which the cavities have been eliminated: the ground states $\ket{1}_A$ and $\ket{1}_B$ representing the state of the atom in cavity $A$ and $B$ are effectively coupled with time-dependent coupling strengths $\tilde{g}_{A}(t)$ and $\tilde{g}_{B}(t)$ to the modes of the fiber $c_{0},c_{\pm 1}$.
(b) Effectively coupled atom-`single fiber mode'-atom description, in which the atoms only couple to a single fiber mode $c_0$. The detuned fiber modes $c_{\pm 1}$ have been eliminated, leading to additional ground state dynamics indicated by the red arrows.
The alternating sign $(-1)^n$ is not depicted.} 
\label{fig:APscheme} 
\end{figure}

We consider the regime in which the decay of the cavity $\kappa$ is much larger than the atom-cavity coupling $G_{A/B}$. We focus here on fiber losses and assume therefore $\gamma_\cav=0$ throughout this section.
In this regime, we can adiabatically eliminate the cavity and obtain an effectively coupled atom-fiber-atom system; see \figref{fig:APscheme}a. In \appref{sec:purelyphotonic} we discuss the very similar case of a purely photonic model in which the state of the atom is mapped rapidly onto the cavity, followed by a state transfer in the coupled cavity-fiber-cavity system.

A full description of the problem would include all fiber modes $c_n$ with $n\in(-\infty, +\infty)$. However, the importance of going beyond the single mode description (\appref{app:atomiclimit}), even in the regime in which $g_{A/B}\ll\FSR_\fib$ and $\mathfrak{n}\ll 1$, is well illustrated by including the first pair of far detuned fiber modes $c_{\pm 1}$ in our analytical derivation of the success probability of the adiabatic state transfer.
The equations of motion for the wave function can be derived from \eeqref{eq:EOM} by eliminating the cavity modes. We consider here three fiber modes such that the atom-fiber-atom system is described by
\begin{align}
\dot{c}_{A}&=-i \tilde{g}_A(t) \left( c_{c_0}+c_{c_{-1}} +c_{c_{+1}} \right), \label{eq:TMLEOM} \\
\dot{c}_{c_{-1}}&=-i \tilde{g}_A(t) c_{A}+i \tilde{g}_B(t) c_{B}-\left(\gamma_\fib/2 -i \FSR_\fib \right) c_{c_{-1}},\notag \\
\dot{c}_{c_0}&=-i \tilde{g}_A(t) c_{A}-i \tilde{g}_B(t) c_{B}-(\gamma_\fib/2) c_{c_0},\notag \\
\dot{c}_{c_{+1}}&=-i \tilde{g}_A(t) c_{A}+i \tilde{g}_B(t) c_{B}-\left(\gamma_\fib/2 +i \FSR_\fib \right) c_{c_{+1}},\notag \\
\dot{c}_{B}&=-i \tilde{g}_B(t) \left( c_{c_0}-c_{c_{-1}} -c_{c_{+1}} \right), \notag
\end{align}
where $\tilde{g}_{A/B}(t)= g_{A/B} (G_{A/B}(t)/\kappa)$ is the effective atom-to-fiber coupling strength (see \figref{fig:APM}b).
The level scheme analogue for this specific example is given in \figref{fig:APscheme}a.
As introduced in \secref{sec:basicmodel} even and odd modes (\figref{fig:spectrum}c) couple in \eeqref{eq:TMLEOM} with a different sign to the cavity $B$ due to the factor $(-1)^n$ in \eeqref{eq:HAPM}. 
This sign is the crucial difference with respect to STIRAP in atoms. In atoms, where all excited states couple with the same strength and phase to the ground states, there exists a dark state with respect to the whole manifold of excited states. In our case, however, due to the alternating sign of the coupling in \eeqref{eq:HAPM}, it is impossible to find a dark state with respect to the whole manifold of excited states because a superposition state that is dark with respect to the even modes couples to the odd modes and vice versa~\cite{vanEnk1999}.
It is thus apparent that some fiber modes (even or odd) have to be populated during the state transfer, and that consequently losses due to absorption in the fiber are unavoidable and affect the success probability of the transfer.

We consider here the particular case in which the temporal profiles of the classical driving fields in cavities $A$ and $B$ are given by a sine and a cosine function respectively (see \appref{app:beyondSML} for details). This specific choice allows us to derive a simple analytical expression for the state transfer success probability $F$ as defined in \eeqref{eq:Fidelity}. To this end, we adiabatically eliminate the far detuned modes $c_{\pm 1}$ in \eeqref{eq:TMLEOM} under the assumption $\gamma_\fib^2/4+\FSR_\fib^2\gg \tilde{g}_{A/B}^2$. The presence of $c_{\pm 1}$ leads to effective dynamics of the modes $c_{0}$, $c_A$, and $c_B$ that include loss terms acting on the qubits in cavities $A$ and $B$ (\figref{fig:APscheme}b), resulting in 
\begin{align}
F_\textnormal{fib}
&=\exp\left(- \frac{\gamma_\fib \pi}{2}\left[\frac{ 1}{{g_0^2 T}}+\frac{{g_0^2 T}}{ \FSR_\fib^2} \right] \right),
\label{eq:fidelity3ML}
\end{align}
where the maximal values of the coupling strengths $\tilde{g}_{A/B}(t)$ are chosen to be equal: $ g_0\equiv \textnormal{max}(\tilde{g}_A)=\textnormal{max}(\tilde{g}_B)$.
The first summand in \eeqref{eq:fidelity3ML} yields the result as obtained from STIRAP and hence the naively truncated Hamiltonian (cf. \eeqref{eq:fidelitystirap} and \appref{app:atomiclimit}), resulting in a success probability of unity in the limit $g_0^2 T\gg \gamma_\fib$.
However, the second summand, which is due to the presence of the far detuned fiber modes $c_{\pm 1}$, compensates for this effect. More specifically, increasing $g_0^2 T$ also increases the effect of the effective decay terms on the qubit ground states. Due to this trade-off, there is an optimal value $g_0^2 T = \FSR_\fib$ which balances the effects of non-adiabaticity (first summand) and the effects due to the coupling to the off-resonant fiber modes (second summand), leading to 
\begin{align}
F_\textnormal{fib}^\textrm{opt}&=\exp\left(- \gamma_\fib \pi/\FSR_\fib \right)=\exp\left(- \gamma_\fib L/c_\f \right),
\label{eq:fidelityMAX}
\end{align}
which coincides with the transmission probability $P_\fib$ of a photon through a fiber of length $L$. This result also agrees with our numerical simulations in \secref{sec:fibernumerics}. 
 
%%%%%%%%%%%%%%%%%%%%%%%%%%%%%%%%%%%%
\section{Cavity Losses}
\label{sec:cavitylosses}
%%%%%%%%%%%%%%%%%%%%%%%%%%%%%%%%%%%%

In this section, we show that in contrast to the restrictions due to fiber losses, the problem of cavity losses, which limits wave packet shaping, can be overcome to a significant extent in current experimental settings by using adiabatic passages. We derive an approximate analytical solution for the state transfer success probability that can be achieved by performing adiabatic state transfers (\secref{sec:cavitylossesanalytics}) and provide a numerical analysis for both methods (\secref{sec:cavitylossesnumerics}).

%%%%%%%%%%%%%%%%%%%%%%%%%%%%%%%%%%%%
\subsection{Approximate Analytical Treatment for Adiabatic State Transfers}
\label{sec:cavitylossesanalytics}
%%%%%%%%%%%%%%%%%%%%%%%%%%%%%%%%%%%%

In the following, we extend the analytical example provided in \secref{sec:fiberanalytics}, which models a quantum state transfer by adiabatic passage for $\gamma_\cav=0$,  to cover the effect of both cavity and fiber losses. 
To this end, we adiabatically eliminate the cavity modes in \eeqref{eq:EOM} in the limit $\kappa\gg G_{A/B}$. As in the previous section, we only include the fiber mode resonant with the cavity $c_0$ and the first pair of detuned fiber modes $c_{\pm}$. The resulting equations of motion describe the interaction between the qubits and the fiber modes with an effective coupling strength $\tilde{g}_{A/B}(t)= g_{A/B} (G_{A/B}(t)/\kappa)$ as shown in \figref{fig:APM}b and are given by \eeqref{eq:TMLEOM}, with the first and last equation modified to
\begin{align}
\dot{c}_{A}&=-i \tilde{g}_A(t) \left( c_{c_0}+c_{c_{-1}}+c_{c_{+1}}\right) -\frac{\tilde{\gamma}^{A}_\cav(t)}{2} c_{A}, \label{eq:TMLEOMCAV} \\
\dot{c}_{B}&=-i \tilde{g}_B(t) \left( c_{c_0}-c_{c_{-1}} -c_{c_{+1}} \right)-\frac{\tilde{\gamma}^{B}_\cav(t)}{2} c_{B}. \notag
\end{align}
In this description, cavity losses lead to an effective decay that acts on the qubits with rate $\tilde{\gamma}^{A/B}_\cav(t)=\gamma_\cav  (G_{A/B}(t)/\kappa)^2$. Using these equations of motion, 
and assuming classical driving fields of sine- and cosine shape (see \secref{sec:fiberanalytics} and \appref{app:beyondSML}), the initially complex problem involving time-dependent couplings and decay rates can be cast into a simpler form. This simplified description allows us to derive an approximate solution of the success probability of state transfer by adiabatic passage. As detailed in \appref{app:beyondSML}, this solution takes the form%
\begin{align}
F_\textnormal{1}&=\exp\left(- \frac{ \pi}{2}\left[\frac{\gamma_\fib}{{g_0^2 T}}+\frac{{\gamma_\fib g_0^2 T }}{ \FSR_\fib^2} +\frac{\tilde{\gamma}_\cav T}{4} \right] \right).\label{eq:PAPunopt}
\end{align}
By using the definitions of $\tilde{\gamma}_\cav$ and $g_0$ given above along with \eeqref{eq:gAB}, 
we can optimize \eeqref{eq:PAPunopt} with respect to the pulse length $T$, resulting in
\begin{align}\label{eq:PAP}
F_\textnormal{AP}&\equiv F_1^\textnormal{opt}=\exp\left(- \frac{ \gamma_\fib L}{c_\f}\sqrt{1+ \frac{\pi^2 c_\f}{2\gamma_\fib L} \frac{\left(1-P_\textnormal{out}\right)}{P_\textnormal{out}} } \right).
\end{align}
We find that this analytical expression agrees well with the full numerical simulation of the achievable state transfer success probability presented in the following \secref{sec:cavitylossesnumerics} (see \figref{fig:cavitylosses} and \figref{fig:Fall1MMLvsFall2SML}). Note that in the case of vanishing cavity losses ($\gamma_\cav=0$, i.e., $P_\textnormal{out} =1$),  \eeqref{eq:PAP}  reduces to \eeqref{eq:fidelityMAX}.

\begin{figure}[h]
\centering
\includegraphics[width=0.80\columnwidth, angle=0]{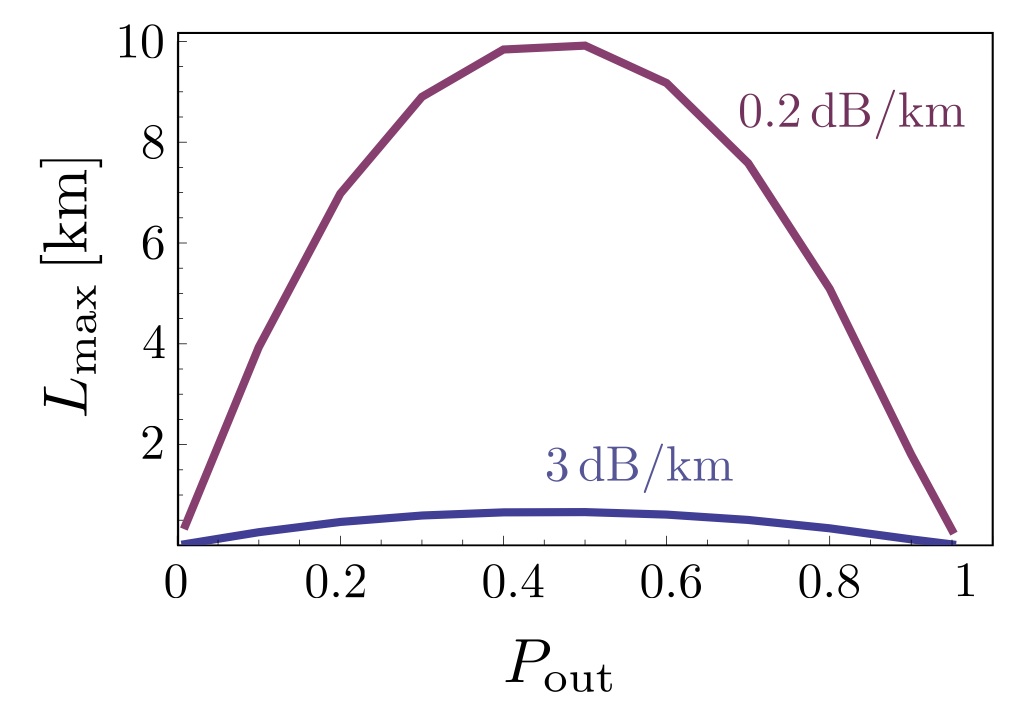}
\caption{Fiber length $L_\text{max}$ up to which the adiabatic state transfer success probability $F_\text{AP}$ given in \eeqref{eq:PAP} exceeds the state transfer success probability achievable by wave packet shaping $P_1$ by more than $5$\%. $L_\text{max}$ is shown as a function of the photon outcoupling probability $P_\textnormal{out}$ given in \eeqref{eq:Pout0} for two fiber attenuation coefficients, $0.2$\,dB/km (magenta) and $3$\,dB/km (purple).} 
\label{fig:Lmax} 
\end{figure}

In \figref{fig:Lmax} we display the maximal length $L_\text{max}$ for which the state transfer success probability achievable by adiabatic passage $F_\text{AP}$ given by \eeqref{eq:PAP} surpasses the state transfer success probability achievable by wave packet shaping $P_1$ given by \eeqref{eq:Ploss} by more than $5$\%, i.e., $F_\text{AP}(L_\text{max})=P_1+0.05$, as a function of $P_\text{out}$.
We plot the results for the maximal length $L_\text{max}$ for two different fiber attenuation coefficients, $0.2$\,dB/km and $3$\,dB/km.

%%%%%%%%%%%%%%%%%%%%%%%%%%%%%%%%%%%%
\subsection{Numerical Treatment}
\label{sec:cavitylossesnumerics}
%%%%%%%%%%%%%%%%%%%%%%%%%%%%%%%%%%%%

In the ideal case in which fiber losses are absent, adiabatic passages allow one to bypass cavity losses completely~\cite{Pellizzari1997} and to achieve a state transfer success probability $F=1$ in the absence of other imperfections.
In the following, we numerically study the achievable quantum state transfer success probability under more realistic conditions by increasing gradually the relative weight of fiber losses to the overall photon loss rate (atomic losses and other imperfections will be included in \secref{sec:robustness}).  
To this end, we take the full photonic mode structure into account and consider the following regimes:
cavity loss dominated ($\gamma_\cav \gg \gamma_\fib$), equal losses ($\gamma_\fib \sim \gamma_\cav$) and fiber loss dominated ($\gamma_\fib \gg \gamma_\cav$). For comparison, we also include the extremal regime in which the fiber losses are absent ($\gamma_\fib=0$).

We find that in {\it all} parameter regimes, improvements of the state transfer success probability $F$, as defined in \eeqref{eq:Fidelity}, with respect to the limit $P_1$ are possible for adiabatic passages in accordance with \eeqref{eq:PAP}. The success probability of state transfer by wave packet shaping, however, is limited by $P_1$ in all regimes, i.e., $F\lesssim P_1$. 
Below, we illustrate this result with numerical simulations for a mirror transmission of the inner mirrors (M2 and M3 in \figref{fig:mainresults}) of $|\mathfrak{t}|^2=13$\,ppm and a cavity length of $l=0.02$\,m as in Ref.~\cite{Stute2012}. 
In addition, we compare the numerical results to the analytical estimate $F_\textnormal{AP}$ for the state transfer success probability by adiabatic passage, as defined in \eeqref{eq:PAP} in \secref{sec:cavitylossesanalytics}.
The cavity decay rate  $\gamma_\cav$ depends on the loss coefficient $|\ell|^2$ as defined in \eeqref{eq:kappagoodandbad}. 
The optimization for both methods is equivalent to the one described in \secref{sec:fibernumerics}. \\
\begin{figure}[h]
\centering
\includegraphics[width=01.0\columnwidth, angle=0]{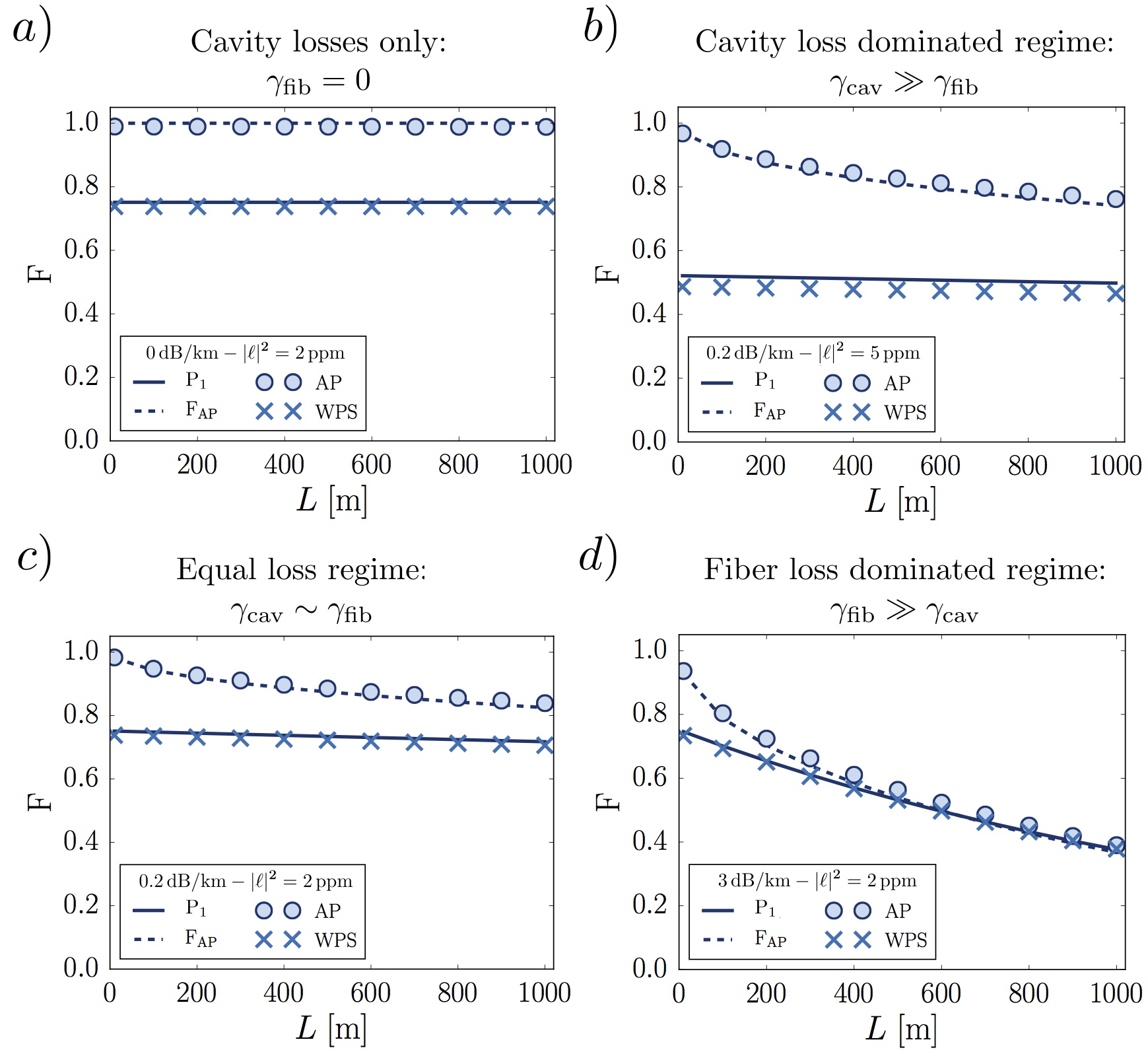}
\caption{State transfer success probability $F$ as a function of the fiber length $L$ 
for different values of the photon loss rates in the cavity and the fiber, $\gamma_\cav$ and $\gamma_\fib$. The state transfer performed by wave packet shaping (WPS, crosses) 
is limited by $P_1$ given in \eeqref{eq:Ploss} for all regimes (solid lines), while state transfer by adiabatic passage (AP, filled circles) surpasses this limit. The analytical estimate $F_\textnormal{AP}$ (dashed lines) for the success probability of state transfer by adiabatic passage in \eeqref{eq:PAP} agrees with the numerical results in all regimes.
Simulations presented here are for a mirror transmission of $|\mathfrak{t}|^2=13$\,ppm and cavity length $l=0.02$\,m.
(a) Cavity losses only: $\gamma_\fib=0$ (with cavity loss coefficient $\sabs{\ell}^2=2$\,ppm, fiber attenuation $0$\,dB/km). In the absence of fiber losses, adiabatic passages reach a success probability of unity. In the presence of fiber losses, adiabatic passages outperform wave packet shaping (b,c,d) up to a certain fiber length at which the success probability converges to $P_1$.
(b) Cavity loss dominated regime: $\gamma_\cav  \gg \gamma_\fib$ (with $\sabs{\ell}^2=5$\,ppm, $0.2$\,dB/km).  (c) Equal loss regime: $\gamma_\fib  \sim \gamma_\cav$ (with $\sabs{\ell}^2=2$\,ppm, $0.2$\,dB/km). (d) Fiber loss dominated regime: $\gamma_\fib \gg \gamma_\cav$ (with $\sabs{\ell}^2=2$\,ppm, $3$\,dB/km).} 
\label{fig:cavitylosses} 
\end{figure}
\\ {\bf Cavity loss dominated regime} $(\gamma_\cav \gg \gamma_\fib)$:  For most current optical realizations (see \tabref{tab:exp}) the cavity loss dominated regime is the relevant regime. In \figref{fig:cavitylosses}b we depict the state transfer success probability $F$ as a function of the fiber length $L$ for both state transfer methods. 
We consider here a cavity loss coefficient of $|\ell|^2=5$\,ppm (per round-trip) resulting in a loss rate of $\gamma_\cav/2\pi=6$\,kHz and a fiber attenuation coefficient of $0.2$\,dB/km, i.e., $\gamma_\fib^{0.2 \dB}/2\pi=1.5$\,kHz.
\figref{fig:cavitylosses}b shows how the success probability for a state transfer by adiabatic passage (filled circles) surpasses the limit $P_1$. \\ 
\\
\\ {\bf Equal loss regime} $(\gamma_\fib \sim \gamma_\cav)$:
In this regime fiber and cavity losses are approximately balanced. In \figref{fig:cavitylosses}c we depict the success probability $F$ as a function of the fiber length $L$ for both state transfer methods. 
To approximately balance both rates, 
we choose a cavity loss coefficient of $|\ell|^2=2$\,ppm, which corresponds to $\gamma_\cav/2\pi=2.4$kHz, and a fiber attenuation coefficient of $0.2$\,dB/km, i.e., $\gamma_\fib^{0.2 \dB}/2\pi=1.5$kHz.
As shown in \figref{fig:cavitylosses}c, the adiabatic passage state transfer (filled circles) once again surpasses the limit $P_1$ (solid line). In the limit of long fiber lengths (not shown here), the state transfer success probability $F$ converges to the limit $P_1$.\\
\\ {\bf Fiber loss dominated regime} $(\gamma_\fib \gg \gamma_\cav)$:
In this regime fiber losses dominate over cavity losses. In \figref{fig:cavitylosses}d we depict the success probability $F$ as a function of the fiber length $L$ for both state transfer methods. 
Here we choose a cavity loss coefficient of $|\ell|^2=2$\,ppm, which corresponds to a loss rate of $\gamma_\cav/2\pi=2.4$\,kHz, and a fiber attenuation coefficient of $3$\,dB/km with a corresponding loss rate of $\gamma_\fib^{3 \dB}/2\pi=22$\,kHz.
\figref{fig:cavitylosses}d shows that the success probability of state transfer by adiabatic passage (filled circles) continues to exceed  the limit $P_1$ (solid line) for smaller distances, while in the limit of longer fibers, the success probability converges to $P_1$. This behavior can also be seen in \figref{fig:Lmax}, in which the maximal fiber length until which adiabatic passage exceeds the  limit $P_1$ by at least $5$\% is displayed. \\
\\ Each of the plots in \figref{fig:cavitylosses} has been obtained for a specific parameter set ($\sabs{\mathfrak{t}}^2, \sabs{\ell}^2, l$). However, our numerical results show that the achievable quantum state transfer success probability for a given fiber length $L$ and fiber loss rate $\gamma_\fib$ is solely determined by the outcoupling probability $P_\textnormal{out}=\sabs{\mathfrak{t}}^2/(\sabs{\mathfrak{t}}^2+ \sabs{\ell}^2)$, in accordance with the analytical solution given by \eeqref{eq:PAP}. For a given value of $\gamma_\fib$, each of the plots in \figref{fig:cavitylosses} corresponds therefore to a whole class of parameter sets ($\sabs{\mathfrak{t}}^2, \sabs{\ell}^2, l$) that is characterized by the initial drop of the success probability to $P_1$ at length $L=0$ given by $P_\textnormal{out}^2$, cf. \eeqref{eq:Ploss}.\\
For illustration, \figref{fig:Fall1MMLvsFall2SML} displays the achievable state transfer success probability for a cavity length $l_1 = 2$\,cm and for mirror transmission and loss coefficients $\sabs{\mathfrak{t}}^2 =5$\,ppm and $\sabs{\ell}^2=2$\,ppm, resulting in $P_\textnormal{out}^2\approx51 \%$.  
The same plot is also obtained for a much shorter cavity of length $l_2=0.5$\,mm with equal transmission and loss coefficients $\sabs{\mathfrak{t}}^2 =5$\,ppm and $\sabs{\ell}^2=2$\,ppm (and therefore equal $P_\textnormal{out}^2\approx51 \%$). Even though the cavities of lengths $l_1$  and $l_2$ have very different cavity decay rates of $\kappa_{\cav,1} /2\pi \approx 6$\,kHz and $\kappa_{\cav,2}/2\pi \approx 240$\,kHz, and despite the very different single mode parameters $\mathfrak{n}_1= 0.17 \ll 1$ and $\mathfrak{n}_2= 6.7$ for the maximally considered fiber length of $L=1000$\,m, the numerical solution of the state transfer success probability for both cavities result in the same plot depicted in \figref{fig:Fall1MMLvsFall2SML}. \\
\begin{figure}[h]
\centering
\includegraphics[width=1.0\columnwidth, angle=0]{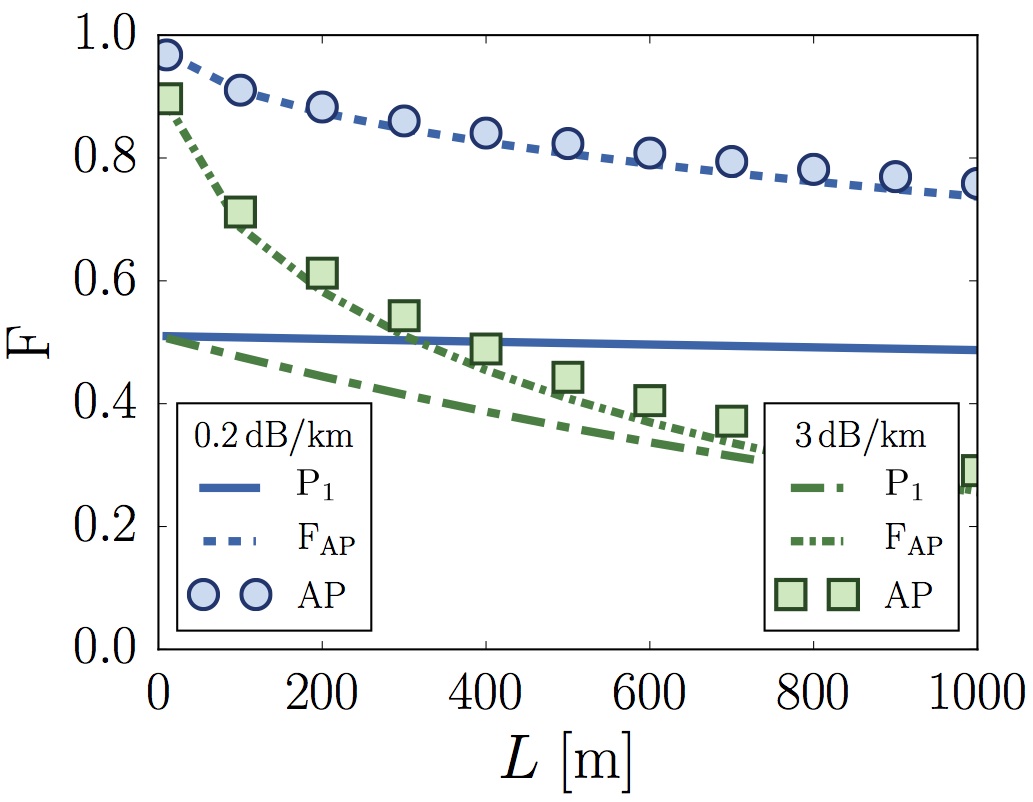}
\caption{The state transfer success probability $F$ that can be achieved by adiabatic passages is shown for two different fiber attenuation coefficients: $0.2$\,dB/km (blue circles) and $3$\,dB/km (green squares). These numerical results are compared to $P_1$ given in \eeqref{eq:Ploss}, which marks the maximum state transfer success probability that can be achieved with wave packet shaping (solid blue line: $0.2$\,dB/km, dashed-dotted green line:  $3$\,dB/km). In addition, the results are also compared to the analytical expression $F_\textnormal{AP}$ in \eeqref{eq:PAP} of the state transfer probability by adiabatic passage (dashed blue line: $0.2$\,dB/km, small dashed-dotted green line:  $3$\,dB/km). The transmission and loss coefficients are $\sabs{\mathfrak{t}}^2 =5$\,ppm and $\sabs{\ell}^2=2$\,ppm for both fiber attenuation coefficients. The cavity length is given by $l_1=2$\,cm.} 
\label{fig:Fall1MMLvsFall2SML} 
\end{figure}

Summarizing, we find that the single mode parameter is not a relevant figure of merit.  The relevant regime for performing quantum state transfer by adiabatic passage goes {\it beyond} the single mode limit and is given by the (more general) long photon limit $L\le L_\textnormal{ph}$, in which the length of the photon is at least on the order of the fiber length. In the ideal case of a lossless fiber, increasing the photon length (which is equivalent to a slower driving of the classical fields) leads to improved state transfer success probabilities. In the presence of fiber losses however, a trade-off exists between preventing cavity losses by slowly driving the classical fields on the one hand and avoiding fiber losses through multiple reflections of the photons in the fiber on the other, as described by \eeqref{eq:PAPunopt}.
 
%%%%%%%%%%%%%%%%%%%%%%%%%%%%%%%%%%%%
\section{Robustness}
\label{sec:robustness}
%%%%%%%%%%%%%%%%%%%%%%%%%%%%%%%%%%%%

In the following, we discuss the influence of atomic decay (\secref{sec:robustatom}) and other imperfections (\secref{sec:robustother}) on the achievable quantum state transfer success probability.

%%%%%%%%%%%%%%%%%%%%%%%%%%%%%%%%%%%%
\subsection{Atomic Decay}
\label{sec:robustatom}
%%%%%%%%%%%%%%%%%%%%%%%%%%%%%%%%%%%%

The role of atomic decay depends on the specific setup under consideration as the spontaneous decay rate $\Gamma$ varies for different atom and ion species.
The presence of atomic losses leads to a modification of the first and last equation of motion in \eeqref{eq:EOM},
\begin{align}
i \dot{c}_{A}&= G_A(t)  c_{a}- i(\tilde{\Gamma}_{A}(t)/2)  c_{A}, \\
i\dot{c}_{B}&= G_B(t) c_{b}- i(\tilde{\Gamma}_{B}(t)/2)  c_{B}, \notag
\end{align}
where the effective decay rate $\tilde{\Gamma}_{A/B}(t)=\Gamma (\Omega_{A/B}(t)/\Delta_\textnormal{at})^2$ results from the elimination of the excited state $\ket{E}$ of the atom in \figref{fig:APM}, with $\Gamma$ the spontaneous emission rate of the excited state.

The probability of a photon to be emitted from the cavity into the desired output mode is altered accordingly in the presence of atomic losses,
\begin{align}
\tilde{P}_\textnormal{out}=\frac{\Gamma_{A/B}^\textnormal{des}}{\Gamma_{A/B}^\textnormal{des}+\Gamma_{A/B}^\textnormal{und}},
\label{eq:Poutatom}
\end{align}
where $\Gamma_{A/B}^\textnormal{des}$ is the (desired) rate of photons leaving the cavity into the fiber
\begin{align}
\Gamma_{A/B}^\textnormal{des}=\left(\frac{G_{A/B}}{ \kappa} \right)^2 \kappa_\cav, \notag
\end{align}
and the (undesired) rate of photon loss due to atomic decay or due to cavity losses is $\Gamma_{A/B}^\textnormal{und}=\tilde{\Gamma}_{A/B}+\gamma_\cav (G_{A/B}/ \kappa)^2$. 
The probability $\tilde{P}_\textnormal{out}$ in \eeqref{eq:Poutatom} can be simplified to 
\begin{align}
\tilde{P}_\textnormal{out}=\frac{\mathcal{C}}{1/4+\mathcal{C}} P_\textnormal{out}, \notag
\end{align}
where we define the cooperativity of the system as
\begin{align}
\mathcal{C}=\frac{\left(G_{A/B}(t)\right)^2}{(2 \kappa) \, \tilde{\Gamma}(t)}=\frac{\left(g^{A/B}_\textnormal{at-c}\right)^2}{(2 \kappa) \, \Gamma}. \notag
\end{align}
Note that the optimal efficiency for transferring the quantum state stored in the emitter to a photon in the desired output channel (or vice versa) is given by $\mathcal{C}/(1/4+\mathcal{C})$ independent of the retrieval (storage) technique~\cite{Gorshkov2007}. 
Hence, the expected limit for transferring a photon when considering cavity, fiber and atomic losses leads to a modified version of \eeqref{eq:Ploss}, 
\begin{align}
\tilde{P}_{1}= \left(\frac{\mathcal{C}}{1/4+\mathcal{C}}\right)^2 \cdot P_1.
\label{eq:Plosstilde}
\end{align}

\begin{figure}[h]
\centering
\includegraphics[width=1.0\columnwidth, angle=0]{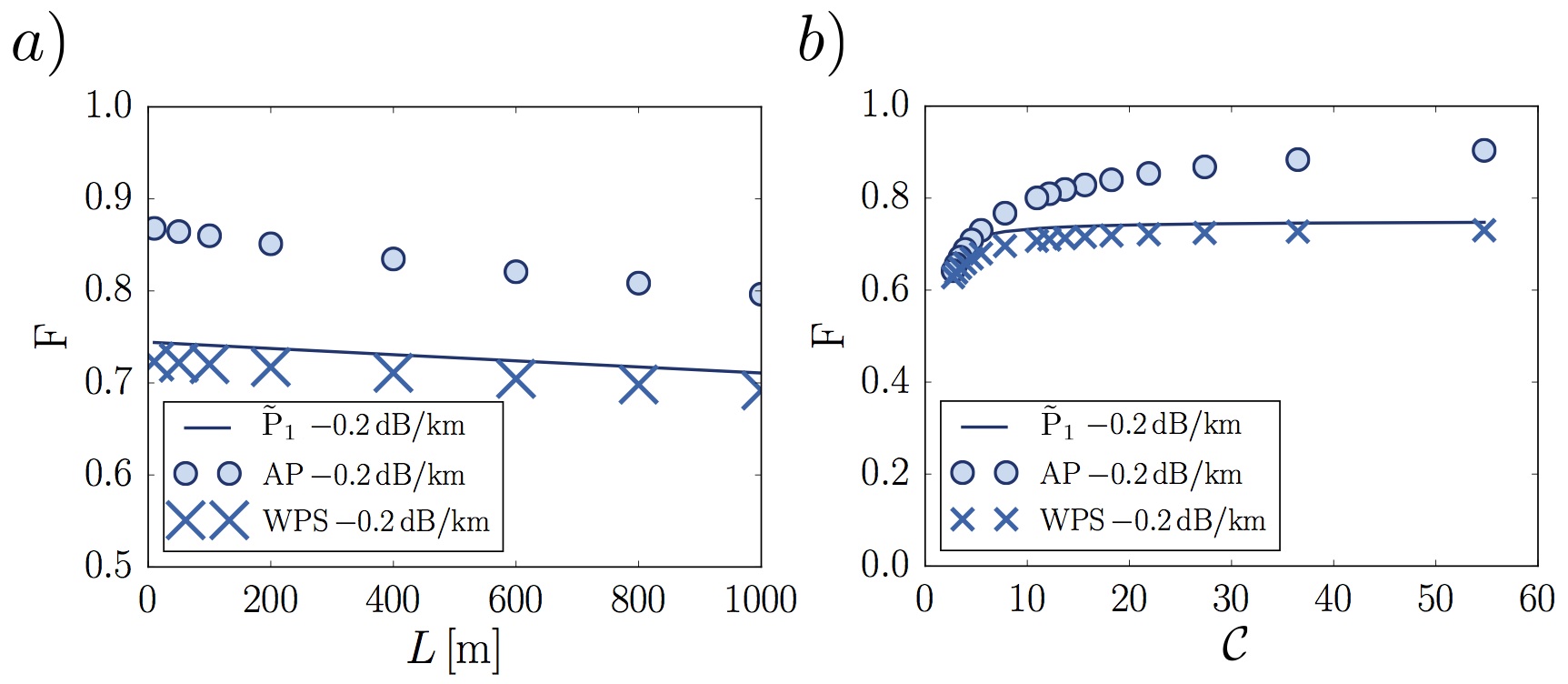}
\caption{
Achievable state transfer success probabilities in the presence of atomic decay.
(a) State transfer success probability $F$ as a function of the fiber length $L$ for a fixed atomic decay $\Gamma$. While wave packet shaping (WPS, crosses) is limited by $\tilde{P}_\textnormal{1}$ (solid line) given in \eeqref{eq:Plosstilde}, state transfer by adiabatic passage (AP, filled circles) surpasses this limit. 
(b) State transfer success probability $F$ as a function of the cooperativity $\mathcal{C}$ for both methods.} 
\label{fig:atomicdecay} 
\end{figure}

In \figref{fig:atomicdecay}, we show the state transfer success probability $F$, as defined in \eeqref{eq:Fidelity}, in the presence of atomic losses for both adiabatic passage and wave packet shaping. As a concrete example, we use here the parameters considered in  \figref{fig:cavitylosses}c, i.e., $\sabs{\mathfrak{t}}^2=13$\,ppm, $l=0.02$\,m, $\sabs{\ell}^2=2$\,ppm and a fiber absorption coefficient of $0.2$\,dB/km. \figref{fig:atomicdecay}a displays the state transfer success probability as a function of the fiber length for a cooperativity of $\mathcal{C}= 27$. 
Experiments for single atoms reach cooperativities of, e.g., $\mathcal{C}=82$~\cite{Hood2000} and $\mathcal{C}= 11$~\cite{Hamsen2016}. 
For experiments with atomic ensembles in cavities, the cooperativity $\mathcal{C}_N=N \mathcal{C}$ is enhanced by using a large number of atoms $N$~\cite{Gorshkov2007}. Hence, high cooperativities can be reached, e.g., $\mathcal{C}_N= 73$~\cite{Colombe2007}. The results in this work apply also to this case.
As shown in \figref{fig:atomicdecay}a, the state transfer by wave packet shaping is limited by $\tilde{P}_\textnormal{1}$, while adiabatic passages provide an advantage also in the presence of atomic losses (see \figref{fig:cavitylosses}c for the corresponding state transfer success probability for $\Gamma=0$).

\figref{fig:atomicdecay}b shows how the state transfer success probability of both methods improves with increasing cooperativity. More specifically, this plot displays the achievable state transfer success probability for very short distances (evaluated at $L=10$\,m, which corresponds to the first point/cross in \figref{fig:atomicdecay}a). While state transfer by wave packet shaping is always limited by $\tilde{P}_\textnormal{1}$, adiabatic passages surpass this bound with a gain in success probability $F-\tilde{P}_\textnormal{1}$ that increases with increasing cooperativity $\mathcal{C}$.

%%%%%%%%%%%%%%%%%%%%%%%%%%%%%%%%%%%%
\subsection{Other Imperfections}
\label{sec:robustother}
\begin{figure}[h]
\centering
\includegraphics[width=1.0\columnwidth, angle=0]{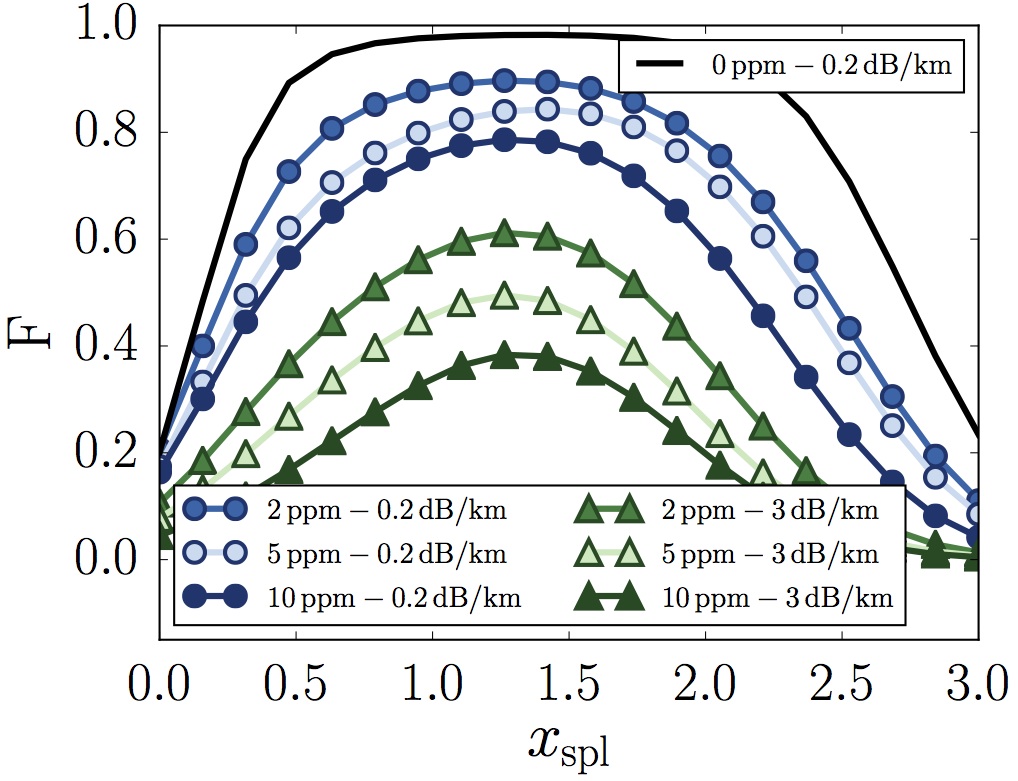}
\caption{Robustness with respect to timing mismatch in adiabatic passage. 
The state transfer success probability $F$ is shown as a function of the temporal separation $x_\textnormal{spl}$ of the two pulses. For a fiber attenuation coefficient of $0.2$\,dB/km (black, blue) mirror absorptions of $\sabs{\ell}^2=0$\,ppm (black, solid line) and $2$, $5$ and $10$\,ppm (blue circles) are depicted. The results for fiber absorptions of $3$\,dB/km (green) are shown for mirror loss coefficients of $\sabs{\ell}^2=2$\,ppm as well as $5$\,ppm and $10$\,ppm (green triangles).} 
\label{fig:robustness} 
\end{figure}

%%%%%%%%%%%%%%%%%%%%%%%%%%%%%%%%%%%%
%
In the following we address in- and outcoupling losses of the fiber and timing errors of the adiabatic passage.\\ 
%%
%%%%%%%%%%%%%%%%%%%%%%%%%%%%%%%%%%%%
%
\\
{\bf In-/outcoupling losses:}
In- and outcoupling losses refer to imperfect coupling of the light field between the cavities and the fiber. 
These imperfections can be included into our model in the fiber loss rate $\gamma_\fib$.
For optical cavities, optimized efficiencies for coupling in or out of a single-mode fiber can exceed $90\%$~\cite{Steiner2014}.
For fiber-integrated cavity systems, there is an additional multiplicative factor due to imperfect overlap between the fiber and the cavity modes. This mode overlap may be as high as $90\%$~\cite{Steiner2014} but drops off for longer cavities~\cite{Hunger2010}.
 \\
%
%%%%%%%%%%%%%%%%%%%%%%%%%%%%%%%%%%%%
%
\\
 {\bf Timing errors:}
State transfer by adiabatic passage depends on optimizing the temporal separation of the two pulses. 
In \figref{fig:robustness} we depict as an example the success probability of a state transfer by adiabatic passage as a function of the relative temporal separation $x_\textnormal{spl}$ of the two Gaussian pulses (see \secref{sec:AP}). 
 Here we use the same parameters as considered for \figref{fig:cavitylosses}, i.e., a mirror transmission $|\mathfrak{t}|^2=13$\,ppm and cavity length $l=0.02$\,m.
 Additionally, we choose a fixed fiber length of $L=400$\,m and optimize the pulse width $T$ and the atom-cavity coupling $G_{A/B}$ accordingly.
We vary the mirror losses $|\ell|^2$ from $0$\,ppm (black, solid line) to $10$\,ppm (green, blue) for the two different fiber attenuations, $0.2$\,dB/km (circles) and $3$\,dB/km (triangles).
The resulting moderate increase of the timing sensitivity with increasing losses is shown in \figref{fig:robustness}.
 \\
%

%%%%%%%%%%%%%%%%%%%%%%%%%%%%%%%%%%%%
\section{Conclusions and Outlook}
\label{sec:outlook}
%%%%%%%%%%%%%%%%%%%%%%%%%%%%%%%%%%%%

We have investigated deterministic quantum state transfer between remote qubits in cavities by studying the standard method of wave packet shaping and the use of adiabatic passages. We have provided an analysis for both methods beyond the single mode limit, taking the full photonic mode structure into account. This analysis has allowed us to assess the potential and limitations of these approaches for future developments of quantum networks. We have also discussed the role of the relevant cavity parameters in view of experimental-design decisions.

In particular, we have clarified that fiber transmission losses cannot be overcome using either of the two methods, and we have shown that cavity losses can be mitigated in a far greater parameter regime than previously known.

We note that the model considered is very general and not limited to optical setups and that the results can be used to evaluate the achievable performance of other experimental platforms such as superconducting qubits in microwave resonators~\cite{Wenner2014,Pfaff2016} or solid-state systems such as color centers or quantum dots coupled to photonic crystal nanocavities~\cite{Riedrich2014,Sipahigil2016,Yoshie2004,Hennessy2007,Englund2007}.

Our results apply to quantum networks on both short and long length scales. Relating to the latter, we have shown for links up to $1000$\,m that adiabatic passages can substantially improve the state transfer success probability for current experimental setups by overcoming limitations due to photon losses in the cavities. Regarding photon losses during the transmission, future networking implementations can in addition resort to quantum error correction techniques that rely on the transmission of multi-photon states and allow for the deterministic detection and correction of loss errors~\cite{Grassl1997,Ofek2016,Michael2016,Vermersch2016,Xiang2017}.

In future work, it would be interesting to study the application of adiabatic passages to enable time-continuous protocols~\cite{ContTeleportation,Hofer2013,Vollbrecht2011,Muschik2011}. Moreover, it will be useful to compare the potential and application range of adiabatic passages to other deterministic quantum state transfer methods such as dissipative schemes~\cite{Kraus2004},  approaches to mitigate cavity losses using quantum error correcting codes~\cite{Grassl1997,Ofek2016} and combinations of deterministic and heralded state transfer techniques.

%%%%%%%%%%%%%%%%%%%%%%%%%%%%%%%%%%%%
\section*{Acknowledgments}
We thank F. Reiter, K. M{\o}lmer, G. Kirchmair and P.-O. Guimond for helpful discussions. 
Research was sponsored by the Army Research Laboratory and was accomplished under Cooperative Agreement Number W911NF-15-2-0060. The views and conclusions contained in this document are those of the authors and should not be interpreted as representing the official policies, either expressed or implied, of the Army Research Laboratory or the U.S. Government. This work was also supported by
the European Research Council (ERC) Synergy Grant UQUAM, and by the Austrian Science Fund through SFB FOQUS (FWF Project No. F4016-N23). 
T. E. Northup acknowledges support from the Austrian Science Fund (FWF) Projects No. F4019-N23 and No. V252.
B. P. Lanyon acknowledges the Austrian Science Fund (FWF) through a START
grant under Project No. Y849-N20.

%%%%%%%%%%%%%%%%%%%%%%%%%%%%%%%%%%%%
\appendix
%%%%%%%%%%%%%%%%%%%%%%%%%%%%%%%%%%%%
%%%%%%%%%%%%%%%%%%%%%%%%%%%%%%%%%%%%
\section{Quantum State Transfer by Adiabatic Passage Including Multiple Fiber Modes}
\label{app:beyondSML}
%%%%%%%%%%%%%%%%%%%%%%%%%%%%%%%%%%%%
In the following, we derive the success probability $F_\text{AP}$ of state transfer by adiabatic passage in \eqsref{eq:FAPsecII} and \eqref{eq:PAP} in the presence of cavity and fiber losses by means of an analytical model. The analytical example in \secref{sec:fiberanalytics} corresponds to the special case of vanishing cavity losses, $\gamma_\cav=0$.

By adiabatically eliminating the cavity modes in the equations of motion for the full system given by \eqsref{eq:EOM} under the condition $\kappa \gg G_{A/B}$, we obtain an effective atom-fiber-atom model (cf. \figref{fig:APscheme}a for the case $\gamma_\cav=0$). In this description, the qubits in cavities A and B are coupled to the fiber modes with an effective time-dependent coupling strength $\tilde{g}_{A/B}(t)=g_{A/B}(G_{A/B}/\kappa)$ and are subject to an effective time-dependent decay that acts with a rate $\tilde{\gamma}_\cav^{A/B}=\gamma_\cav (G_{A/B} /\kappa)^2$. The equations of motion for the two qubits, the resonant fiber mode $c_0$ and the first pair of detuned fiber modes $c_{\pm 1}$ are given by
\begin{align}
\dot{c}_{A}&=-i \tilde{g}_A(t) \left( c_{c_0}+c_{c_{-1}}+c_{c_{+1}}\right) -\frac{\tilde{\gamma}^{A}_\cav(t)}{2} c_{A},\notag\\
\dot{c}_{c_{-1}}&=-i \tilde{g}_A(t) c_{A}+i \tilde{g}_B(t) c_{B}-\left(\gamma_\fib/2 -i \, \FSR_\fib \right) c_{c_{-1}},\notag \\
\dot{c}_{c_0}&=-i \tilde{g}_A(t) c_{A}-i \tilde{g}_B(t) c_{B}-(\gamma_\fib/2) c_{c_0},\notag \\
\dot{c}_{c_{+1}}&=-i \tilde{g}_A(t) c_{A}+i \tilde{g}_B(t) c_{B}-\left(\gamma_\fib/2 +i \, \FSR_\fib \right) c_{c_{+1}},\notag \\
\dot{c}_{B}&=-i \tilde{g}_B(t) \left( c_{c_0}-c_{c_{-1}} -c_{c_{+1}} \right)-\frac{\tilde{\gamma}^{B}_\cav(t)}{2} c_{B}. \notag
\end{align}%
We proceed by adiabatically eliminating the dynamics of the off-resonant fiber modes $c_{\pm 1}$, which evolve on a much faster time scale than the other modes for $\gamma_\fib^2/4+\delta^2\gg \tilde{g}_{A/B}^2$. Here, we introduce the abbreviations $\delta=\FSR_\fib$, $\gamma_{g_{AB}}=\frac{2 \gamma_\fib \tilde{g}_{A} \tilde{g}_{B}}{\gamma_\fib^2/4+\delta^2}$ and $\gamma_{g_{A/B}}=\frac{2 \gamma_\fib \tilde{g}_{A/B}^2 }{\gamma_\fib^2/4+\delta^2}$. The resulting effective three mode system (atom-fiber-atom) is depicted in \figref{fig:APscheme}b and described by
\begin{align}
\left(\begin{array}{c}
 i \dot{c_A}  \\
i\dot{c_0} \\
  i \dot{c_B}
   \end{array} \right) &=
 \left(\begin{array}{c c c}
-i \frac{\gamma_{g_A} +\tilde{\gamma}_\cav^{A}}{2} & \tilde{g}_A  & i \ \gamma_{g_{AB}} \\
  \tilde{g}_A & - i \frac{\gamma_\fib}{2}& \tilde{g}_B \\
   i \gamma_{g_{AB}} & \tilde{g}_B  &  -i \frac{\gamma_{g_B}+\tilde{\gamma}_\cav^{B}}{2}
    \end{array}\right) \left( \begin{array}{c}
  c_A  \\
  c_0 \\
  c_B
\end{array} \right).
\label{eq:analyticmodel3M}
\end{align}
Note that the couplings to the fiber modes $c_+$ and $c_-$ lead to Stark shifts ($\sim \frac{i \delta \tilde{g}_{A/B}^2}{\gamma_\fib^2/4+\delta^2}$) on the ground states with equal magnitude and opposite sign, which cancel. The same is true for the coherent couplings between the ground states that result from the coupling to these fiber modes ($\sim \frac{i \delta \tilde{g}_{A} \tilde{g}_B}{\gamma_\fib^2/4+\delta^2}$). 
Furthermore, the off-diagonal corner entries of the matrix in \eeqref{eq:analyticmodel3M} have different signs with respect to the diagonal corner entries due to the alternating sign of the coupling to cavity $B$ as discussed in the main text, cf. \eeqref{eq:Hcavfib}.

The Hamiltonian represented by the matrix in \eeqref{eq:analyticmodel3M} can be split into two parts: a coherent part corresponding to the Hamiltonian $H_\textrm{coh}=\hbar\left(\tilde{g}_A a^\dag+\tilde{g}_B b^\dag\right) c_0 + \hc$, which involves only the two qubits and the resonant fiber mode $c_0$ (as discussed in detail in \appref{app:atomiclimit}), and a dissipative part, which will be labelled $H_\textrm{diss}$. With this \eeqref{eq:analyticmodel3M} becomes: $i \hbar \,\dot{c} = [H_\textrm{coh}+H_\textrm{diss}] c$, where $c$ denotes the vector $c=(c_A,c_0,c_B)$.

In the absence of losses (i.e., $\gamma_\fib=\gamma_\cav=0$), the dissipate part $H_\textrm{diss}$ vanishes and the adiabatic instantaneous eigenstates are given by the ones of $H_\textrm{coh}$, i.e., $|\pm\rangle$ and $|D\rangle$ as defined in \eqsref{eq:SMLdark}-\eqref{eq:SMLbright2}. 

We transform the Hamiltonian into the basis of the adiabatic states $|\pm\rangle, |D\rangle$ with amplitudes $a_\pm$ and $a_D$ (see \cite{Vitanov1997}) and proceed by choosing the coupling functions $\tilde{g}_A=g_0 \sin(t/T)$ and $\tilde{g}_B=g_0 \cos(t/T)$. This choice renders the resulting Hamiltonian in the adiabatic representation partially time-independent and thus greatly simplifies the problem. More specifically, the Schr{\"o}dinger equation in the adiabatic representation is given by
\begin{align}
i \left(\begin{array}{c}
  \dot{a}_+  \\
  \dot{a}_D \\
   \dot{a}_-
   \end{array} \right) &=
 \left[H_\textrm{adia}+H_\textrm{corr}\right] \left( \begin{array}{c}
  a_+  \\
  a_D \\
  a_-
\end{array} \right),
\label{eq:adiabaticSE}
\end{align}
where the adiabatic Hamiltonian $H_\textrm{adia}$ describes the dynamics of the resonant fiber mode and is given by
\begin{align}
H_\textrm{adia} &=
 \left(\begin{array}{c c c}
 g_0- \frac{i (\gamma_\fib+\tilde{\gamma}_\cav)}{4}  & \frac{i}{\sqrt{2} T}  &  \frac{i \gamma_\fib }{4}  \\
   \frac{-i}{\sqrt{2} T} & 0 &  \frac{-i}{\sqrt{2} T} \\
   \frac{i \gamma_\fib }{4} &  \frac{i}{\sqrt{2} T}   &  -g_0- \frac{i (\gamma_\fib+\tilde{\gamma}_\cav)}{4} 
 \end{array}\right),
\notag
\end{align}
where $g_0=\sqrt{\tilde{g}_A^2+\tilde{g}_B^2}$ is the maximal coupling and $\tilde{\gamma}_\cav=\tilde{\gamma}_\cav^A=\tilde{\gamma}_\cav^B$. Note that we set the effective cavity decay $\tilde{\gamma}_\cav$ time-independent and equal for both cavities. With the chosen coupling functions, the Hamiltonian $H_\textrm{adia}$ is time-independent. 
In contrast, the second Hamiltonian $H_\textrm{corr}$ in \eeqref{eq:adiabaticSE} is time-dependent and arises from the adiabatic elimination of the two off-resonant fiber modes $c_{\pm 1}$, representing the effect of the detuned fiber modes on the success probability of state transfer. The Hamiltonian
$H_\textrm{corr}$ is given by
\begin{equation}
H_\textrm{corr}=\frac{-i \gamma_\fib g_0^2}{\frac{\gamma_\fib^2}{2}+2\delta^2}
 \left(\begin{array}{c c c}
\cos(\frac{2t}{T})^2  & -\frac{\sin(\frac{4t}{T})}{\sqrt{2}} & \cos(\frac{2t}{T})^2  \\
  -\frac{\sin(\frac{4t}{T})}{\sqrt{2}} & 2 \sin(\frac{2t}{T})^2 &  -\frac{\sin(\frac{4t}{T})}{\sqrt{2} }\\
 \cos(\frac{2t}{T})^2 &  -\frac{\sin(\frac{4t}{T})}{\sqrt{2}}  & \cos(\frac{2t}{T})^2 
 \end{array}\right).
\notag % \label{eq:Hcorr}
\end{equation}
In the limit $\gamma_\fib^2/4+\delta^2\gg g_0^2$ and $g_0\gg\gamma_\fib,\tilde{\gamma}_\cav$, the dynamics of the bright states can be eliminated, resulting in a slow effective decay of the dark state. Assuming $\delta \gg\gamma_\fib,\tilde{\gamma}_\cav$, we obtain
\begin{align}
a_D(t)=\exp \left( -\frac{\gamma_\fib t}{2 g_0^2 T^2}-\frac{\gamma_\fib g_0^2 t}{ 2 \delta^2}-\frac{\tilde{\gamma}_\cav t}{8} \right).
\label{eq:a0oft}
\end{align}
Due to the chosen coupling functions we evaluate \eeqref{eq:a0oft} at the final time of $t=\pi \,T/2$, and hence the final population of the dark state $a_D$ is given by $|a_D(\pi \,T/2)|^2$.
The state transfer success probability for adiabatic passage $F_1=|a_D(\pi \, T/2)|^2$ in the presence of cavity and fiber losses results in \eeqref{eq:PAPunoptsecII} and \eeqref{eq:PAPunopt}. 

%%%%%%%%%%%%%%%%%%%%%%%%%%%%%%%%%%%%

%%%%%%%%%%%%%%%%%%%%%%%%%%%%%%%%%%%%
\section{Quantum State Transfer by Adiabatic Passage for a Single Excited State}
\label{app:atomiclimit}
%%%%%%%%%%%%%%%%%%%%%%%%%%%%%%%%%%%%

Here, we review the derivation of the success probability of an adiabatic state transfer via a decaying state in general. Subsequently, we map the derivations to a coupled
atom-fiber-atom system as assumed in Refs.~\cite{Serafini2006, Yin2007,Chen2007,Ye2008,Lu2008,Zhou2009,Clader2014,Chen2015,Hua2015,Huang2016}.

We consider a system consisting of two ground states $\ket{A}$ and $\ket{B}$ and an excited state $\ket{C}$.
The states $\ket{A}$ and $\ket{C}$ are coupled with a time-dependent coupling strength $\mathcal{G}_A(t)$, and the states $\ket{B}$ and $\ket{C}$ are coupled with strength $\mathcal{G}_B(t)$. 
The state of the whole system in the standard basis is given by
\begin{align}
|\Psi\rangle_\textnormal{}=c_A \ket{A} +c_B \ket{B}+c_C \ket{C},
\notag
\end{align}
with probability amplitudes $c_A$, $c_B$ and $c_C$. In the adiabatic basis the state vector of the system is given by
\begin{align}
|\Psi\rangle_\textnormal{adia}=a_D \ket{D} +a_+ \ket{+}+a_- \ket{-}, \notag
\end{align}
with probability amplitudes $a_D$, $a_-$ and $a_+$.
The adiabatic states are connected to the standard basis via
\begin{align}
|D\rangle&= \frac{\mathcal{G}_B}{\sqrt{\mathcal{G}_A^2+\mathcal{G}_B^2}} \ket{A}-\frac{\mathcal{G}_A}{\sqrt{\mathcal{G}_A^2+\mathcal{G}_B^2}} \ket{B}, \label{eq:SMLdark}\\
|\pm\rangle&=\frac{1}{\sqrt{2}} \left(\frac{\mathcal{G}_A \ket{A}+\mathcal{G}_B \ket{B}}{\sqrt{\mathcal{G}_A^2+\mathcal{G}_B^2}} \pm \ket{C}\right).  \label{eq:SMLbright2}
\end{align}
$|D\rangle$ is a dark state with respect to the coupling $\mathcal{G}_{A/B}$, while $|\pm\rangle$ are so-called bright states. We are interested in performing an adiabatic state transfer from the ground state $\ket{A}$ to the ground state $\ket{B}$ using the dark state $\ket{D}$. 
The success probability $F$ of this state transfer is given by the final population of the dark state at $t=t_\textnormal{fin}$, i.e., $F=\sabs{a_D(t_\textnormal{fin})}^2$.

In the following, we introduce a decay of the excited state $\ket{C}$ at rate $\Gamma_C$. This decay is included in the derivation as described in \secref{sec:modeldissipation} such that the Schr{\"o}dinger equation for the probability amplitudes reads
\begin{align}
i \left(\begin{array}{c}
  \dot{c_A}  \\
  \dot{c_C} \\
   \dot{c_B}
   \end{array} \right) &=
 \left(\begin{array}{c c c}
  0  & \mathcal{G}_A (t) &0 \\
 \mathcal{G}_A(t) & - i \Gamma_C/2 & \mathcal{G}_B(t) \\
   0 & \mathcal{G}_B(t)  & 0
 \end{array}\right) \left( \begin{array}{c}
  c_A  \\
  c_C \\
  c_B
\end{array} \right).
\label{eq:SMLEOM}
\end{align}
In order to perform a state transfer by using the dark state $\ket{D}$, the coupling strengths $\mathcal{G}_A(t)$ and $\mathcal{G}_B(t)$ have to fulfil the conditions in \eeqref{eq:adiabaticcondition}. In particular, we choose the coupling strengths to be 
\begin{align}
\mathcal{G}_A(t)= \mathcal{G} \sin(t/T), \qquad \mathcal{G}_B(t)= \mathcal{G} \cos(t/T),
\label{eq:timedep}
\end{align}
for times $t\in{[0,T\pi/2]}$ with temporal length $T$ of the coupling and amplitude $\mathcal{G}$.

We transform \eeqref{eq:SMLEOM}  into the adiabatic basis, yielding the evolution of the amplitudes $a_D$, $a_-$ and $a_+$ (see Ref.~\cite{Vitanov1997}).
In the limit $\mathcal{G} \gg \Gamma_C$, the evolution of the amplitudes of the bright states $a_{\pm}$ is much faster than the evolution of the dark state $a_D$. 
We therefore first solve for the amplitudes of the bright states $a_{\pm}$ and subsequently derive the amplitude of the dark state $a_D$. 
The success probability $F$ of state transfer is given by the population of the dark state 
\begin{align}
F_{\text{}}=\sabs{a_D(t_\text{fin})}^2=\exp\left(-\frac{ \Gamma_C}{{\mathcal{G}^2 T}} \frac{\pi}{2}\right)
\label{eq:fidelitysml}
\end{align}
at time $t_\text{fin}=T\pi/2$, the final time of the coupling sequence.
In the adiabatic limit $\mathcal{G}^2 T \gg \Gamma_C$, the success probability in \eeqref{eq:fidelitysml} reaches unity and a perfect state transfer can be achieved. 
Note that this derivation can directly be mapped to an adiabatic state transfer in atoms (STIRAP)~\cite{RMPAP2016}. \\

In Refs.~\cite{Serafini2006, Yin2007,Chen2007,Ye2008,Lu2008,Zhou2009,Clader2014,Chen2015,Hua2015,Huang2016}, the Hamiltonian in \eeqref{eq:HAPM} has been truncated, which results in a description in which all fiber modes except the resonant mode $c_0$ are neglected.
This case can be mapped to the situation described above, where the ground states $\ket{A}$ and $\ket{B}$ represent the qubit states in cavities $A$ and $B$. The excited state $\ket{C}$ corresponds to the state of the fiber mode $c_0$ with associated decay rate $\Gamma_C = \gamma_\fib$.
The time-dependent coupling strengths $\mathcal{G}_{A/B}(t)$ translate into the effective atom-to-fiber coupling strengths $\tilde{g}_{A/B}(t)$ as depicted in \figref{fig:APM}b and defined in \secref{sec:fiberanalytics}.
This mapping, using the truncated Hamiltonian, in which only a single fiber mode (sfm) is considered, results in the success probability
\begin{align}
F_{\text{sfm}}=\exp\left(-\frac{ \gamma_\fib}{{g_0^2 T}} \frac{\pi}{2}\right),
\label{eq:fidelitynth}
\end{align}
where $g_0$ is the maximal atom-to-fiber coupling strength. By choosing a large pulse area $g_0^2 T \gg \gamma_\fib$, the success probability in \eeqref{eq:fidelitynth} reaches unity. In this limit, it seems that a perfect state transfer via a decaying fiber is possible. However, as we show in \secref{sec:fiberlosses}, the naive truncation of the Hamiltonian as done in Refs.~\cite{Serafini2006, Yin2007,Chen2007,Ye2008,Lu2008,Zhou2009,Clader2014,Chen2015,Hua2015,Huang2016} is not valid, cf. \appref{app:beyondSML}.

%%%%%%%%%%%%%%%%%%%%%%%%%%%%%
\section{Purely Photonic Description}
\label{sec:purelyphotonic}
%%%%%%%%%%%%%%%%%%%%%%%%%%%%%

The derivation of the analytical example in \secref{sec:fiberanalytics} can also be considered in a purely photonic context. Here, we consider the same setup as in \figref{fig:APM} but in a regime in which we first map the atomic state onto the cavity field by a fast swapping laser pulse. Subsequently we consider the purely photonic state transfer from cavity $A$ to cavity $B$ via fiber $C$. 

This cavity-fiber-cavity system can be described by the second line of the Hamiltonian in \eeqref{eq:HAPM} given by
\begin{align} 
H_\textnormal{cfc}=& \hbar \, \sum_n \,n \, \FSR_\fib \ c_n^{\dag}c_n \label{eq:HPP}  \\
&+\hbar \, \sum_{n}\Big[g_{A}(t)\, a^{\dag}+ g_{B}(t) (-1)^n \, b^{\dag}\Big]c_n+\hc \, . \notag
\end{align}
In contrast to the Hamiltonian in \eeqref{eq:HAPM}, here we have introduced time-variable cavity-fiber couplings $g_{A/B}(t)$ with a time dependence in analogy to \eeqref{eq:APpulsesequence} and \figref{fig:mainresults}c. 

In order to map the arguments from the atom-fiber-atom system described in \secref{sec:fiberanalytics} onto the purely photonic model, we replace the effective atom-fiber coupling strengths $\tilde{g}_{A/B}(t)$ by the (now time-dependent) cavity-fiber coupling strengths $g_{A/B}(t)$ as defined in \eeqref{eq:gAB}. With this, we recover the equations of motion as in \eeqref{eq:TMLEOM} for the purely photonic model and hence also the limited success probability as given by \eeqref{eq:fidelityMAX}.

The variation of the cavity-fiber coupling $g_{A/B}(t)$ in a time-dependent coupling sequence (see \secref{sec:AP}) is not straightforward for optical cavities but has been realized for superconducting resonators \cite{Yin2013} and photonic crystal nanocavities~\cite{Sato2012}.

%%%%%%%%%%%%%%%%%%%%%%%%%%%%%
\section{Description of Coupled Cavity-Fiber-Cavity System}
\label{app:methodCFC}
%
%%%%%%%%%%%%%%%%%%%%%%%%%%%%

In this appendix, we discuss the choice of basis states for the electric field modes in the coupled cavity-fiber-cavity setting. 
In the main text, we use independent field modes for the two outer cavities $a$, $b$ and for the fiber $c_n$ that are linearly coupled as described by \eeqref{eq:Hcavfib}.
In the following we relate this approach (which is generally valid in the case of high finesse cavities and for time scales that are long compared to the round-trip time $2\tau$ of a photon) to an alternative description that is based on the derivation of the electromagnetic field eigenmodes in the second quantization for the whole cavity-fiber-cavity system (see Ref.~\cite{vanEnk1999}).\\

\begin{figure}[h]
\centering
\includegraphics[width=1.0\columnwidth, angle=0]{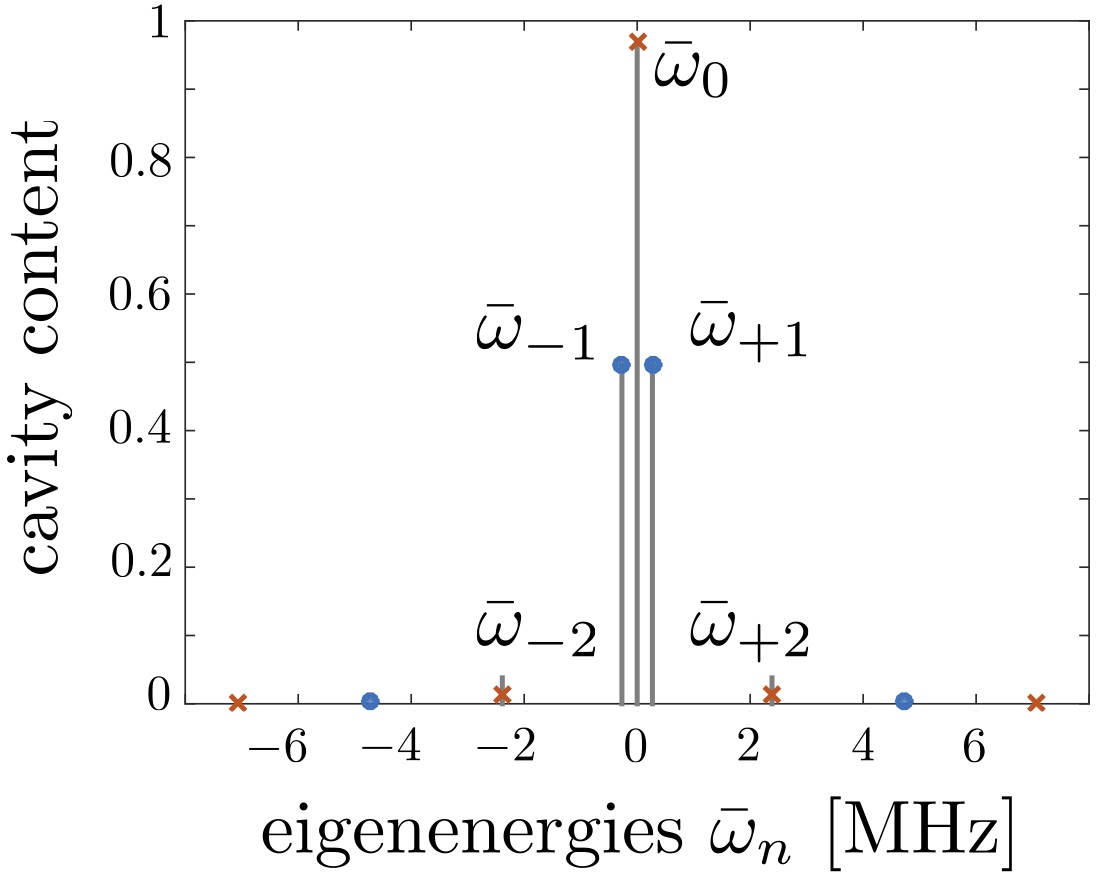}
\caption{Cavity content as a function of the eigenenergies $\bar{\omega}_n$. A coupled cavity-fiber-cavity system with transmission of M2 and M3 given by $\sabs{\mathfrak{t}}^2=13$\,ppm, cavity length $l=0.02$\,m and fiber length $L=400$\,m is considered. The x-position of the red crosses (odd modes) and the blue dots (even modes) indicates the energy of the associated mode, while the y-position indicates the cavity content of that mode. The grey lines indicate the positions of the energies given by \eeqref{eq:eigenenergies}. }
\label{fig:phphbasis}
\end{figure}

The eigenmodes $\bar{c}_n$ of the complete optical system consisting of two perfectly reflecting outer mirrors M1 and M4 and two identical partially transmitting mirrors M2 and M3 (see \figref{fig:mainresults}a) can be calculated as shown in Ref.~\cite{Ley1987} for the mirrors M2 and M3, together with an additional boundary condition at the positions of the outer mirrors M1 and M4. The corresponding eigenenergies $\bar{\omega}_n$ of the whole system can be inferred by solving Eq.~(2) in Ref.~\cite{vanEnk1999}, such that the Hamiltonian in  \eeqref{eq:HAPM}  can be expressed as
\begin{align} 
H_\textnormal{hyb}= &\hbar \, \sum_n \, \bar{\omega}_n \ \bar{c}_n^{\dag}\bar{c}_n +\hbar \, \sum_{n}\Big[G_A(t) \, \sigma_A^+ \sqrt{CC_n} \, \bar{c}_n+\hc \Big]  \label{eq:Hhyb}  \\
&+ \hbar \, \sum_{n}\Big[G_B(t) (-1)^n \, \sigma_B^+ \sqrt{CC_n} \, \bar{c}_n+\hc \Big],\notag
\end{align}
where the coupling between atom and field modes is weighted with the cavity content $CC_n$. The cavity content $CC_n$ quantifies the fraction of the population in mode $n$ that populates the cavities, as defined in Eq.~(5) in Ref.~\cite{vanEnk1999} (the fiber content of mode $n$ is given by $FC_n=1-CC_n$). 
The losses for the hybrid cavity-fiber-cavity modes $\bar{c}_n$ are modeled as a weighted combination of the loss processes discussed in \secref{sec:modeldissipation}, such that each hybrid mode $\bar{c}_n$ decays with
\begin{align}
\bar{\gamma}_n=CC_n \, \gamma_\cav + FC_n \,  \gamma_\fib.
\end{align}
By calculating the eigenenergies $\bar{\omega}_n$ and cavity contents $CC_n$ using the methods mentioned above and 
deriving the equations of motion as described in \secref{sec:modelEOM}, the numerical results shown in the main text can be (and have been) reproduced using \eeqref{eq:Hhyb}. The eigenenergies and cavity content for the parameter set used in \figref{fig:cavitylosses} are shown in \figref{fig:phphbasis}.

We illustrate the basis transformation that relates the modes $a$, $b$ and $c_n$ used in the main text and the hybrid modes $\bar{c}_n$ for the truncated mode set (involving only three fiber modes) that is used in the analytical example discussed in \secref{sec:fiberanalytics} and \secref{sec:cavitylossesanalytics}.
By diagonalizing the cavity and fiber Hamiltonians in \eeqref{eq:Hfib} and in \eeqref{eq:Hcavfib} for three fiber modes, we obtain the eigenfrequencies $\bar{\omega}_n$ of the hybrid cavity-fiber-cavity modes as
\begin{align}
&\bar{\omega}_0=0, \qquad \bar{\omega}_{\pm 1}=\pm \sqrt{2} g_{A/B} , \label{eq:eigenenergies}\\
& \qquad \bar{\omega}_{\pm 2}=\pm \sqrt{4 g^2_{A/B}+ \FSR_\fib^2}, \notag \\
\end{align}
where $g_{A/B}$ is defined in \eeqref{eq:gAB}. \figref{fig:phphbasis} shows that even for the truncated mode set, these values (grey lines) are very close to the data points (red crosses and blue dots) that are obtained by numerically solving Eq.~(2) in Ref.~\cite{vanEnk1999} or by solving the problem using transfer matrices. The height of each data point indicates the cavity content, i.e. the fraction of that mode that is populating the cavities. The grey lines indicate the positions of the eigenenergies as derived in \eeqref{eq:eigenenergies}.

The corresponding eigenstates of the hybrid modes $\bar{c}_n$ can be written as a superposition of the original basis of cavity $a$ and $b$ and fiber modes $c_n$ as used in the main text
\begin{align}
\begin{psmallmatrix}
  \bar{c}_0  \\
  \bar{c}_{+1} \\
    \bar{c}_{-1}  \\
      \bar{c}_{+2}  \\
        \bar{c}_{-2} 
   \end{psmallmatrix}  &=\mathfrak{M}
  \begin{psmallmatrix}
   a  \\
  b \\
     c_{+1}\\
   c_0\\
   c_{-1} \\
 \end{psmallmatrix} ,
\end{align}
with $\mathfrak{M}$ given by
\begin{align}  \mathcal{N} 
\begin{psmallmatrix}
   \frac{\FSR_\fib}{\sqrt{2} \bar{\omega}_{\pm 1}^2} & - \frac{\FSR_\fib}{\sqrt{2}  \bar{\omega}_{\pm 1}^2} & -1 & 0 & 1 \\
   -\frac{1}{\sqrt{2}} & -\frac{1}{\sqrt{2}}&0&1&0 \\
     \frac{1}{\sqrt{2}} & \frac{1}{\sqrt{2}}&0&1&0 \\
      \frac{\FSR_\fib+\bar{\omega}_{-2}}{\sqrt{2} \bar{\omega}_{\pm 1}^2} &    \frac{-\FSR_\fib+\bar{\omega}_{+2}}{\sqrt{2} \bar{\omega}_{\pm 1}^2} & 
           1 +  \frac{\FSR_\fib (\FSR_\fib+\bar{\omega}_{-2})}{\bar{\omega}_{\pm 1}^2} &0&1\\
      \frac{\FSR_\fib+\bar{\omega}_{+2}}{\sqrt{2} \bar{\omega}_{\pm 1}^2} &    \frac{-\FSR_\fib+\bar{\omega}_{-2}}{\sqrt{2} \bar{\omega}_{\pm 1}^2} & 
           1 +  \frac{\FSR_\fib (\FSR_\fib+\bar{\omega}_{+2})}{\bar{\omega}_{\pm 1}^2} &0&1
 \end{psmallmatrix},
 \end{align}
where $\mathcal{N}$ indicates the proper normalization (not shown here).

%%%%%%%%%%%%%%%%%%%%%%%%%%%%%%%%%%%%%%%%%

\bibliography{longdistancestatetransfer}

\begin{thebibliography}{93}%
\makeatletter
\providecommand \@ifxundefined [1]{%
 \@ifx{#1\undefined}
}%
\providecommand \@ifnum [1]{%
 \ifnum #1\expandafter \@firstoftwo
 \else \expandafter \@secondoftwo
 \fi
}%
\providecommand \@ifx [1]{%
 \ifx #1\expandafter \@firstoftwo
 \else \expandafter \@secondoftwo
 \fi
}%
\providecommand \natexlab [1]{#1}%
\providecommand \enquote  [1]{``#1''}%
\providecommand \bibnamefont  [1]{#1}%
\providecommand \bibfnamefont [1]{#1}%
\providecommand \citenamefont [1]{#1}%
\providecommand \href@noop [0]{\@secondoftwo}%
\providecommand \href [0]{\begingroup \@sanitize@url \@href}%
\providecommand \@href[1]{\@@startlink{#1}\@@href}%
\providecommand \@@href[1]{\endgroup#1\@@endlink}%
\providecommand \@sanitize@url [0]{\catcode `\\12\catcode `\$12\catcode
  `\&12\catcode `\#12\catcode `\^12\catcode `\_12\catcode `\%12\relax}%
\providecommand \@@startlink[1]{}%
\providecommand \@@endlink[0]{}%
\providecommand \url  [0]{\begingroup\@sanitize@url \@url }%
\providecommand \@url [1]{\endgroup\@href {#1}{\urlprefix }}%
\providecommand \urlprefix  [0]{URL }%
\providecommand \Eprint [0]{\href }%
\providecommand \doibase [0]{http://dx.doi.org/}%
\providecommand \selectlanguage [0]{\@gobble}%
\providecommand \bibinfo  [0]{\@secondoftwo}%
\providecommand \bibfield  [0]{\@secondoftwo}%
\providecommand \translation [1]{[#1]}%
\providecommand \BibitemOpen [0]{}%
\providecommand \bibitemStop [0]{}%
\providecommand \bibitemNoStop [0]{.\EOS\space}%
\providecommand \EOS [0]{\spacefactor3000\relax}%
\providecommand \BibitemShut  [1]{\csname bibitem#1\endcsname}%
\let\auto@bib@innerbib\@empty
%</preamble>
\bibitem [{\citenamefont {Kimble}(2008)}]{KimbleReview}%
  \BibitemOpen
  \bibfield  {author} {\bibinfo {author} {\bibfnamefont {H.~J.}\ \bibnamefont
  {Kimble}},\ }\href {http://dx.doi.org/10.1038/nature07127} {\bibfield
  {journal} {\bibinfo  {journal} {Nature}\ }\textbf {\bibinfo {volume} {453}},\
  \bibinfo {pages} {1023} (\bibinfo {year} {2008})}\BibitemShut {NoStop}%
\bibitem [{\citenamefont {Northup}\ and\ \citenamefont
  {Blatt}(2014)}]{TracyReview}%
  \BibitemOpen
  \bibfield  {author} {\bibinfo {author} {\bibfnamefont {T.~E.}\ \bibnamefont
  {Northup}}\ and\ \bibinfo {author} {\bibfnamefont {R.}~\bibnamefont
  {Blatt}},\ }\href {http://dx.doi.org/10.1038/nphoton.2014.53} {\bibfield
  {journal} {\bibinfo  {journal} {Nat Photon}\ }\textbf {\bibinfo {volume}
  {8}},\ \bibinfo {pages} {356} (\bibinfo {year} {2014})}\BibitemShut {NoStop}%
\bibitem [{\citenamefont {Gibney}(2016)}]{Manifesto}%
  \BibitemOpen
  \bibfield  {author} {\bibinfo {author} {\bibfnamefont {E.}~\bibnamefont
  {Gibney}},\ }\href
  {http://www.nature.com/news/europe-plans-giant-billion-euro-quantum-technologies-project-1.19796}
  {\bibfield  {journal} {\bibinfo  {journal} {Nature}\ }\textbf {\bibinfo
  {volume} {532}},\ \bibinfo {pages} {426} (\bibinfo {year}
  {2016})}\BibitemShut {NoStop}%
\bibitem [{\citenamefont {Bru{\ss}}\ and\ \citenamefont
  {L{\"u}tkenhaus}(2000)}]{Bruss2000}%
  \BibitemOpen
  \bibfield  {author} {\bibinfo {author} {\bibfnamefont {D.}~\bibnamefont
  {Bru{\ss}}}\ and\ \bibinfo {author} {\bibfnamefont {N.}~\bibnamefont
  {L{\"u}tkenhaus}},\ }\href {\doibase 10.1007/s002000050137} {\bibfield
  {journal} {\bibinfo  {journal} {Applicable Algebra in Engineering,
  Communication and Computing}\ }\textbf {\bibinfo {volume} {10}},\ \bibinfo
  {pages} {383} (\bibinfo {year} {2000})}\BibitemShut {NoStop}%
\bibitem [{\citenamefont {Lo}\ \emph {et~al.}(2014)\citenamefont {Lo},
  \citenamefont {Curty},\ and\ \citenamefont {Tamaki}}]{Lo2014}%
  \BibitemOpen
  \bibfield  {author} {\bibinfo {author} {\bibfnamefont {H.-K.}\ \bibnamefont
  {Lo}}, \bibinfo {author} {\bibfnamefont {M.}~\bibnamefont {Curty}}, \ and\
  \bibinfo {author} {\bibfnamefont {K.}~\bibnamefont {Tamaki}},\ }\href
  {http://dx.doi.org/10.1038/nphoton.2014.149} {\bibfield  {journal} {\bibinfo
  {journal} {Nat Photon}\ }\textbf {\bibinfo {volume} {8}},\ \bibinfo {pages}
  {595} (\bibinfo {year} {2014})}\BibitemShut {NoStop}%
\bibitem [{\citenamefont {Beals}\ \emph {et~al.}(2013)\citenamefont {Beals},
  \citenamefont {Brierley}, \citenamefont {Gray}, \citenamefont {Harrow},
  \citenamefont {Kutin}, \citenamefont {Linden}, \citenamefont {Shepherd},\
  and\ \citenamefont {Stather}}]{Beals2013}%
  \BibitemOpen
  \bibfield  {author} {\bibinfo {author} {\bibfnamefont {R.}~\bibnamefont
  {Beals}}, \bibinfo {author} {\bibfnamefont {S.}~\bibnamefont {Brierley}},
  \bibinfo {author} {\bibfnamefont {O.}~\bibnamefont {Gray}}, \bibinfo {author}
  {\bibfnamefont {A.~W.}\ \bibnamefont {Harrow}}, \bibinfo {author}
  {\bibfnamefont {S.}~\bibnamefont {Kutin}}, \bibinfo {author} {\bibfnamefont
  {N.}~\bibnamefont {Linden}}, \bibinfo {author} {\bibfnamefont
  {D.}~\bibnamefont {Shepherd}}, \ and\ \bibinfo {author} {\bibfnamefont
  {M.}~\bibnamefont {Stather}},\ }\href
  {http://rspa.royalsocietypublishing.org/content/469/2153/20120686.abstract}
  {\bibfield  {journal} {\bibinfo  {journal} {Proceedings of the Royal Society
  A: Mathematical, Physical and Engineering Science}\ }\textbf {\bibinfo
  {volume} {469}} (\bibinfo {year} {2013})}\BibitemShut {NoStop}%
\bibitem [{\citenamefont {Buhrman}\ \emph {et~al.}(2001)\citenamefont
  {Buhrman}, \citenamefont {Cleve}, \citenamefont {Watrous},\ and\
  \citenamefont {de~Wolf}}]{QuFingerprinting1}%
  \BibitemOpen
  \bibfield  {author} {\bibinfo {author} {\bibfnamefont {H.}~\bibnamefont
  {Buhrman}}, \bibinfo {author} {\bibfnamefont {R.}~\bibnamefont {Cleve}},
  \bibinfo {author} {\bibfnamefont {J.}~\bibnamefont {Watrous}}, \ and\
  \bibinfo {author} {\bibfnamefont {R.}~\bibnamefont {de~Wolf}},\ }\href
  {\doibase 10.1103/PhysRevLett.87.167902} {\bibfield  {journal} {\bibinfo
  {journal} {Phys. Rev. Lett.}\ }\textbf {\bibinfo {volume} {87}},\ \bibinfo
  {pages} {167902} (\bibinfo {year} {2001})}\BibitemShut {NoStop}%
\bibitem [{\citenamefont {Xu}\ \emph {et~al.}(2015)\citenamefont {Xu},
  \citenamefont {Arrazola}, \citenamefont {Wei}, \citenamefont {Wang},
  \citenamefont {Palacios-Avila}, \citenamefont {Feng}, \citenamefont {Sajeed},
  \citenamefont {L{\"u}tkenhaus},\ and\ \citenamefont
  {Lo}}]{QuFingerprinting2}%
  \BibitemOpen
  \bibfield  {author} {\bibinfo {author} {\bibfnamefont {F.}~\bibnamefont
  {Xu}}, \bibinfo {author} {\bibfnamefont {J.~M.}\ \bibnamefont {Arrazola}},
  \bibinfo {author} {\bibfnamefont {K.}~\bibnamefont {Wei}}, \bibinfo {author}
  {\bibfnamefont {W.}~\bibnamefont {Wang}}, \bibinfo {author} {\bibfnamefont
  {P.}~\bibnamefont {Palacios-Avila}}, \bibinfo {author} {\bibfnamefont
  {C.}~\bibnamefont {Feng}}, \bibinfo {author} {\bibfnamefont {S.}~\bibnamefont
  {Sajeed}}, \bibinfo {author} {\bibfnamefont {N.}~\bibnamefont
  {L{\"u}tkenhaus}}, \ and\ \bibinfo {author} {\bibfnamefont {H.-K.}\
  \bibnamefont {Lo}},\ }\href {http://dx.doi.org/10.1038/ncomms9735} {\bibfield
   {journal} {\bibinfo  {journal} {Nature Communications}\ }\textbf {\bibinfo
  {volume} {6}},\ \bibinfo {pages} {8735} (\bibinfo {year} {2015})}\BibitemShut
  {NoStop}%
\bibitem [{\citenamefont {Goorden}\ \emph {et~al.}(2014)\citenamefont
  {Goorden}, \citenamefont {Horstmann}, \citenamefont {Mosk}, \citenamefont
  {{\v S}kori{\'c}},\ and\ \citenamefont {Pinkse}}]{QuCreditCards}%
  \BibitemOpen
  \bibfield  {author} {\bibinfo {author} {\bibfnamefont {S.~A.}\ \bibnamefont
  {Goorden}}, \bibinfo {author} {\bibfnamefont {M.}~\bibnamefont {Horstmann}},
  \bibinfo {author} {\bibfnamefont {A.~P.}\ \bibnamefont {Mosk}}, \bibinfo
  {author} {\bibfnamefont {B.}~\bibnamefont {{\v S}kori{\'c}}}, \ and\ \bibinfo
  {author} {\bibfnamefont {P.~W.~H.}\ \bibnamefont {Pinkse}},\ }\bibfield
  {booktitle} {\emph {\bibinfo {booktitle} {Optica}},\ }\href {\doibase
  10.1364/OPTICA.1.000421} {\bibfield  {journal} {\bibinfo  {journal} {Optica}\
  }\textbf {\bibinfo {volume} {1}},\ \bibinfo {pages} {421} (\bibinfo {year}
  {2014})}\BibitemShut {NoStop}%
\bibitem [{\citenamefont {Hillery}\ \emph {et~al.}(2006)\citenamefont
  {Hillery}, \citenamefont {Ziman}, \citenamefont {Bu{\v z}ek},\ and\
  \citenamefont {Bielikov{\'a}}}]{QuSecretVoting}%
  \BibitemOpen
  \bibfield  {author} {\bibinfo {author} {\bibfnamefont {M.}~\bibnamefont
  {Hillery}}, \bibinfo {author} {\bibfnamefont {M.}~\bibnamefont {Ziman}},
  \bibinfo {author} {\bibfnamefont {V.}~\bibnamefont {Bu{\v z}ek}}, \ and\
  \bibinfo {author} {\bibfnamefont {M.}~\bibnamefont {Bielikov{\'a}}},\ }\href
  {\doibase http://dx.doi.org/10.1016/j.physleta.2005.09.010} {\bibfield
  {journal} {\bibinfo  {journal} {Physics Letters A}\ }\textbf {\bibinfo
  {volume} {349}},\ \bibinfo {pages} {75} (\bibinfo {year} {2006})}\BibitemShut
  {NoStop}%
\bibitem [{\citenamefont {Hillery}\ \emph {et~al.}(1999)\citenamefont
  {Hillery}, \citenamefont {Bu\ifmmode~\check{z}\else \v{z}\fi{}ek},\ and\
  \citenamefont {Berthiaume}}]{QSecretSharing}%
  \BibitemOpen
  \bibfield  {author} {\bibinfo {author} {\bibfnamefont {M.}~\bibnamefont
  {Hillery}}, \bibinfo {author} {\bibfnamefont {V.}~\bibnamefont
  {Bu\ifmmode~\check{z}\else \v{z}\fi{}ek}}, \ and\ \bibinfo {author}
  {\bibfnamefont {A.}~\bibnamefont {Berthiaume}},\ }\href {\doibase
  10.1103/PhysRevA.59.1829} {\bibfield  {journal} {\bibinfo  {journal} {Phys.
  Rev. A}\ }\textbf {\bibinfo {volume} {59}},\ \bibinfo {pages} {1829}
  (\bibinfo {year} {1999})}\BibitemShut {NoStop}%
\bibitem [{\citenamefont {Broadbent}\ \emph {et~al.}(2009)\citenamefont
  {Broadbent}, \citenamefont {Fitzsimons},\ and\ \citenamefont
  {Kashefi}}]{QuCloudComputing1}%
  \BibitemOpen
  \bibfield  {author} {\bibinfo {author} {\bibfnamefont {A.}~\bibnamefont
  {Broadbent}}, \bibinfo {author} {\bibfnamefont {J.}~\bibnamefont
  {Fitzsimons}}, \ and\ \bibinfo {author} {\bibfnamefont {E.}~\bibnamefont
  {Kashefi}},\ }in\ \href {\doibase 10.1109/FOCS.2009.36} {\emph {\bibinfo
  {booktitle} {2009 50th Annual IEEE Symposium on Foundations of Computer
  Science}}}\ (\bibinfo {year} {2009})\ pp.\ \bibinfo {pages}
  {517--526}\BibitemShut {NoStop}%
\bibitem [{\citenamefont {Barz}\ \emph {et~al.}(2012)\citenamefont {Barz},
  \citenamefont {Kashefi}, \citenamefont {Broadbent}, \citenamefont
  {Fitzsimons}, \citenamefont {Zeilinger},\ and\ \citenamefont
  {Walther}}]{QuCloudComputing2}%
  \BibitemOpen
  \bibfield  {author} {\bibinfo {author} {\bibfnamefont {S.}~\bibnamefont
  {Barz}}, \bibinfo {author} {\bibfnamefont {E.}~\bibnamefont {Kashefi}},
  \bibinfo {author} {\bibfnamefont {A.}~\bibnamefont {Broadbent}}, \bibinfo
  {author} {\bibfnamefont {J.~F.}\ \bibnamefont {Fitzsimons}}, \bibinfo
  {author} {\bibfnamefont {A.}~\bibnamefont {Zeilinger}}, \ and\ \bibinfo
  {author} {\bibfnamefont {P.}~\bibnamefont {Walther}},\ }\href
  {http://science.sciencemag.org/content/335/6066/303.abstract} {\bibfield
  {journal} {\bibinfo  {journal} {Science}\ }\textbf {\bibinfo {volume}
  {335}},\ \bibinfo {pages} {303} (\bibinfo {year} {2012})}\BibitemShut
  {NoStop}%
\bibitem [{\citenamefont {Komar}\ \emph {et~al.}(2014)\citenamefont {Komar},
  \citenamefont {Kessler}, \citenamefont {Bishof}, \citenamefont {Jiang},
  \citenamefont {Sorensen}, \citenamefont {Ye},\ and\ \citenamefont
  {Lukin}}]{Komar14}%
  \BibitemOpen
  \bibfield  {author} {\bibinfo {author} {\bibfnamefont {P.}~\bibnamefont
  {Komar}}, \bibinfo {author} {\bibfnamefont {E.~M.}\ \bibnamefont {Kessler}},
  \bibinfo {author} {\bibfnamefont {M.}~\bibnamefont {Bishof}}, \bibinfo
  {author} {\bibfnamefont {L.}~\bibnamefont {Jiang}}, \bibinfo {author}
  {\bibfnamefont {A.~S.}\ \bibnamefont {Sorensen}}, \bibinfo {author}
  {\bibfnamefont {J.}~\bibnamefont {Ye}}, \ and\ \bibinfo {author}
  {\bibfnamefont {M.~D.}\ \bibnamefont {Lukin}},\ }\href
  {http://dx.doi.org/10.1038/nphys3000} {\bibfield  {journal} {\bibinfo
  {journal} {Nat Phys}\ }\textbf {\bibinfo {volume} {10}},\ \bibinfo {pages}
  {582} (\bibinfo {year} {2014})}\BibitemShut {NoStop}%
\bibitem [{\citenamefont {Bell}(2004)}]{BellTest1}%
  \BibitemOpen
  \bibfield  {author} {\bibinfo {author} {\bibfnamefont {J.~S.}\ \bibnamefont
  {Bell}},\ }\href@noop {} {\emph {\bibinfo {title} {Speakable and Unspeakable
  in Quantum Mechanics}}}\ (\bibinfo  {publisher} {Cambridge University
  Press},\ \bibinfo {year} {2004})\BibitemShut {NoStop}%
\bibitem [{\citenamefont {Hensen}\ \emph {et~al.}(2015)\citenamefont {Hensen},
  \citenamefont {Bernien}, \citenamefont {Dreau}, \citenamefont {Reiserer},
  \citenamefont {Kalb}, \citenamefont {Blok}, \citenamefont {Ruitenberg},
  \citenamefont {Vermeulen}, \citenamefont {Schouten}, \citenamefont {Abellan},
  \citenamefont {Amaya}, \citenamefont {Pruneri}, \citenamefont {Mitchell},
  \citenamefont {Markham}, \citenamefont {Twitchen}, \citenamefont {Elkouss},
  \citenamefont {Wehner}, \citenamefont {Taminiau},\ and\ \citenamefont
  {Hanson}}]{BellTest2}%
  \BibitemOpen
  \bibfield  {author} {\bibinfo {author} {\bibfnamefont {B.}~\bibnamefont
  {Hensen}}, \bibinfo {author} {\bibfnamefont {H.}~\bibnamefont {Bernien}},
  \bibinfo {author} {\bibfnamefont {A.~E.}\ \bibnamefont {Dreau}}, \bibinfo
  {author} {\bibfnamefont {A.}~\bibnamefont {Reiserer}}, \bibinfo {author}
  {\bibfnamefont {N.}~\bibnamefont {Kalb}}, \bibinfo {author} {\bibfnamefont
  {M.~S.}\ \bibnamefont {Blok}}, \bibinfo {author} {\bibfnamefont
  {J.}~\bibnamefont {Ruitenberg}}, \bibinfo {author} {\bibfnamefont {R.~F.~L.}\
  \bibnamefont {Vermeulen}}, \bibinfo {author} {\bibfnamefont {R.~N.}\
  \bibnamefont {Schouten}}, \bibinfo {author} {\bibfnamefont {C.}~\bibnamefont
  {Abellan}}, \bibinfo {author} {\bibfnamefont {W.}~\bibnamefont {Amaya}},
  \bibinfo {author} {\bibfnamefont {V.}~\bibnamefont {Pruneri}}, \bibinfo
  {author} {\bibfnamefont {M.~W.}\ \bibnamefont {Mitchell}}, \bibinfo {author}
  {\bibfnamefont {M.}~\bibnamefont {Markham}}, \bibinfo {author} {\bibfnamefont
  {D.~J.}\ \bibnamefont {Twitchen}}, \bibinfo {author} {\bibfnamefont
  {D.}~\bibnamefont {Elkouss}}, \bibinfo {author} {\bibfnamefont
  {S.}~\bibnamefont {Wehner}}, \bibinfo {author} {\bibfnamefont {T.~H.}\
  \bibnamefont {Taminiau}}, \ and\ \bibinfo {author} {\bibfnamefont
  {R.}~\bibnamefont {Hanson}},\ }\href {http://dx.doi.org/10.1038/nature15759}
  {\bibfield  {journal} {\bibinfo  {journal} {Nature}\ }\textbf {\bibinfo
  {volume} {526}},\ \bibinfo {pages} {682} (\bibinfo {year}
  {2015})}\BibitemShut {NoStop}%
\bibitem [{\citenamefont {Monroe}\ \emph {et~al.}(2014)\citenamefont {Monroe},
  \citenamefont {Raussendorf}, \citenamefont {Ruthven}, \citenamefont {Brown},
  \citenamefont {Maunz}, \citenamefont {Duan},\ and\ \citenamefont
  {Kim}}]{ScalableArchitecture1}%
  \BibitemOpen
  \bibfield  {author} {\bibinfo {author} {\bibfnamefont {C.}~\bibnamefont
  {Monroe}}, \bibinfo {author} {\bibfnamefont {R.}~\bibnamefont {Raussendorf}},
  \bibinfo {author} {\bibfnamefont {A.}~\bibnamefont {Ruthven}}, \bibinfo
  {author} {\bibfnamefont {K.~R.}\ \bibnamefont {Brown}}, \bibinfo {author}
  {\bibfnamefont {P.}~\bibnamefont {Maunz}}, \bibinfo {author} {\bibfnamefont
  {L.-M.}\ \bibnamefont {Duan}}, \ and\ \bibinfo {author} {\bibfnamefont
  {J.}~\bibnamefont {Kim}},\ }\href {\doibase 10.1103/PhysRevA.89.022317}
  {\bibfield  {journal} {\bibinfo  {journal} {Phys. Rev. A}\ }\textbf {\bibinfo
  {volume} {89}},\ \bibinfo {pages} {022317} (\bibinfo {year}
  {2014})}\BibitemShut {NoStop}%
\bibitem [{\citenamefont {Cabrillo}\ \emph {et~al.}(1999)\citenamefont
  {Cabrillo}, \citenamefont {Cirac}, \citenamefont {Garc\'{\i}a-Fern\'andez},\
  and\ \citenamefont {Zoller}}]{Cabrillo99}%
  \BibitemOpen
  \bibfield  {author} {\bibinfo {author} {\bibfnamefont {C.}~\bibnamefont
  {Cabrillo}}, \bibinfo {author} {\bibfnamefont {J.~I.}\ \bibnamefont {Cirac}},
  \bibinfo {author} {\bibfnamefont {P.}~\bibnamefont
  {Garc\'{\i}a-Fern\'andez}}, \ and\ \bibinfo {author} {\bibfnamefont
  {P.}~\bibnamefont {Zoller}},\ }\href {\doibase 10.1103/PhysRevA.59.1025}
  {\bibfield  {journal} {\bibinfo  {journal} {Phys. Rev. A}\ }\textbf {\bibinfo
  {volume} {59}},\ \bibinfo {pages} {1025} (\bibinfo {year}
  {1999})}\BibitemShut {NoStop}%
\bibitem [{\citenamefont {Chou}\ \emph {et~al.}(2005)\citenamefont {Chou},
  \citenamefont {de~Riedmatten}, \citenamefont {Felinto}, \citenamefont
  {Polyakov}, \citenamefont {van Enk},\ and\ \citenamefont
  {Kimble}}]{Chou2005}%
  \BibitemOpen
  \bibfield  {author} {\bibinfo {author} {\bibfnamefont {C.~W.}\ \bibnamefont
  {Chou}}, \bibinfo {author} {\bibfnamefont {H.}~\bibnamefont {de~Riedmatten}},
  \bibinfo {author} {\bibfnamefont {D.}~\bibnamefont {Felinto}}, \bibinfo
  {author} {\bibfnamefont {S.~V.}\ \bibnamefont {Polyakov}}, \bibinfo {author}
  {\bibfnamefont {S.~J.}\ \bibnamefont {van Enk}}, \ and\ \bibinfo {author}
  {\bibfnamefont {H.~J.}\ \bibnamefont {Kimble}},\ }\href
  {http://dx.doi.org/10.1038/nature04353} {\bibfield  {journal} {\bibinfo
  {journal} {Nature}\ }\textbf {\bibinfo {volume} {438}},\ \bibinfo {pages}
  {828} (\bibinfo {year} {2005})}\BibitemShut {NoStop}%
\bibitem [{\citenamefont {Moehring}\ \emph {et~al.}(2007)\citenamefont
  {Moehring}, \citenamefont {Maunz}, \citenamefont {Olmschenk}, \citenamefont
  {Younge}, \citenamefont {Matsukevich}, \citenamefont {Duan},\ and\
  \citenamefont {Monroe}}]{Moehring07}%
  \BibitemOpen
  \bibfield  {author} {\bibinfo {author} {\bibfnamefont {D.~L.}\ \bibnamefont
  {Moehring}}, \bibinfo {author} {\bibfnamefont {P.}~\bibnamefont {Maunz}},
  \bibinfo {author} {\bibfnamefont {S.}~\bibnamefont {Olmschenk}}, \bibinfo
  {author} {\bibfnamefont {K.~C.}\ \bibnamefont {Younge}}, \bibinfo {author}
  {\bibfnamefont {D.~N.}\ \bibnamefont {Matsukevich}}, \bibinfo {author}
  {\bibfnamefont {L.~M.}\ \bibnamefont {Duan}}, \ and\ \bibinfo {author}
  {\bibfnamefont {C.}~\bibnamefont {Monroe}},\ }\href
  {http://dx.doi.org/10.1038/nature06118} {\bibfield  {journal} {\bibinfo
  {journal} {Nature}\ }\textbf {\bibinfo {volume} {449}},\ \bibinfo {pages}
  {68} (\bibinfo {year} {2007})}\BibitemShut {NoStop}%
\bibitem [{\citenamefont {Hofmann}\ \emph {et~al.}(2012)\citenamefont
  {Hofmann}, \citenamefont {Krug}, \citenamefont {Ortegel}, \citenamefont
  {G{\'e}rard}, \citenamefont {Weber}, \citenamefont {Rosenfeld},\ and\
  \citenamefont {Weinfurter}}]{Hofmann2012}%
  \BibitemOpen
  \bibfield  {author} {\bibinfo {author} {\bibfnamefont {J.}~\bibnamefont
  {Hofmann}}, \bibinfo {author} {\bibfnamefont {M.}~\bibnamefont {Krug}},
  \bibinfo {author} {\bibfnamefont {N.}~\bibnamefont {Ortegel}}, \bibinfo
  {author} {\bibfnamefont {L.}~\bibnamefont {G{\'e}rard}}, \bibinfo {author}
  {\bibfnamefont {M.}~\bibnamefont {Weber}}, \bibinfo {author} {\bibfnamefont
  {W.}~\bibnamefont {Rosenfeld}}, \ and\ \bibinfo {author} {\bibfnamefont
  {H.}~\bibnamefont {Weinfurter}},\ }\href
  {http://science.sciencemag.org/content/337/6090/72.abstract} {\bibfield
  {journal} {\bibinfo  {journal} {Science}\ }\textbf {\bibinfo {volume}
  {337}},\ \bibinfo {pages} {72} (\bibinfo {year} {2012})}\BibitemShut
  {NoStop}%
\bibitem [{\citenamefont {Bernien}\ \emph {et~al.}(2013)\citenamefont
  {Bernien}, \citenamefont {Hensen}, \citenamefont {Pfaff}, \citenamefont
  {Koolstra}, \citenamefont {Blok}, \citenamefont {Robledo}, \citenamefont
  {Taminiau}, \citenamefont {Markham}, \citenamefont {Twitchen}, \citenamefont
  {Childress},\ and\ \citenamefont {Hanson}}]{Bernien2013}%
  \BibitemOpen
  \bibfield  {author} {\bibinfo {author} {\bibfnamefont {H.}~\bibnamefont
  {Bernien}}, \bibinfo {author} {\bibfnamefont {B.}~\bibnamefont {Hensen}},
  \bibinfo {author} {\bibfnamefont {W.}~\bibnamefont {Pfaff}}, \bibinfo
  {author} {\bibfnamefont {G.}~\bibnamefont {Koolstra}}, \bibinfo {author}
  {\bibfnamefont {M.~S.}\ \bibnamefont {Blok}}, \bibinfo {author}
  {\bibfnamefont {L.}~\bibnamefont {Robledo}}, \bibinfo {author} {\bibfnamefont
  {T.~H.}\ \bibnamefont {Taminiau}}, \bibinfo {author} {\bibfnamefont
  {M.}~\bibnamefont {Markham}}, \bibinfo {author} {\bibfnamefont {D.~J.}\
  \bibnamefont {Twitchen}}, \bibinfo {author} {\bibfnamefont {L.}~\bibnamefont
  {Childress}}, \ and\ \bibinfo {author} {\bibfnamefont {R.}~\bibnamefont
  {Hanson}},\ }\href {http://dx.doi.org/10.1038/nature12016} {\bibfield
  {journal} {\bibinfo  {journal} {Nature}\ }\textbf {\bibinfo {volume} {497}},\
  \bibinfo {pages} {86} (\bibinfo {year} {2013})}\BibitemShut {NoStop}%
\bibitem [{\citenamefont {Olmschenk}\ \emph {et~al.}(2009)\citenamefont
  {Olmschenk}, \citenamefont {Matsukevich}, \citenamefont {Maunz},
  \citenamefont {Hayes}, \citenamefont {Duan},\ and\ \citenamefont
  {Monroe}}]{Olmschenk2009}%
  \BibitemOpen
  \bibfield  {author} {\bibinfo {author} {\bibfnamefont {S.}~\bibnamefont
  {Olmschenk}}, \bibinfo {author} {\bibfnamefont {D.~N.}\ \bibnamefont
  {Matsukevich}}, \bibinfo {author} {\bibfnamefont {P.}~\bibnamefont {Maunz}},
  \bibinfo {author} {\bibfnamefont {D.}~\bibnamefont {Hayes}}, \bibinfo
  {author} {\bibfnamefont {L.~M.}\ \bibnamefont {Duan}}, \ and\ \bibinfo
  {author} {\bibfnamefont {C.}~\bibnamefont {Monroe}},\ }\href
  {http://science.sciencemag.org/content/323/5913/486.abstract} {\bibfield
  {journal} {\bibinfo  {journal} {Science}\ }\textbf {\bibinfo {volume}
  {323}},\ \bibinfo {pages} {486} (\bibinfo {year} {2009})}\BibitemShut
  {NoStop}%
\bibitem [{\citenamefont {Bao}\ \emph {et~al.}(2012)\citenamefont {Bao},
  \citenamefont {Xu}, \citenamefont {Li}, \citenamefont {Yuan}, \citenamefont
  {Lu},\ and\ \citenamefont {Pan}}]{Bao2012}%
  \BibitemOpen
  \bibfield  {author} {\bibinfo {author} {\bibfnamefont {X.-H.}\ \bibnamefont
  {Bao}}, \bibinfo {author} {\bibfnamefont {X.-F.}\ \bibnamefont {Xu}},
  \bibinfo {author} {\bibfnamefont {C.-M.}\ \bibnamefont {Li}}, \bibinfo
  {author} {\bibfnamefont {Z.-S.}\ \bibnamefont {Yuan}}, \bibinfo {author}
  {\bibfnamefont {C.-Y.}\ \bibnamefont {Lu}}, \ and\ \bibinfo {author}
  {\bibfnamefont {J.-W.}\ \bibnamefont {Pan}},\ }\href
  {http://www.pnas.org/content/109/50/20347} {\bibfield  {journal} {\bibinfo
  {journal} {Proceedings of the National Academy of Sciences}\ }\textbf
  {\bibinfo {volume} {109}},\ \bibinfo {pages} {20347} (\bibinfo {year}
  {2012})}\BibitemShut {NoStop}%
\bibitem [{\citenamefont {Krauter}\ \emph {et~al.}(2013)\citenamefont
  {Krauter}, \citenamefont {Salart}, \citenamefont {Muschik}, \citenamefont
  {Petersen}, \citenamefont {Shen}, \citenamefont {Fernholz},\ and\
  \citenamefont {Polzik}}]{Krauter2013}%
  \BibitemOpen
  \bibfield  {author} {\bibinfo {author} {\bibfnamefont {H.}~\bibnamefont
  {Krauter}}, \bibinfo {author} {\bibfnamefont {D.}~\bibnamefont {Salart}},
  \bibinfo {author} {\bibfnamefont {C.~A.}\ \bibnamefont {Muschik}}, \bibinfo
  {author} {\bibfnamefont {J.~M.}\ \bibnamefont {Petersen}}, \bibinfo {author}
  {\bibfnamefont {H.}~\bibnamefont {Shen}}, \bibinfo {author} {\bibfnamefont
  {T.}~\bibnamefont {Fernholz}}, \ and\ \bibinfo {author} {\bibfnamefont
  {E.~S.}\ \bibnamefont {Polzik}},\ }\href
  {http://dx.doi.org/10.1038/nphys2631} {\bibfield  {journal} {\bibinfo
  {journal} {Nat Phys}\ }\textbf {\bibinfo {volume} {9}},\ \bibinfo {pages}
  {400} (\bibinfo {year} {2013})}\BibitemShut {NoStop}%
\bibitem [{\citenamefont {N\"olleke}\ \emph {et~al.}(2013)\citenamefont
  {N\"olleke}, \citenamefont {Neuzner}, \citenamefont {Reiserer}, \citenamefont
  {Hahn}, \citenamefont {Rempe},\ and\ \citenamefont {Ritter}}]{Noelleke2013}%
  \BibitemOpen
  \bibfield  {author} {\bibinfo {author} {\bibfnamefont {C.}~\bibnamefont
  {N\"olleke}}, \bibinfo {author} {\bibfnamefont {A.}~\bibnamefont {Neuzner}},
  \bibinfo {author} {\bibfnamefont {A.}~\bibnamefont {Reiserer}}, \bibinfo
  {author} {\bibfnamefont {C.}~\bibnamefont {Hahn}}, \bibinfo {author}
  {\bibfnamefont {G.}~\bibnamefont {Rempe}}, \ and\ \bibinfo {author}
  {\bibfnamefont {S.}~\bibnamefont {Ritter}},\ }\href {\doibase
  10.1103/PhysRevLett.110.140403} {\bibfield  {journal} {\bibinfo  {journal}
  {Phys. Rev. Lett.}\ }\textbf {\bibinfo {volume} {110}},\ \bibinfo {pages}
  {140403} (\bibinfo {year} {2013})}\BibitemShut {NoStop}%
\bibitem [{\citenamefont {Pfaff}\ \emph {et~al.}(2014)\citenamefont {Pfaff},
  \citenamefont {Hensen}, \citenamefont {Bernien}, \citenamefont {van Dam},
  \citenamefont {Blok}, \citenamefont {Taminiau}, \citenamefont {Tiggelman},
  \citenamefont {Schouten}, \citenamefont {Markham}, \citenamefont {Twitchen},\
  and\ \citenamefont {Hanson}}]{Pfaff2014}%
  \BibitemOpen
  \bibfield  {author} {\bibinfo {author} {\bibfnamefont {W.}~\bibnamefont
  {Pfaff}}, \bibinfo {author} {\bibfnamefont {B.~J.}\ \bibnamefont {Hensen}},
  \bibinfo {author} {\bibfnamefont {H.}~\bibnamefont {Bernien}}, \bibinfo
  {author} {\bibfnamefont {S.~B.}\ \bibnamefont {van Dam}}, \bibinfo {author}
  {\bibfnamefont {M.~S.}\ \bibnamefont {Blok}}, \bibinfo {author}
  {\bibfnamefont {T.~H.}\ \bibnamefont {Taminiau}}, \bibinfo {author}
  {\bibfnamefont {M.~J.}\ \bibnamefont {Tiggelman}}, \bibinfo {author}
  {\bibfnamefont {R.~N.}\ \bibnamefont {Schouten}}, \bibinfo {author}
  {\bibfnamefont {M.}~\bibnamefont {Markham}}, \bibinfo {author} {\bibfnamefont
  {D.~J.}\ \bibnamefont {Twitchen}}, \ and\ \bibinfo {author} {\bibfnamefont
  {R.}~\bibnamefont {Hanson}},\ }\href
  {http://science.sciencemag.org/content/345/6196/532.abstract} {\bibfield
  {journal} {\bibinfo  {journal} {Science}\ }\textbf {\bibinfo {volume}
  {345}},\ \bibinfo {pages} {532} (\bibinfo {year} {2014})}\BibitemShut
  {NoStop}%
\bibitem [{det()}]{detST}%
  \BibitemOpen
  \href@noop {} {}\bibinfo {note} {The process is called deterministic since a
  state transfer is completed in every attempt. Photon losses contribute here
  to the state transfer infidelity.}\BibitemShut {Stop}%
\bibitem [{\citenamefont {Cirac}\ \emph {et~al.}(1997)\citenamefont {Cirac},
  \citenamefont {Zoller}, \citenamefont {Kimble},\ and\ \citenamefont
  {Mabuchi}}]{Mabuchi1997}%
  \BibitemOpen
  \bibfield  {author} {\bibinfo {author} {\bibfnamefont {J.~I.}\ \bibnamefont
  {Cirac}}, \bibinfo {author} {\bibfnamefont {P.}~\bibnamefont {Zoller}},
  \bibinfo {author} {\bibfnamefont {H.~J.}\ \bibnamefont {Kimble}}, \ and\
  \bibinfo {author} {\bibfnamefont {H.}~\bibnamefont {Mabuchi}},\ }\href
  {\doibase 10.1103/PhysRevLett.78.3221} {\bibfield  {journal} {\bibinfo
  {journal} {Phys. Rev. Lett.}\ }\textbf {\bibinfo {volume} {78}},\ \bibinfo
  {pages} {3221} (\bibinfo {year} {1997})}\BibitemShut {NoStop}%
\bibitem [{\citenamefont {Ritter}\ \emph {et~al.}(2012)\citenamefont {Ritter},
  \citenamefont {Nolleke}, \citenamefont {Hahn}, \citenamefont {Reiserer},
  \citenamefont {Neuzner}, \citenamefont {Uphoff}, \citenamefont {Mucke},
  \citenamefont {Figueroa}, \citenamefont {Bochmann},\ and\ \citenamefont
  {Rempe}}]{ExpDetTransfer}%
  \BibitemOpen
  \bibfield  {author} {\bibinfo {author} {\bibfnamefont {S.}~\bibnamefont
  {Ritter}}, \bibinfo {author} {\bibfnamefont {C.}~\bibnamefont {Nolleke}},
  \bibinfo {author} {\bibfnamefont {C.}~\bibnamefont {Hahn}}, \bibinfo {author}
  {\bibfnamefont {A.}~\bibnamefont {Reiserer}}, \bibinfo {author}
  {\bibfnamefont {A.}~\bibnamefont {Neuzner}}, \bibinfo {author} {\bibfnamefont
  {M.}~\bibnamefont {Uphoff}}, \bibinfo {author} {\bibfnamefont
  {M.}~\bibnamefont {Mucke}}, \bibinfo {author} {\bibfnamefont
  {E.}~\bibnamefont {Figueroa}}, \bibinfo {author} {\bibfnamefont
  {J.}~\bibnamefont {Bochmann}}, \ and\ \bibinfo {author} {\bibfnamefont
  {G.}~\bibnamefont {Rempe}},\ }\href {http://dx.doi.org/10.1038/nature11023}
  {\bibfield  {journal} {\bibinfo  {journal} {Nature}\ }\textbf {\bibinfo
  {volume} {484}},\ \bibinfo {pages} {195} (\bibinfo {year}
  {2012})}\BibitemShut {NoStop}%
\bibitem [{\citenamefont {Muschik}\ \emph {et~al.}(2013)\citenamefont
  {Muschik}, \citenamefont {Hammerer}, \citenamefont {Polzik},\ and\
  \citenamefont {Cirac}}]{ContTeleportation}%
  \BibitemOpen
  \bibfield  {author} {\bibinfo {author} {\bibfnamefont {C.~A.}\ \bibnamefont
  {Muschik}}, \bibinfo {author} {\bibfnamefont {K.}~\bibnamefont {Hammerer}},
  \bibinfo {author} {\bibfnamefont {E.~S.}\ \bibnamefont {Polzik}}, \ and\
  \bibinfo {author} {\bibfnamefont {I.~J.}\ \bibnamefont {Cirac}},\ }\href
  {\doibase 10.1103/PhysRevLett.111.020501} {\bibfield  {journal} {\bibinfo
  {journal} {Phys. Rev. Lett.}\ }\textbf {\bibinfo {volume} {111}},\ \bibinfo
  {pages} {020501} (\bibinfo {year} {2013})}\BibitemShut {NoStop}%
\bibitem [{\citenamefont {Hofer}\ \emph {et~al.}(2013)\citenamefont {Hofer},
  \citenamefont {Vasilyev}, \citenamefont {Aspelmeyer},\ and\ \citenamefont
  {Hammerer}}]{Hofer2013}%
  \BibitemOpen
  \bibfield  {author} {\bibinfo {author} {\bibfnamefont {S.~G.}\ \bibnamefont
  {Hofer}}, \bibinfo {author} {\bibfnamefont {D.~V.}\ \bibnamefont {Vasilyev}},
  \bibinfo {author} {\bibfnamefont {M.}~\bibnamefont {Aspelmeyer}}, \ and\
  \bibinfo {author} {\bibfnamefont {K.}~\bibnamefont {Hammerer}},\ }\href
  {https://link.aps.org/doi/10.1103/PhysRevLett.111.170404} {\bibfield
  {journal} {\bibinfo  {journal} {Physical Review Letters}\ }\textbf {\bibinfo
  {volume} {111}},\ \bibinfo {pages} {170404} (\bibinfo {year}
  {2013})}\BibitemShut {NoStop}%
\bibitem [{\citenamefont {Vollbrecht}\ \emph {et~al.}(2011)\citenamefont
  {Vollbrecht}, \citenamefont {Muschik},\ and\ \citenamefont
  {Cirac}}]{Vollbrecht2011}%
  \BibitemOpen
  \bibfield  {author} {\bibinfo {author} {\bibfnamefont {K.~G.~H.}\
  \bibnamefont {Vollbrecht}}, \bibinfo {author} {\bibfnamefont {C.~A.}\
  \bibnamefont {Muschik}}, \ and\ \bibinfo {author} {\bibfnamefont {J.~I.}\
  \bibnamefont {Cirac}},\ }\href
  {https://link.aps.org/doi/10.1103/PhysRevLett.107.120502} {\bibfield
  {journal} {\bibinfo  {journal} {Physical Review Letters}\ }\textbf {\bibinfo
  {volume} {107}},\ \bibinfo {pages} {120502} (\bibinfo {year}
  {2011})}\BibitemShut {NoStop}%
\bibitem [{\citenamefont {Duan}\ and\ \citenamefont {Monroe}(2010)}]{Duan2010}%
  \BibitemOpen
  \bibfield  {author} {\bibinfo {author} {\bibfnamefont {L.~M.}\ \bibnamefont
  {Duan}}\ and\ \bibinfo {author} {\bibfnamefont {C.}~\bibnamefont {Monroe}},\
  }\href {http://link.aps.org/doi/10.1103/RevModPhys.82.1209} {\bibfield
  {journal} {\bibinfo  {journal} {Reviews of Modern Physics}\ }\textbf
  {\bibinfo {volume} {82}},\ \bibinfo {pages} {1209} (\bibinfo {year}
  {2010})}\BibitemShut {NoStop}%
\bibitem [{\citenamefont {Pellizzari}(1997)}]{Pellizzari1997}%
  \BibitemOpen
  \bibfield  {author} {\bibinfo {author} {\bibfnamefont {T.}~\bibnamefont
  {Pellizzari}},\ }\href {\doibase 10.1103/PhysRevLett.79.5242} {\bibfield
  {journal} {\bibinfo  {journal} {Phys. Rev. Lett.}\ }\textbf {\bibinfo
  {volume} {79}},\ \bibinfo {pages} {5242} (\bibinfo {year}
  {1997})}\BibitemShut {NoStop}%
\bibitem [{\citenamefont {Serafini}\ \emph {et~al.}(2006)\citenamefont
  {Serafini}, \citenamefont {Mancini},\ and\ \citenamefont
  {Bose}}]{Serafini2006}%
  \BibitemOpen
  \bibfield  {author} {\bibinfo {author} {\bibfnamefont {A.}~\bibnamefont
  {Serafini}}, \bibinfo {author} {\bibfnamefont {S.}~\bibnamefont {Mancini}}, \
  and\ \bibinfo {author} {\bibfnamefont {S.}~\bibnamefont {Bose}},\ }\href
  {\doibase 10.1103/PhysRevLett.96.010503} {\bibfield  {journal} {\bibinfo
  {journal} {Phys. Rev. Lett.}\ }\textbf {\bibinfo {volume} {96}},\ \bibinfo
  {pages} {010503} (\bibinfo {year} {2006})}\BibitemShut {NoStop}%
\bibitem [{\citenamefont {Yin}\ and\ \citenamefont {Li}(2007)}]{Yin2007}%
  \BibitemOpen
  \bibfield  {author} {\bibinfo {author} {\bibfnamefont {Z.-Q.}\ \bibnamefont
  {Yin}}\ and\ \bibinfo {author} {\bibfnamefont {F.-L.}\ \bibnamefont {Li}},\
  }\href {http://link.aps.org/doi/10.1103/PhysRevA.75.012324} {\bibfield
  {journal} {\bibinfo  {journal} {Physical Review A}\ }\textbf {\bibinfo
  {volume} {75}},\ \bibinfo {pages} {012324} (\bibinfo {year}
  {2007})}\BibitemShut {NoStop}%
\bibitem [{\citenamefont {Chen}\ \emph {et~al.}(2007)\citenamefont {Chen},
  \citenamefont {Ye}, \citenamefont {Lin}, \citenamefont {Du},\ and\
  \citenamefont {Lin}}]{Chen2007}%
  \BibitemOpen
  \bibfield  {author} {\bibinfo {author} {\bibfnamefont {L.-B.}\ \bibnamefont
  {Chen}}, \bibinfo {author} {\bibfnamefont {M.-Y.}\ \bibnamefont {Ye}},
  \bibinfo {author} {\bibfnamefont {G.-W.}\ \bibnamefont {Lin}}, \bibinfo
  {author} {\bibfnamefont {Q.-H.}\ \bibnamefont {Du}}, \ and\ \bibinfo {author}
  {\bibfnamefont {X.-M.}\ \bibnamefont {Lin}},\ }\href
  {http://link.aps.org/doi/10.1103/PhysRevA.76.062304} {\bibfield  {journal}
  {\bibinfo  {journal} {Physical Review A}\ }\textbf {\bibinfo {volume} {76}},\
  \bibinfo {pages} {062304} (\bibinfo {year} {2007})}\BibitemShut {NoStop}%
\bibitem [{\citenamefont {Ye}\ \emph {et~al.}(2008)\citenamefont {Ye},
  \citenamefont {Zhong},\ and\ \citenamefont {Zheng}}]{Ye2008}%
  \BibitemOpen
  \bibfield  {author} {\bibinfo {author} {\bibfnamefont {S.-Y.}\ \bibnamefont
  {Ye}}, \bibinfo {author} {\bibfnamefont {Z.-R.}\ \bibnamefont {Zhong}}, \
  and\ \bibinfo {author} {\bibfnamefont {S.-B.}\ \bibnamefont {Zheng}},\ }\href
  {http://link.aps.org/doi/10.1103/PhysRevA.77.014303} {\bibfield  {journal}
  {\bibinfo  {journal} {Physical Review A}\ }\textbf {\bibinfo {volume} {77}},\
  \bibinfo {pages} {014303} (\bibinfo {year} {2008})}\BibitemShut {NoStop}%
\bibitem [{\citenamefont {L{\"u}}\ \emph {et~al.}(2008)\citenamefont {L{\"u}},
  \citenamefont {Liu}, \citenamefont {Ding},\ and\ \citenamefont
  {Li}}]{Lu2008}%
  \BibitemOpen
  \bibfield  {author} {\bibinfo {author} {\bibfnamefont {X.-Y.}\ \bibnamefont
  {L{\"u}}}, \bibinfo {author} {\bibfnamefont {J.-B.}\ \bibnamefont {Liu}},
  \bibinfo {author} {\bibfnamefont {C.-L.}\ \bibnamefont {Ding}}, \ and\
  \bibinfo {author} {\bibfnamefont {J.~H.}\ \bibnamefont {Li}},\ }\href
  {http://link.aps.org/doi/10.1103/PhysRevA.78.032305} {\bibfield  {journal}
  {\bibinfo  {journal} {Physical Review A}\ }\textbf {\bibinfo {volume} {78}},\
  \bibinfo {pages} {032305} (\bibinfo {year} {2008})}\BibitemShut {NoStop}%
\bibitem [{\citenamefont {Zhou}\ \emph {et~al.}(2009)\citenamefont {Zhou},
  \citenamefont {Wang}, \citenamefont {Liang},\ and\ \citenamefont
  {Li}}]{Zhou2009}%
  \BibitemOpen
  \bibfield  {author} {\bibinfo {author} {\bibfnamefont {Y.~L.}\ \bibnamefont
  {Zhou}}, \bibinfo {author} {\bibfnamefont {Y.~M.}\ \bibnamefont {Wang}},
  \bibinfo {author} {\bibfnamefont {L.~M.}\ \bibnamefont {Liang}}, \ and\
  \bibinfo {author} {\bibfnamefont {C.~Z.}\ \bibnamefont {Li}},\ }\href
  {http://link.aps.org/doi/10.1103/PhysRevA.79.044304} {\bibfield  {journal}
  {\bibinfo  {journal} {Physical Review A}\ }\textbf {\bibinfo {volume} {79}},\
  \bibinfo {pages} {044304} (\bibinfo {year} {2009})}\BibitemShut {NoStop}%
\bibitem [{\citenamefont {Clader}(2014)}]{Clader2014}%
  \BibitemOpen
  \bibfield  {author} {\bibinfo {author} {\bibfnamefont {B.~D.}\ \bibnamefont
  {Clader}},\ }\href {\doibase 10.1103/PhysRevA.90.012324} {\bibfield
  {journal} {\bibinfo  {journal} {Phys. Rev. A}\ }\textbf {\bibinfo {volume}
  {90}},\ \bibinfo {pages} {012324} (\bibinfo {year} {2014})}\BibitemShut
  {NoStop}%
\bibitem [{\citenamefont {Chen}\ \emph {et~al.}(2015)\citenamefont {Chen},
  \citenamefont {Xia}, \citenamefont {Chen},\ and\ \citenamefont
  {Song}}]{Chen2015}%
  \BibitemOpen
  \bibfield  {author} {\bibinfo {author} {\bibfnamefont {Y.-H.}\ \bibnamefont
  {Chen}}, \bibinfo {author} {\bibfnamefont {Y.}~\bibnamefont {Xia}}, \bibinfo
  {author} {\bibfnamefont {Q.-Q.}\ \bibnamefont {Chen}}, \ and\ \bibinfo
  {author} {\bibfnamefont {J.}~\bibnamefont {Song}},\ }\href
  {http://link.aps.org/doi/10.1103/PhysRevA.91.012325} {\bibfield  {journal}
  {\bibinfo  {journal} {Physical Review A}\ }\textbf {\bibinfo {volume} {91}},\
  \bibinfo {pages} {012325} (\bibinfo {year} {2015})}\BibitemShut {NoStop}%
\bibitem [{\citenamefont {Hua}\ \emph {et~al.}(2015)\citenamefont {Hua},
  \citenamefont {Tao},\ and\ \citenamefont {Deng}}]{Hua2015}%
  \BibitemOpen
  \bibfield  {author} {\bibinfo {author} {\bibfnamefont {M.}~\bibnamefont
  {Hua}}, \bibinfo {author} {\bibfnamefont {M.-J.}\ \bibnamefont {Tao}}, \ and\
  \bibinfo {author} {\bibfnamefont {F.-G.}\ \bibnamefont {Deng}},\ }\href
  {https://arxiv.org/abs/1511.00090} {\bibfield  {journal} {\bibinfo  {journal}
  {arXiv:1511.00090 [quant-ph]}\ } (\bibinfo {year} {2015})}\BibitemShut
  {NoStop}%
\bibitem [{\citenamefont {Huang}\ \emph {et~al.}(2016)\citenamefont {Huang},
  \citenamefont {Chen},\ and\ \citenamefont {Wang}}]{Huang2016}%
  \BibitemOpen
  \bibfield  {author} {\bibinfo {author} {\bibfnamefont {X.-B.}\ \bibnamefont
  {Huang}}, \bibinfo {author} {\bibfnamefont {Y.-H.}\ \bibnamefont {Chen}}, \
  and\ \bibinfo {author} {\bibfnamefont {Z.}~\bibnamefont {Wang}},\ }\href
  {http://dx.doi.org/10.1038/srep25707} {\bibfield  {journal} {\bibinfo
  {journal} {Scientific Reports}\ }\textbf {\bibinfo {volume} {6}},\ \bibinfo
  {pages} {25707 EP } (\bibinfo {year} {2016})}\BibitemShut {NoStop}%
\bibitem [{\citenamefont {van Enk}\ \emph {et~al.}(1999)\citenamefont {van
  Enk}, \citenamefont {Kimble}, \citenamefont {Cirac},\ and\ \citenamefont
  {Zoller}}]{vanEnk1999}%
  \BibitemOpen
  \bibfield  {author} {\bibinfo {author} {\bibfnamefont {S.~J.}\ \bibnamefont
  {van Enk}}, \bibinfo {author} {\bibfnamefont {H.~J.}\ \bibnamefont {Kimble}},
  \bibinfo {author} {\bibfnamefont {J.~I.}\ \bibnamefont {Cirac}}, \ and\
  \bibinfo {author} {\bibfnamefont {P.}~\bibnamefont {Zoller}},\ }\href
  {\doibase 10.1103/PhysRevA.59.2659} {\bibfield  {journal} {\bibinfo
  {journal} {Phys. Rev. A}\ }\textbf {\bibinfo {volume} {59}},\ \bibinfo
  {pages} {2659} (\bibinfo {year} {1999})}\BibitemShut {NoStop}%
\bibitem [{\citenamefont {Boozer}\ \emph {et~al.}(2007)\citenamefont {Boozer},
  \citenamefont {Boca}, \citenamefont {Miller}, \citenamefont {Northup},\ and\
  \citenamefont {Kimble}}]{Boozer2007}%
  \BibitemOpen
  \bibfield  {author} {\bibinfo {author} {\bibfnamefont {A.~D.}\ \bibnamefont
  {Boozer}}, \bibinfo {author} {\bibfnamefont {A.}~\bibnamefont {Boca}},
  \bibinfo {author} {\bibfnamefont {R.}~\bibnamefont {Miller}}, \bibinfo
  {author} {\bibfnamefont {T.~E.}\ \bibnamefont {Northup}}, \ and\ \bibinfo
  {author} {\bibfnamefont {H.~J.}\ \bibnamefont {Kimble}},\ }\href {\doibase
  10.1103/PhysRevLett.98.193601} {\bibfield  {journal} {\bibinfo  {journal}
  {Phys. Rev. Lett.}\ }\textbf {\bibinfo {volume} {98}},\ \bibinfo {pages}
  {193601} (\bibinfo {year} {2007})}\BibitemShut {NoStop}%
\bibitem [{pol()}]{polEncoding}%
  \BibitemOpen
  \href@noop {} {}\bibinfo {note} {If polarization qubits are used, a state
  transfer protocol can be employed, where photon losses do or do not lead to a
  result outside of the computational subspace. We are interested in the latter
  case. The state transfer given in \eeqref{eq:statetransfer} can for example
  be performed by mapping the emitter qubit to horizontally (H) and vertically
  (V) polarized photon states ($\ket{0}_A\rightarrow \ket{H}_P$,
  $\ket{1}_A\rightarrow \ket{V}_P$) and transferring the receiving qubit from
  its initial state $\ket{0}_B$ to the state $\ket{1}_B$, in case the
  transmitted photon is vertically polarized ($\ket{V}_P\rightarrow
  \ket{1}_B$). The considerations in this work also apply for this encoding. In
  a broader context, alternative polarization encoding strategies can also be
  implemented, e.g., as illustrated in Fig.~1 in
  Ref.~\cite{TracyReview}.}\BibitemShut {Stop}%
\bibitem [{\citenamefont {Stannigel}\ \emph {et~al.}(2011)\citenamefont
  {Stannigel}, \citenamefont {Rabl}, \citenamefont {S\o{}rensen}, \citenamefont
  {Lukin},\ and\ \citenamefont {Zoller}}]{Stannigel2011}%
  \BibitemOpen
  \bibfield  {author} {\bibinfo {author} {\bibfnamefont {K.}~\bibnamefont
  {Stannigel}}, \bibinfo {author} {\bibfnamefont {P.}~\bibnamefont {Rabl}},
  \bibinfo {author} {\bibfnamefont {A.~S.}\ \bibnamefont {S\o{}rensen}},
  \bibinfo {author} {\bibfnamefont {M.~D.}\ \bibnamefont {Lukin}}, \ and\
  \bibinfo {author} {\bibfnamefont {P.}~\bibnamefont {Zoller}},\ }\href
  {\doibase 10.1103/PhysRevA.84.042341} {\bibfield  {journal} {\bibinfo
  {journal} {Phys. Rev. A}\ }\textbf {\bibinfo {volume} {84}},\ \bibinfo
  {pages} {042341} (\bibinfo {year} {2011})}\BibitemShut {NoStop}%
\bibitem [{\citenamefont {Gorshkov}\ \emph {et~al.}(2007)\citenamefont
  {Gorshkov}, \citenamefont {Andr{\'e}}, \citenamefont {Lukin},\ and\
  \citenamefont {S{\o}rensen}}]{Gorshkov2007}%
  \BibitemOpen
  \bibfield  {author} {\bibinfo {author} {\bibfnamefont {A.~V.}\ \bibnamefont
  {Gorshkov}}, \bibinfo {author} {\bibfnamefont {A.}~\bibnamefont {Andr{\'e}}},
  \bibinfo {author} {\bibfnamefont {M.~D.}\ \bibnamefont {Lukin}}, \ and\
  \bibinfo {author} {\bibfnamefont {A.~S.}\ \bibnamefont {S{\o}rensen}},\
  }\href {https://link.aps.org/doi/10.1103/PhysRevA.76.033804} {\bibfield
  {journal} {\bibinfo  {journal} {Physical Review A}\ }\textbf {\bibinfo
  {volume} {76}},\ \bibinfo {pages} {033804} (\bibinfo {year}
  {2007})}\BibitemShut {NoStop}%
\bibitem [{\citenamefont {Fleischhauer}\ \emph {et~al.}(2000)\citenamefont
  {Fleischhauer}, \citenamefont {Yelin},\ and\ \citenamefont
  {Lukin}}]{Fleischhauer2000}%
  \BibitemOpen
  \bibfield  {author} {\bibinfo {author} {\bibfnamefont {M.}~\bibnamefont
  {Fleischhauer}}, \bibinfo {author} {\bibfnamefont {S.~F.}\ \bibnamefont
  {Yelin}}, \ and\ \bibinfo {author} {\bibfnamefont {M.~D.}\ \bibnamefont
  {Lukin}},\ }\href {\doibase http://dx.doi.org/10.1016/S0030-4018(99)00679-3}
  {\bibfield  {journal} {\bibinfo  {journal} {Optics Communications}\ }\textbf
  {\bibinfo {volume} {179}},\ \bibinfo {pages} {395} (\bibinfo {year}
  {2000})}\BibitemShut {NoStop}%
\bibitem [{\citenamefont {Dilley}\ \emph {et~al.}(2012)\citenamefont {Dilley},
  \citenamefont {Nisbet-Jones}, \citenamefont {Shore},\ and\ \citenamefont
  {Kuhn}}]{Dilley2012}%
  \BibitemOpen
  \bibfield  {author} {\bibinfo {author} {\bibfnamefont {J.}~\bibnamefont
  {Dilley}}, \bibinfo {author} {\bibfnamefont {P.}~\bibnamefont
  {Nisbet-Jones}}, \bibinfo {author} {\bibfnamefont {B.~W.}\ \bibnamefont
  {Shore}}, \ and\ \bibinfo {author} {\bibfnamefont {A.}~\bibnamefont {Kuhn}},\
  }\href {https://link.aps.org/doi/10.1103/PhysRevA.85.023834} {\bibfield
  {journal} {\bibinfo  {journal} {Physical Review A}\ }\textbf {\bibinfo
  {volume} {85}},\ \bibinfo {pages} {023834} (\bibinfo {year}
  {2012})}\BibitemShut {NoStop}%
\bibitem [{\citenamefont {Vitanov}\ \emph {et~al.}(2017)\citenamefont
  {Vitanov}, \citenamefont {Rangelov}, \citenamefont {Shore},\ and\
  \citenamefont {Bergmann}}]{RMPAP2016}%
  \BibitemOpen
  \bibfield  {author} {\bibinfo {author} {\bibfnamefont {N.~V.}\ \bibnamefont
  {Vitanov}}, \bibinfo {author} {\bibfnamefont {A.~A.}\ \bibnamefont
  {Rangelov}}, \bibinfo {author} {\bibfnamefont {B.~W.}\ \bibnamefont {Shore}},
  \ and\ \bibinfo {author} {\bibfnamefont {K.}~\bibnamefont {Bergmann}},\
  }\href {https://link.aps.org/doi/10.1103/RevModPhys.89.015006} {\bibfield
  {journal} {\bibinfo  {journal} {Reviews of Modern Physics}\ }\textbf
  {\bibinfo {volume} {89}},\ \bibinfo {pages} {015006} (\bibinfo {year}
  {2017})}\BibitemShut {NoStop}%
\bibitem [{\citenamefont {Pfister}\ \emph {et~al.}(2016)\citenamefont
  {Pfister}, \citenamefont {Salz}, \citenamefont {Hettrich}, \citenamefont
  {Poschinger},\ and\ \citenamefont {Schmidt-Kaler}}]{Pfister2016}%
  \BibitemOpen
  \bibfield  {author} {\bibinfo {author} {\bibfnamefont {A.~D.}\ \bibnamefont
  {Pfister}}, \bibinfo {author} {\bibfnamefont {M.}~\bibnamefont {Salz}},
  \bibinfo {author} {\bibfnamefont {M.}~\bibnamefont {Hettrich}}, \bibinfo
  {author} {\bibfnamefont {U.~G.}\ \bibnamefont {Poschinger}}, \ and\ \bibinfo
  {author} {\bibfnamefont {F.}~\bibnamefont {Schmidt-Kaler}},\ }\href {\doibase
  10.1007/s00340-016-6362-7} {\bibfield  {journal} {\bibinfo  {journal}
  {Applied Physics B}\ }\textbf {\bibinfo {volume} {122}},\ \bibinfo {pages}
  {89} (\bibinfo {year} {2016})}\BibitemShut {NoStop}%
\bibitem [{\citenamefont {Stute}\ \emph {et~al.}(2012)\citenamefont {Stute},
  \citenamefont {Casabone}, \citenamefont {Schindler}, \citenamefont {Monz},
  \citenamefont {Schmidt}, \citenamefont {Brandstatter}, \citenamefont
  {Northup},\ and\ \citenamefont {Blatt}}]{Stute2012}%
  \BibitemOpen
  \bibfield  {author} {\bibinfo {author} {\bibfnamefont {A.}~\bibnamefont
  {Stute}}, \bibinfo {author} {\bibfnamefont {B.}~\bibnamefont {Casabone}},
  \bibinfo {author} {\bibfnamefont {P.}~\bibnamefont {Schindler}}, \bibinfo
  {author} {\bibfnamefont {T.}~\bibnamefont {Monz}}, \bibinfo {author}
  {\bibfnamefont {P.~O.}\ \bibnamefont {Schmidt}}, \bibinfo {author}
  {\bibfnamefont {B.}~\bibnamefont {Brandstatter}}, \bibinfo {author}
  {\bibfnamefont {T.~E.}\ \bibnamefont {Northup}}, \ and\ \bibinfo {author}
  {\bibfnamefont {R.}~\bibnamefont {Blatt}},\ }\href
  {http://dx.doi.org/10.1038/nature11120} {\bibfield  {journal} {\bibinfo
  {journal} {Nature}\ }\textbf {\bibinfo {volume} {485}},\ \bibinfo {pages}
  {482} (\bibinfo {year} {2012})}\BibitemShut {NoStop}%
\bibitem [{\citenamefont {Hunger}\ \emph {et~al.}(2010)\citenamefont {Hunger},
  \citenamefont {Steinmetz}, \citenamefont {Colombe}, \citenamefont {Deutsch},
  \citenamefont {H\"{a}nsch},\ and\ \citenamefont {Reichel}}]{Hunger2010}%
  \BibitemOpen
  \bibfield  {author} {\bibinfo {author} {\bibfnamefont {D.}~\bibnamefont
  {Hunger}}, \bibinfo {author} {\bibfnamefont {T.}~\bibnamefont {Steinmetz}},
  \bibinfo {author} {\bibfnamefont {Y.}~\bibnamefont {Colombe}}, \bibinfo
  {author} {\bibfnamefont {C.}~\bibnamefont {Deutsch}}, \bibinfo {author}
  {\bibfnamefont {T.~W.}\ \bibnamefont {H\"{a}nsch}}, \ and\ \bibinfo {author}
  {\bibfnamefont {J.}~\bibnamefont {Reichel}},\ }\href
  {http://stacks.iop.org/1367-2630/12/i=6/a=065038} {\bibfield  {journal}
  {\bibinfo  {journal} {New Journal of Physics}\ }\textbf {\bibinfo {volume}
  {12}},\ \bibinfo {pages} {065038} (\bibinfo {year} {2010})}\BibitemShut
  {NoStop}%
\bibitem [{\citenamefont {Steiner}\ \emph {et~al.}(2014)\citenamefont
  {Steiner}, \citenamefont {Meyer}, \citenamefont {Reichel},\ and\
  \citenamefont {K\"ohl}}]{Steiner2014}%
  \BibitemOpen
  \bibfield  {author} {\bibinfo {author} {\bibfnamefont {M.}~\bibnamefont
  {Steiner}}, \bibinfo {author} {\bibfnamefont {H.~M.}\ \bibnamefont {Meyer}},
  \bibinfo {author} {\bibfnamefont {J.}~\bibnamefont {Reichel}}, \ and\
  \bibinfo {author} {\bibfnamefont {M.}~\bibnamefont {K\"ohl}},\ }\href
  {\doibase 10.1103/PhysRevLett.113.263003} {\bibfield  {journal} {\bibinfo
  {journal} {Phys. Rev. Lett.}\ }\textbf {\bibinfo {volume} {113}},\ \bibinfo
  {pages} {263003} (\bibinfo {year} {2014})}\BibitemShut {NoStop}%
\bibitem [{\citenamefont {Hood}\ \emph {et~al.}(1998)\citenamefont {Hood},
  \citenamefont {Chapman}, \citenamefont {Lynn},\ and\ \citenamefont
  {Kimble}}]{Hood1998}%
  \BibitemOpen
  \bibfield  {author} {\bibinfo {author} {\bibfnamefont {C.~J.}\ \bibnamefont
  {Hood}}, \bibinfo {author} {\bibfnamefont {M.~S.}\ \bibnamefont {Chapman}},
  \bibinfo {author} {\bibfnamefont {T.~W.}\ \bibnamefont {Lynn}}, \ and\
  \bibinfo {author} {\bibfnamefont {H.~J.}\ \bibnamefont {Kimble}},\ }\href
  {\doibase 10.1103/PhysRevLett.80.4157} {\bibfield  {journal} {\bibinfo
  {journal} {Phys. Rev. Lett.}\ }\textbf {\bibinfo {volume} {80}},\ \bibinfo
  {pages} {4157} (\bibinfo {year} {1998})}\BibitemShut {NoStop}%
\bibitem [{\citenamefont {Hamsen}\ \emph {et~al.}(2017)\citenamefont {Hamsen},
  \citenamefont {Tolazzi}, \citenamefont {Wilk},\ and\ \citenamefont
  {Rempe}}]{Hamsen2016}%
  \BibitemOpen
  \bibfield  {author} {\bibinfo {author} {\bibfnamefont {C.}~\bibnamefont
  {Hamsen}}, \bibinfo {author} {\bibfnamefont {K.~N.}\ \bibnamefont {Tolazzi}},
  \bibinfo {author} {\bibfnamefont {T.}~\bibnamefont {Wilk}}, \ and\ \bibinfo
  {author} {\bibfnamefont {G.}~\bibnamefont {Rempe}},\ }\href
  {https://link.aps.org/doi/10.1103/PhysRevLett.118.133604} {\bibfield
  {journal} {\bibinfo  {journal} {Physical Review Letters}\ }\textbf {\bibinfo
  {volume} {118}},\ \bibinfo {pages} {133604} (\bibinfo {year}
  {2017})}\BibitemShut {NoStop}%
\bibitem [{\citenamefont {Chibani}(2016)}]{Chibani2016}%
  \BibitemOpen
  \bibfield  {author} {\bibinfo {author} {\bibfnamefont {H.}~\bibnamefont
  {Chibani}},\ }\emph {\bibinfo {title} {Photon Blockade with Memory and Slow
  Light using a Single Atom in an Optical Cavity}},\ \href
  {https://mediatum.ub.tum.de/doc/1293639/1293639.pdf} {\bibinfo {type}
  {Dissertation}},\ \bibinfo  {school} {Technische Universit{\"a}t
  M{\"u}nchen}, \bibinfo {address} {M{\"u}nchen} (\bibinfo {year}
  {2016})\BibitemShut {NoStop}%
\bibitem [{\citenamefont {Gallego}\ \emph {et~al.}(2016)\citenamefont
  {Gallego}, \citenamefont {Ghosh}, \citenamefont {Alavi}, \citenamefont {Alt},
  \citenamefont {Martinez-Dorantes}, \citenamefont {Meschede},\ and\
  \citenamefont {Ratschbacher}}]{Gallego2016}%
  \BibitemOpen
  \bibfield  {author} {\bibinfo {author} {\bibfnamefont {J.}~\bibnamefont
  {Gallego}}, \bibinfo {author} {\bibfnamefont {S.}~\bibnamefont {Ghosh}},
  \bibinfo {author} {\bibfnamefont {S.~K.}\ \bibnamefont {Alavi}}, \bibinfo
  {author} {\bibfnamefont {W.}~\bibnamefont {Alt}}, \bibinfo {author}
  {\bibfnamefont {M.}~\bibnamefont {Martinez-Dorantes}}, \bibinfo {author}
  {\bibfnamefont {D.}~\bibnamefont {Meschede}}, \ and\ \bibinfo {author}
  {\bibfnamefont {L.}~\bibnamefont {Ratschbacher}},\ }\href {\doibase
  10.1007/s00340-015-6281-z} {\bibfield  {journal} {\bibinfo  {journal}
  {Applied Physics B}\ }\textbf {\bibinfo {volume} {122}},\ \bibinfo {pages}
  {47} (\bibinfo {year} {2016})}\BibitemShut {NoStop}%
\bibitem [{WAl()}]{WAlt}%
  \BibitemOpen
  \href@noop {} {}\bibinfo {note} {W. Alt, personal communication.}\BibitemShut
  {Stop}%
\bibitem [{\citenamefont {Herskind}\ \emph {et~al.}(2008)\citenamefont
  {Herskind}, \citenamefont {Dantan}, \citenamefont {Langkilde-Lauesen},
  \citenamefont {Mortensen}, \citenamefont {S{\o}rensen},\ and\ \citenamefont
  {Drewsen}}]{Herskind2008}%
  \BibitemOpen
  \bibfield  {author} {\bibinfo {author} {\bibfnamefont {P.}~\bibnamefont
  {Herskind}}, \bibinfo {author} {\bibfnamefont {A.}~\bibnamefont {Dantan}},
  \bibinfo {author} {\bibfnamefont {M.~B.}\ \bibnamefont {Langkilde-Lauesen}},
  \bibinfo {author} {\bibfnamefont {A.}~\bibnamefont {Mortensen}}, \bibinfo
  {author} {\bibfnamefont {J.~L.}\ \bibnamefont {S{\o}rensen}}, \ and\ \bibinfo
  {author} {\bibfnamefont {M.}~\bibnamefont {Drewsen}},\ }\href {\doibase
  10.1007/s00340-008-3199-8} {\bibfield  {journal} {\bibinfo  {journal}
  {Applied Physics B}\ }\textbf {\bibinfo {volume} {93}},\ \bibinfo {pages}
  {373} (\bibinfo {year} {2008})}\BibitemShut {NoStop}%
\bibitem [{\citenamefont {Begley}\ \emph {et~al.}(2016)\citenamefont {Begley},
  \citenamefont {Vogt}, \citenamefont {Gulati}, \citenamefont {Takahashi},\
  and\ \citenamefont {Keller}}]{Begley2016}%
  \BibitemOpen
  \bibfield  {author} {\bibinfo {author} {\bibfnamefont {S.}~\bibnamefont
  {Begley}}, \bibinfo {author} {\bibfnamefont {M.}~\bibnamefont {Vogt}},
  \bibinfo {author} {\bibfnamefont {G.~K.}\ \bibnamefont {Gulati}}, \bibinfo
  {author} {\bibfnamefont {H.}~\bibnamefont {Takahashi}}, \ and\ \bibinfo
  {author} {\bibfnamefont {M.}~\bibnamefont {Keller}},\ }\href {\doibase
  10.1103/PhysRevLett.116.223001} {\bibfield  {journal} {\bibinfo  {journal}
  {Phys. Rev. Lett.}\ }\textbf {\bibinfo {volume} {116}},\ \bibinfo {pages}
  {223001} (\bibinfo {year} {2016})}\BibitemShut {NoStop}%
\bibitem [{MKe()}]{MKeller}%
  \BibitemOpen
  \href@noop {} {}\bibinfo {note} {M. Keller, personal
  communication.}\BibitemShut {Stop}%
\bibitem [{\citenamefont {Reiserer}(2014)}]{Reiserer2014}%
  \BibitemOpen
  \bibfield  {author} {\bibinfo {author} {\bibfnamefont {A.~A.}\ \bibnamefont
  {Reiserer}},\ }\emph {\bibinfo {title} {A controlled phase gate between a
  single atom and an optical photon}},\ \href
  {https://mediatum.ub.tum.de/doc/1192216/1192216.pdf} {\bibinfo {type}
  {Dissertation}},\ \bibinfo  {school} {Technische Universit{\"a}t
  M{\"u}nchen}, \bibinfo {address} {M{\"u}nchen} (\bibinfo {year}
  {2014})\BibitemShut {NoStop}%
\bibitem [{\citenamefont {Ramos}\ \emph {et~al.}(2016)\citenamefont {Ramos},
  \citenamefont {Vermersch}, \citenamefont {Hauke}, \citenamefont {Pichler},\
  and\ \citenamefont {Zoller}}]{Ramos2016}%
  \BibitemOpen
  \bibfield  {author} {\bibinfo {author} {\bibfnamefont {T.}~\bibnamefont
  {Ramos}}, \bibinfo {author} {\bibfnamefont {B.}~\bibnamefont {Vermersch}},
  \bibinfo {author} {\bibfnamefont {P.}~\bibnamefont {Hauke}}, \bibinfo
  {author} {\bibfnamefont {H.}~\bibnamefont {Pichler}}, \ and\ \bibinfo
  {author} {\bibfnamefont {P.}~\bibnamefont {Zoller}},\ }\href
  {https://link.aps.org/doi/10.1103/PhysRevA.93.062104} {\bibfield  {journal}
  {\bibinfo  {journal} {Physical Review A}\ }\textbf {\bibinfo {volume} {93}},\
  \bibinfo {pages} {062104} (\bibinfo {year} {2016})}\BibitemShut {NoStop}%
\bibitem [{\citenamefont {Vermersch}\ \emph {et~al.}(2017)\citenamefont
  {Vermersch}, \citenamefont {Guimond}, \citenamefont {Pichler},\ and\
  \citenamefont {Zoller}}]{Vermersch2016}%
  \BibitemOpen
  \bibfield  {author} {\bibinfo {author} {\bibfnamefont {B.}~\bibnamefont
  {Vermersch}}, \bibinfo {author} {\bibfnamefont {P.~O.}\ \bibnamefont
  {Guimond}}, \bibinfo {author} {\bibfnamefont {H.}~\bibnamefont {Pichler}}, \
  and\ \bibinfo {author} {\bibfnamefont {P.}~\bibnamefont {Zoller}},\ }\href
  {https://link.aps.org/doi/10.1103/PhysRevLett.118.133601} {\bibfield
  {journal} {\bibinfo  {journal} {Physical Review Letters}\ }\textbf {\bibinfo
  {volume} {118}},\ \bibinfo {pages} {133601} (\bibinfo {year}
  {2017})}\BibitemShut {NoStop}%
\bibitem [{\citenamefont {Claude Cohen-Tannoudji}(2004)}]{AtomPhotonInt}%
  \BibitemOpen
  \bibfield  {author} {\bibinfo {author} {\bibfnamefont {G.~G.}\ \bibnamefont
  {Claude Cohen-Tannoudji}, \bibfnamefont {Jacques Dupont-Roc}},\ }\href@noop
  {} {\emph {\bibinfo {title} {Atom - Photon Interactions: Basic Process and
  Appilcations}}}\ (\bibinfo  {publisher} {Wiley-VCH Verlag GmbH $\&$ Co.},\
  \bibinfo {year} {2004})\BibitemShut {NoStop}%
\bibitem [{\citenamefont {Pelc}\ \emph {et~al.}(2011)\citenamefont {Pelc},
  \citenamefont {Ma}, \citenamefont {Phillips}, \citenamefont {Zhang},
  \citenamefont {Langrock}, \citenamefont {Slattery}, \citenamefont {Tang},\
  and\ \citenamefont {Fejer}}]{Pelc2011}%
  \BibitemOpen
  \bibfield  {author} {\bibinfo {author} {\bibfnamefont {J.~S.}\ \bibnamefont
  {Pelc}}, \bibinfo {author} {\bibfnamefont {L.}~\bibnamefont {Ma}}, \bibinfo
  {author} {\bibfnamefont {C.~R.}\ \bibnamefont {Phillips}}, \bibinfo {author}
  {\bibfnamefont {Q.}~\bibnamefont {Zhang}}, \bibinfo {author} {\bibfnamefont
  {C.}~\bibnamefont {Langrock}}, \bibinfo {author} {\bibfnamefont
  {O.}~\bibnamefont {Slattery}}, \bibinfo {author} {\bibfnamefont
  {X.}~\bibnamefont {Tang}}, \ and\ \bibinfo {author} {\bibfnamefont {M.~M.}\
  \bibnamefont {Fejer}},\ }\bibfield  {booktitle} {\emph {\bibinfo {booktitle}
  {Optics Express}},\ }\href {\doibase 10.1364/OE.19.021445} {\bibfield
  {journal} {\bibinfo  {journal} {Optics Express}\ }\textbf {\bibinfo {volume}
  {19}},\ \bibinfo {pages} {21445} (\bibinfo {year} {2011})}\BibitemShut
  {NoStop}%
\bibitem [{\citenamefont {Dorner}\ and\ \citenamefont
  {Zoller}(2002)}]{Dorner2002}%
  \BibitemOpen
  \bibfield  {author} {\bibinfo {author} {\bibfnamefont {U.}~\bibnamefont
  {Dorner}}\ and\ \bibinfo {author} {\bibfnamefont {P.}~\bibnamefont
  {Zoller}},\ }\href {http://link.aps.org/doi/10.1103/PhysRevA.66.023816}
  {\bibfield  {journal} {\bibinfo  {journal} {Physical Review A}\ }\textbf
  {\bibinfo {volume} {66}},\ \bibinfo {pages} {023816} (\bibinfo {year}
  {2002})}\BibitemShut {NoStop}%
\bibitem [{\citenamefont {Habraken}\ \emph {et~al.}(2012)\citenamefont
  {Habraken}, \citenamefont {Stannigel}, \citenamefont {Lukin}, \citenamefont
  {Zoller},\ and\ \citenamefont {Rabl}}]{Habraken2012}%
  \BibitemOpen
  \bibfield  {author} {\bibinfo {author} {\bibfnamefont {S.~J.~M.}\
  \bibnamefont {Habraken}}, \bibinfo {author} {\bibfnamefont {K.}~\bibnamefont
  {Stannigel}}, \bibinfo {author} {\bibfnamefont {M.~D.}\ \bibnamefont
  {Lukin}}, \bibinfo {author} {\bibfnamefont {P.}~\bibnamefont {Zoller}}, \
  and\ \bibinfo {author} {\bibfnamefont {P.}~\bibnamefont {Rabl}},\ }\href
  {http://stacks.iop.org/1367-2630/14/i=11/a=115004} {\bibfield  {journal}
  {\bibinfo  {journal} {New Journal of Physics}\ }\textbf {\bibinfo {volume}
  {14}},\ \bibinfo {pages} {115004} (\bibinfo {year} {2012})}\BibitemShut
  {NoStop}%
\bibitem [{\citenamefont {Vitanov}\ and\ \citenamefont
  {Stenholm}(1997{\natexlab{a}})}]{Vitanov1997b}%
  \BibitemOpen
  \bibfield  {author} {\bibinfo {author} {\bibfnamefont {N.~V.}\ \bibnamefont
  {Vitanov}}\ and\ \bibinfo {author} {\bibfnamefont {S.}~\bibnamefont
  {Stenholm}},\ }\href {\doibase 10.1103/PhysRevA.55.648} {\bibfield  {journal}
  {\bibinfo  {journal} {Phys. Rev. A}\ }\textbf {\bibinfo {volume} {55}},\
  \bibinfo {pages} {648} (\bibinfo {year} {1997}{\natexlab{a}})}\BibitemShut
  {NoStop}%
\bibitem [{WPS()}]{WPS}%
  \BibitemOpen
  \href@noop {} {}\bibinfo {note} {Note that wave packet shaping can also be
  evaluated in other regimes~\cite{Mabuchi1997}.}\BibitemShut {Stop}%
\bibitem [{\citenamefont {Hood}\ \emph {et~al.}(2000)\citenamefont {Hood},
  \citenamefont {Lynn}, \citenamefont {Doherty}, \citenamefont {Parkins},\ and\
  \citenamefont {Kimble}}]{Hood2000}%
  \BibitemOpen
  \bibfield  {author} {\bibinfo {author} {\bibfnamefont {C.~J.}\ \bibnamefont
  {Hood}}, \bibinfo {author} {\bibfnamefont {T.~W.}\ \bibnamefont {Lynn}},
  \bibinfo {author} {\bibfnamefont {A.~C.}\ \bibnamefont {Doherty}}, \bibinfo
  {author} {\bibfnamefont {A.~S.}\ \bibnamefont {Parkins}}, \ and\ \bibinfo
  {author} {\bibfnamefont {H.~J.}\ \bibnamefont {Kimble}},\ }\href
  {http://science.sciencemag.org/content/287/5457/1447.abstract} {\bibfield
  {journal} {\bibinfo  {journal} {Science}\ }\textbf {\bibinfo {volume}
  {287}},\ \bibinfo {pages} {1447} (\bibinfo {year} {2000})}\BibitemShut
  {NoStop}%
\bibitem [{\citenamefont {Colombe}\ \emph {et~al.}(2007)\citenamefont
  {Colombe}, \citenamefont {Steinmetz}, \citenamefont {Dubois}, \citenamefont
  {Linke}, \citenamefont {Hunger},\ and\ \citenamefont
  {Reichel}}]{Colombe2007}%
  \BibitemOpen
  \bibfield  {author} {\bibinfo {author} {\bibfnamefont {Y.}~\bibnamefont
  {Colombe}}, \bibinfo {author} {\bibfnamefont {T.}~\bibnamefont {Steinmetz}},
  \bibinfo {author} {\bibfnamefont {G.}~\bibnamefont {Dubois}}, \bibinfo
  {author} {\bibfnamefont {F.}~\bibnamefont {Linke}}, \bibinfo {author}
  {\bibfnamefont {D.}~\bibnamefont {Hunger}}, \ and\ \bibinfo {author}
  {\bibfnamefont {J.}~\bibnamefont {Reichel}},\ }\href
  {http://dx.doi.org/10.1038/nature06331} {\bibfield  {journal} {\bibinfo
  {journal} {Nature}\ }\textbf {\bibinfo {volume} {450}},\ \bibinfo {pages}
  {272} (\bibinfo {year} {2007})}\BibitemShut {NoStop}%
\bibitem [{\citenamefont {Wenner}\ \emph {et~al.}(2014)\citenamefont {Wenner},
  \citenamefont {Yin}, \citenamefont {Chen}, \citenamefont {Barends},
  \citenamefont {Chiaro}, \citenamefont {Jeffrey}, \citenamefont {Kelly},
  \citenamefont {Megrant}, \citenamefont {Mutus}, \citenamefont {Neill},
  \citenamefont {O'Malley}, \citenamefont {Roushan}, \citenamefont {Sank},
  \citenamefont {Vainsencher}, \citenamefont {White}, \citenamefont {Korotkov},
  \citenamefont {Cleland},\ and\ \citenamefont {Martinis}}]{Wenner2014}%
  \BibitemOpen
  \bibfield  {author} {\bibinfo {author} {\bibfnamefont {J.}~\bibnamefont
  {Wenner}}, \bibinfo {author} {\bibfnamefont {Y.}~\bibnamefont {Yin}},
  \bibinfo {author} {\bibfnamefont {Y.}~\bibnamefont {Chen}}, \bibinfo {author}
  {\bibfnamefont {R.}~\bibnamefont {Barends}}, \bibinfo {author} {\bibfnamefont
  {B.}~\bibnamefont {Chiaro}}, \bibinfo {author} {\bibfnamefont
  {E.}~\bibnamefont {Jeffrey}}, \bibinfo {author} {\bibfnamefont
  {J.}~\bibnamefont {Kelly}}, \bibinfo {author} {\bibfnamefont
  {A.}~\bibnamefont {Megrant}}, \bibinfo {author} {\bibfnamefont {J.~Y.}\
  \bibnamefont {Mutus}}, \bibinfo {author} {\bibfnamefont {C.}~\bibnamefont
  {Neill}}, \bibinfo {author} {\bibfnamefont {P.~J.~J.}\ \bibnamefont
  {O'Malley}}, \bibinfo {author} {\bibfnamefont {P.}~\bibnamefont {Roushan}},
  \bibinfo {author} {\bibfnamefont {D.}~\bibnamefont {Sank}}, \bibinfo {author}
  {\bibfnamefont {A.}~\bibnamefont {Vainsencher}}, \bibinfo {author}
  {\bibfnamefont {T.~C.}\ \bibnamefont {White}}, \bibinfo {author}
  {\bibfnamefont {A.~N.}\ \bibnamefont {Korotkov}}, \bibinfo {author}
  {\bibfnamefont {A.~N.}\ \bibnamefont {Cleland}}, \ and\ \bibinfo {author}
  {\bibfnamefont {J.~M.}\ \bibnamefont {Martinis}},\ }\href
  {http://link.aps.org/doi/10.1103/PhysRevLett.112.210501} {\bibfield
  {journal} {\bibinfo  {journal} {Physical Review Letters}\ }\textbf {\bibinfo
  {volume} {112}},\ \bibinfo {pages} {210501} (\bibinfo {year}
  {2014})}\BibitemShut {NoStop}%
\bibitem [{\citenamefont {Pfaff}\ \emph {et~al.}(2017)\citenamefont {Pfaff},
  \citenamefont {Axline}, \citenamefont {Burkhart}, \citenamefont {Vool},
  \citenamefont {Reinhold}, \citenamefont {Frunzio}, \citenamefont {Jiang},
  \citenamefont {Devoret},\ and\ \citenamefont {Schoelkopf}}]{Pfaff2016}%
  \BibitemOpen
  \bibfield  {author} {\bibinfo {author} {\bibfnamefont {W.}~\bibnamefont
  {Pfaff}}, \bibinfo {author} {\bibfnamefont {C.~J.}\ \bibnamefont {Axline}},
  \bibinfo {author} {\bibfnamefont {L.~D.}\ \bibnamefont {Burkhart}}, \bibinfo
  {author} {\bibfnamefont {U.}~\bibnamefont {Vool}}, \bibinfo {author}
  {\bibfnamefont {P.}~\bibnamefont {Reinhold}}, \bibinfo {author}
  {\bibfnamefont {L.}~\bibnamefont {Frunzio}}, \bibinfo {author} {\bibfnamefont
  {L.}~\bibnamefont {Jiang}}, \bibinfo {author} {\bibfnamefont {M.~H.}\
  \bibnamefont {Devoret}}, \ and\ \bibinfo {author} {\bibfnamefont {R.~J.}\
  \bibnamefont {Schoelkopf}},\ }\href {http://dx.doi.org/10.1038/nphys4143}
  {\bibfield  {journal} {\bibinfo  {journal} {Nat Phys}\ }\textbf {\bibinfo
  {volume} {advance online publication}},\  (\bibinfo {year}
  {2017})}\BibitemShut {NoStop}%
\bibitem [{\citenamefont {Riedrich-M{\"o}ller}\ \emph
  {et~al.}(2014)\citenamefont {Riedrich-M{\"o}ller}, \citenamefont {Arend},
  \citenamefont {Pauly}, \citenamefont {M{\"u}cklich}, \citenamefont {Fischer},
  \citenamefont {Gsell}, \citenamefont {Schreck},\ and\ \citenamefont
  {Becher}}]{Riedrich2014}%
  \BibitemOpen
  \bibfield  {author} {\bibinfo {author} {\bibfnamefont {J.}~\bibnamefont
  {Riedrich-M{\"o}ller}}, \bibinfo {author} {\bibfnamefont {C.}~\bibnamefont
  {Arend}}, \bibinfo {author} {\bibfnamefont {C.}~\bibnamefont {Pauly}},
  \bibinfo {author} {\bibfnamefont {F.}~\bibnamefont {M{\"u}cklich}}, \bibinfo
  {author} {\bibfnamefont {M.}~\bibnamefont {Fischer}}, \bibinfo {author}
  {\bibfnamefont {S.}~\bibnamefont {Gsell}}, \bibinfo {author} {\bibfnamefont
  {M.}~\bibnamefont {Schreck}}, \ and\ \bibinfo {author} {\bibfnamefont
  {C.}~\bibnamefont {Becher}},\ }\bibfield  {booktitle} {\emph {\bibinfo
  {booktitle} {Nano Letters}},\ }\href {\doibase 10.1021/nl502327b} {\bibfield
  {journal} {\bibinfo  {journal} {Nano Letters}\ }\textbf {\bibinfo {volume}
  {14}},\ \bibinfo {pages} {5281} (\bibinfo {year} {2014})}\BibitemShut
  {NoStop}%
\bibitem [{\citenamefont {Sipahigil}\ \emph {et~al.}(2016)\citenamefont
  {Sipahigil}, \citenamefont {Evans}, \citenamefont {Sukachev}, \citenamefont
  {Burek}, \citenamefont {Borregaard}, \citenamefont {Bhaskar}, \citenamefont
  {Nguyen}, \citenamefont {Pacheco}, \citenamefont {Atikian}, \citenamefont
  {Meuwly}, \citenamefont {Camacho}, \citenamefont {Jelezko}, \citenamefont
  {Bielejec}, \citenamefont {Park}, \citenamefont {Lon{\v c}ar},\ and\
  \citenamefont {Lukin}}]{Sipahigil2016}%
  \BibitemOpen
  \bibfield  {author} {\bibinfo {author} {\bibfnamefont {A.}~\bibnamefont
  {Sipahigil}}, \bibinfo {author} {\bibfnamefont {R.~E.}\ \bibnamefont
  {Evans}}, \bibinfo {author} {\bibfnamefont {D.~D.}\ \bibnamefont {Sukachev}},
  \bibinfo {author} {\bibfnamefont {M.~J.}\ \bibnamefont {Burek}}, \bibinfo
  {author} {\bibfnamefont {J.}~\bibnamefont {Borregaard}}, \bibinfo {author}
  {\bibfnamefont {M.~K.}\ \bibnamefont {Bhaskar}}, \bibinfo {author}
  {\bibfnamefont {C.~T.}\ \bibnamefont {Nguyen}}, \bibinfo {author}
  {\bibfnamefont {J.~L.}\ \bibnamefont {Pacheco}}, \bibinfo {author}
  {\bibfnamefont {H.~A.}\ \bibnamefont {Atikian}}, \bibinfo {author}
  {\bibfnamefont {C.}~\bibnamefont {Meuwly}}, \bibinfo {author} {\bibfnamefont
  {R.~M.}\ \bibnamefont {Camacho}}, \bibinfo {author} {\bibfnamefont
  {F.}~\bibnamefont {Jelezko}}, \bibinfo {author} {\bibfnamefont
  {E.}~\bibnamefont {Bielejec}}, \bibinfo {author} {\bibfnamefont
  {H.}~\bibnamefont {Park}}, \bibinfo {author} {\bibfnamefont {M.}~\bibnamefont
  {Lon{\v c}ar}}, \ and\ \bibinfo {author} {\bibfnamefont {M.~D.}\ \bibnamefont
  {Lukin}},\ }\href
  {http://science.sciencemag.org/content/354/6314/847.abstract} {\bibfield
  {journal} {\bibinfo  {journal} {Science}\ }\textbf {\bibinfo {volume}
  {354}},\ \bibinfo {pages} {847} (\bibinfo {year} {2016})}\BibitemShut
  {NoStop}%
\bibitem [{\citenamefont {Yoshie}\ \emph {et~al.}(2004)\citenamefont {Yoshie},
  \citenamefont {Scherer}, \citenamefont {Hendrickson}, \citenamefont
  {Khitrova}, \citenamefont {Gibbs}, \citenamefont {Rupper}, \citenamefont
  {Ell}, \citenamefont {Shchekin},\ and\ \citenamefont {Deppe}}]{Yoshie2004}%
  \BibitemOpen
  \bibfield  {author} {\bibinfo {author} {\bibfnamefont {T.}~\bibnamefont
  {Yoshie}}, \bibinfo {author} {\bibfnamefont {A.}~\bibnamefont {Scherer}},
  \bibinfo {author} {\bibfnamefont {J.}~\bibnamefont {Hendrickson}}, \bibinfo
  {author} {\bibfnamefont {G.}~\bibnamefont {Khitrova}}, \bibinfo {author}
  {\bibfnamefont {H.~M.}\ \bibnamefont {Gibbs}}, \bibinfo {author}
  {\bibfnamefont {G.}~\bibnamefont {Rupper}}, \bibinfo {author} {\bibfnamefont
  {C.}~\bibnamefont {Ell}}, \bibinfo {author} {\bibfnamefont {O.~B.}\
  \bibnamefont {Shchekin}}, \ and\ \bibinfo {author} {\bibfnamefont {D.~G.}\
  \bibnamefont {Deppe}},\ }\href {http://dx.doi.org/10.1038/nature03119}
  {\bibfield  {journal} {\bibinfo  {journal} {Nature}\ }\textbf {\bibinfo
  {volume} {432}},\ \bibinfo {pages} {200} (\bibinfo {year}
  {2004})}\BibitemShut {NoStop}%
\bibitem [{\citenamefont {Hennessy}\ \emph {et~al.}(2007)\citenamefont
  {Hennessy}, \citenamefont {Badolato}, \citenamefont {Winger}, \citenamefont
  {Gerace}, \citenamefont {Atature}, \citenamefont {Gulde}, \citenamefont
  {Falt}, \citenamefont {Hu},\ and\ \citenamefont {Imamoglu}}]{Hennessy2007}%
  \BibitemOpen
  \bibfield  {author} {\bibinfo {author} {\bibfnamefont {K.}~\bibnamefont
  {Hennessy}}, \bibinfo {author} {\bibfnamefont {A.}~\bibnamefont {Badolato}},
  \bibinfo {author} {\bibfnamefont {M.}~\bibnamefont {Winger}}, \bibinfo
  {author} {\bibfnamefont {D.}~\bibnamefont {Gerace}}, \bibinfo {author}
  {\bibfnamefont {M.}~\bibnamefont {Atature}}, \bibinfo {author} {\bibfnamefont
  {S.}~\bibnamefont {Gulde}}, \bibinfo {author} {\bibfnamefont
  {S.}~\bibnamefont {Falt}}, \bibinfo {author} {\bibfnamefont {E.~L.}\
  \bibnamefont {Hu}}, \ and\ \bibinfo {author} {\bibfnamefont {A.}~\bibnamefont
  {Imamoglu}},\ }\href {http://dx.doi.org/10.1038/nature05586} {\bibfield
  {journal} {\bibinfo  {journal} {Nature}\ }\textbf {\bibinfo {volume} {445}},\
  \bibinfo {pages} {896} (\bibinfo {year} {2007})}\BibitemShut {NoStop}%
\bibitem [{\citenamefont {Englund}\ \emph {et~al.}(2007)\citenamefont
  {Englund}, \citenamefont {Faraon}, \citenamefont {Fushman}, \citenamefont
  {Stoltz}, \citenamefont {Petroff},\ and\ \citenamefont
  {Vuckovic}}]{Englund2007}%
  \BibitemOpen
  \bibfield  {author} {\bibinfo {author} {\bibfnamefont {D.}~\bibnamefont
  {Englund}}, \bibinfo {author} {\bibfnamefont {A.}~\bibnamefont {Faraon}},
  \bibinfo {author} {\bibfnamefont {I.}~\bibnamefont {Fushman}}, \bibinfo
  {author} {\bibfnamefont {N.}~\bibnamefont {Stoltz}}, \bibinfo {author}
  {\bibfnamefont {P.}~\bibnamefont {Petroff}}, \ and\ \bibinfo {author}
  {\bibfnamefont {J.}~\bibnamefont {Vuckovic}},\ }\href
  {http://dx.doi.org/10.1038/nature06234} {\bibfield  {journal} {\bibinfo
  {journal} {Nature}\ }\textbf {\bibinfo {volume} {450}},\ \bibinfo {pages}
  {857} (\bibinfo {year} {2007})}\BibitemShut {NoStop}%
\bibitem [{\citenamefont {Grassl}\ \emph {et~al.}(1997)\citenamefont {Grassl},
  \citenamefont {Beth},\ and\ \citenamefont {Pellizzari}}]{Grassl1997}%
  \BibitemOpen
  \bibfield  {author} {\bibinfo {author} {\bibfnamefont {M.}~\bibnamefont
  {Grassl}}, \bibinfo {author} {\bibfnamefont {T.}~\bibnamefont {Beth}}, \ and\
  \bibinfo {author} {\bibfnamefont {T.}~\bibnamefont {Pellizzari}},\ }\href
  {https://link.aps.org/doi/10.1103/PhysRevA.56.33} {\bibfield  {journal}
  {\bibinfo  {journal} {Physical Review A}\ }\textbf {\bibinfo {volume} {56}},\
  \bibinfo {pages} {33} (\bibinfo {year} {1997})}\BibitemShut {NoStop}%
\bibitem [{\citenamefont {Ofek}\ \emph {et~al.}(2016)\citenamefont {Ofek},
  \citenamefont {Petrenko}, \citenamefont {Heeres}, \citenamefont {Reinhold},
  \citenamefont {Leghtas}, \citenamefont {Vlastakis}, \citenamefont {Liu},
  \citenamefont {Frunzio}, \citenamefont {Girvin}, \citenamefont {Jiang},
  \citenamefont {Mirrahimi}, \citenamefont {Devoret},\ and\ \citenamefont
  {Schoelkopf}}]{Ofek2016}%
  \BibitemOpen
  \bibfield  {author} {\bibinfo {author} {\bibfnamefont {N.}~\bibnamefont
  {Ofek}}, \bibinfo {author} {\bibfnamefont {A.}~\bibnamefont {Petrenko}},
  \bibinfo {author} {\bibfnamefont {R.}~\bibnamefont {Heeres}}, \bibinfo
  {author} {\bibfnamefont {P.}~\bibnamefont {Reinhold}}, \bibinfo {author}
  {\bibfnamefont {Z.}~\bibnamefont {Leghtas}}, \bibinfo {author} {\bibfnamefont
  {B.}~\bibnamefont {Vlastakis}}, \bibinfo {author} {\bibfnamefont
  {Y.}~\bibnamefont {Liu}}, \bibinfo {author} {\bibfnamefont {L.}~\bibnamefont
  {Frunzio}}, \bibinfo {author} {\bibfnamefont {S.~M.}\ \bibnamefont {Girvin}},
  \bibinfo {author} {\bibfnamefont {L.}~\bibnamefont {Jiang}}, \bibinfo
  {author} {\bibfnamefont {M.}~\bibnamefont {Mirrahimi}}, \bibinfo {author}
  {\bibfnamefont {M.~H.}\ \bibnamefont {Devoret}}, \ and\ \bibinfo {author}
  {\bibfnamefont {R.~J.}\ \bibnamefont {Schoelkopf}},\ }\href
  {http://dx.doi.org/10.1038/nature18949} {\bibfield  {journal} {\bibinfo
  {journal} {Nature}\ }\textbf {\bibinfo {volume} {536}},\ \bibinfo {pages}
  {441} (\bibinfo {year} {2016})}\BibitemShut {NoStop}%
\bibitem [{\citenamefont {Michael}\ \emph {et~al.}(2016)\citenamefont
  {Michael}, \citenamefont {Silveri}, \citenamefont {Brierley}, \citenamefont
  {Albert}, \citenamefont {Salmilehto}, \citenamefont {Jiang},\ and\
  \citenamefont {Girvin}}]{Michael2016}%
  \BibitemOpen
  \bibfield  {author} {\bibinfo {author} {\bibfnamefont {M.~H.}\ \bibnamefont
  {Michael}}, \bibinfo {author} {\bibfnamefont {M.}~\bibnamefont {Silveri}},
  \bibinfo {author} {\bibfnamefont {R.~T.}\ \bibnamefont {Brierley}}, \bibinfo
  {author} {\bibfnamefont {V.~V.}\ \bibnamefont {Albert}}, \bibinfo {author}
  {\bibfnamefont {J.}~\bibnamefont {Salmilehto}}, \bibinfo {author}
  {\bibfnamefont {L.}~\bibnamefont {Jiang}}, \ and\ \bibinfo {author}
  {\bibfnamefont {S.~M.}\ \bibnamefont {Girvin}},\ }\href
  {http://link.aps.org/doi/10.1103/PhysRevX.6.031006} {\bibfield  {journal}
  {\bibinfo  {journal} {Physical Review X}\ }\textbf {\bibinfo {volume} {6}},\
  \bibinfo {pages} {031006} (\bibinfo {year} {2016})}\BibitemShut {NoStop}%
\bibitem [{\citenamefont {Xiang}\ \emph {et~al.}(2017)\citenamefont {Xiang},
  \citenamefont {Zhang}, \citenamefont {Jiang},\ and\ \citenamefont
  {Rabl}}]{Xiang2017}%
  \BibitemOpen
  \bibfield  {author} {\bibinfo {author} {\bibfnamefont {Z.-L.}\ \bibnamefont
  {Xiang}}, \bibinfo {author} {\bibfnamefont {M.}~\bibnamefont {Zhang}},
  \bibinfo {author} {\bibfnamefont {L.}~\bibnamefont {Jiang}}, \ and\ \bibinfo
  {author} {\bibfnamefont {P.}~\bibnamefont {Rabl}},\ }\href
  {https://link.aps.org/doi/10.1103/PhysRevX.7.011035} {\bibfield  {journal}
  {\bibinfo  {journal} {Physical Review X}\ }\textbf {\bibinfo {volume} {7}},\
  \bibinfo {pages} {011035} (\bibinfo {year} {2017})}\BibitemShut {NoStop}%
\bibitem [{\citenamefont {Muschik}\ \emph {et~al.}(2011)\citenamefont
  {Muschik}, \citenamefont {Krauter}, \citenamefont {Hammerer},\ and\
  \citenamefont {Polzik}}]{Muschik2011}%
  \BibitemOpen
  \bibfield  {author} {\bibinfo {author} {\bibfnamefont {C.~A.}\ \bibnamefont
  {Muschik}}, \bibinfo {author} {\bibfnamefont {H.}~\bibnamefont {Krauter}},
  \bibinfo {author} {\bibfnamefont {K.}~\bibnamefont {Hammerer}}, \ and\
  \bibinfo {author} {\bibfnamefont {E.~S.}\ \bibnamefont {Polzik}},\ }\href
  {http://link.springer.com/article/10.1007/s11128-011-0294-2} {\bibfield
  {journal} {\bibinfo  {journal} {Quantum Information Processing}\ }\textbf
  {\bibinfo {volume} {10}},\ \bibinfo {pages} {839} (\bibinfo {year}
  {2011})}\BibitemShut {NoStop}%
\bibitem [{\citenamefont {Kraus}\ and\ \citenamefont
  {Cirac}(2004)}]{Kraus2004}%
  \BibitemOpen
  \bibfield  {author} {\bibinfo {author} {\bibfnamefont {B.}~\bibnamefont
  {Kraus}}\ and\ \bibinfo {author} {\bibfnamefont {J.~I.}\ \bibnamefont
  {Cirac}},\ }\href {http://link.aps.org/doi/10.1103/PhysRevLett.92.013602}
  {\bibfield  {journal} {\bibinfo  {journal} {Physical Review Letters}\
  }\textbf {\bibinfo {volume} {92}},\ \bibinfo {pages} {013602} (\bibinfo
  {year} {2004})}\BibitemShut {NoStop}%
\bibitem [{\citenamefont {Vitanov}\ and\ \citenamefont
  {Stenholm}(1997{\natexlab{b}})}]{Vitanov1997}%
  \BibitemOpen
  \bibfield  {author} {\bibinfo {author} {\bibfnamefont {N.~V.}\ \bibnamefont
  {Vitanov}}\ and\ \bibinfo {author} {\bibfnamefont {S.}~\bibnamefont
  {Stenholm}},\ }\href {\doibase 10.1103/PhysRevA.56.1463} {\bibfield
  {journal} {\bibinfo  {journal} {Phys. Rev. A}\ }\textbf {\bibinfo {volume}
  {56}},\ \bibinfo {pages} {1463} (\bibinfo {year}
  {1997}{\natexlab{b}})}\BibitemShut {NoStop}%
\bibitem [{\citenamefont {Yin}\ \emph {et~al.}(2013)\citenamefont {Yin},
  \citenamefont {Chen}, \citenamefont {Sank}, \citenamefont {O'Malley},
  \citenamefont {White}, \citenamefont {Barends}, \citenamefont {Kelly},
  \citenamefont {Lucero}, \citenamefont {Mariantoni}, \citenamefont {Megrant},
  \citenamefont {Neill}, \citenamefont {Vainsencher}, \citenamefont {Wenner},
  \citenamefont {Korotkov}, \citenamefont {Cleland},\ and\ \citenamefont
  {Martinis}}]{Yin2013}%
  \BibitemOpen
  \bibfield  {author} {\bibinfo {author} {\bibfnamefont {Y.}~\bibnamefont
  {Yin}}, \bibinfo {author} {\bibfnamefont {Y.}~\bibnamefont {Chen}}, \bibinfo
  {author} {\bibfnamefont {D.}~\bibnamefont {Sank}}, \bibinfo {author}
  {\bibfnamefont {P.~J.~J.}\ \bibnamefont {O'Malley}}, \bibinfo {author}
  {\bibfnamefont {T.~C.}\ \bibnamefont {White}}, \bibinfo {author}
  {\bibfnamefont {R.}~\bibnamefont {Barends}}, \bibinfo {author} {\bibfnamefont
  {J.}~\bibnamefont {Kelly}}, \bibinfo {author} {\bibfnamefont
  {E.}~\bibnamefont {Lucero}}, \bibinfo {author} {\bibfnamefont
  {M.}~\bibnamefont {Mariantoni}}, \bibinfo {author} {\bibfnamefont
  {A.}~\bibnamefont {Megrant}}, \bibinfo {author} {\bibfnamefont
  {C.}~\bibnamefont {Neill}}, \bibinfo {author} {\bibfnamefont
  {A.}~\bibnamefont {Vainsencher}}, \bibinfo {author} {\bibfnamefont
  {J.}~\bibnamefont {Wenner}}, \bibinfo {author} {\bibfnamefont {A.~N.}\
  \bibnamefont {Korotkov}}, \bibinfo {author} {\bibfnamefont {A.~N.}\
  \bibnamefont {Cleland}}, \ and\ \bibinfo {author} {\bibfnamefont {J.~M.}\
  \bibnamefont {Martinis}},\ }\href
  {https://link.aps.org/doi/10.1103/PhysRevLett.110.107001} {\bibfield
  {journal} {\bibinfo  {journal} {Physical Review Letters}\ }\textbf {\bibinfo
  {volume} {110}},\ \bibinfo {pages} {107001} (\bibinfo {year}
  {2013})}\BibitemShut {NoStop}%
\bibitem [{\citenamefont {Sato}\ \emph {et~al.}(2012)\citenamefont {Sato},
  \citenamefont {Tanaka}, \citenamefont {Upham}, \citenamefont {Takahashi},
  \citenamefont {Asano},\ and\ \citenamefont {Noda}}]{Sato2012}%
  \BibitemOpen
  \bibfield  {author} {\bibinfo {author} {\bibfnamefont {Y.}~\bibnamefont
  {Sato}}, \bibinfo {author} {\bibfnamefont {Y.}~\bibnamefont {Tanaka}},
  \bibinfo {author} {\bibfnamefont {J.}~\bibnamefont {Upham}}, \bibinfo
  {author} {\bibfnamefont {Y.}~\bibnamefont {Takahashi}}, \bibinfo {author}
  {\bibfnamefont {T.}~\bibnamefont {Asano}}, \ and\ \bibinfo {author}
  {\bibfnamefont {S.}~\bibnamefont {Noda}},\ }\href
  {http://dx.doi.org/10.1038/nphoton.2011.286} {\bibfield  {journal} {\bibinfo
  {journal} {Nat Photon}\ }\textbf {\bibinfo {volume} {6}},\ \bibinfo {pages}
  {56} (\bibinfo {year} {2012})}\BibitemShut {NoStop}%
\bibitem [{\citenamefont {Ley}\ and\ \citenamefont {Loudon}(1987)}]{Ley1987}%
  \BibitemOpen
  \bibfield  {author} {\bibinfo {author} {\bibfnamefont {M.}~\bibnamefont
  {Ley}}\ and\ \bibinfo {author} {\bibfnamefont {R.}~\bibnamefont {Loudon}},\
  }\href@noop {} {\bibfield  {journal} {\bibinfo  {journal} {Journal of Modern
  Optics}\ }\textbf {\bibinfo {volume} {34}},\ \bibinfo {pages} {227} (\bibinfo
  {year} {1987})}\BibitemShut {NoStop}%
\end{thebibliography}%

\end{document}